# Optimizing XML querying using type-based document projection

Véronique Benzaken
LRI, Université Paris-Sud 11, Orsay, France
Giuseppe Castagna
PPS, Université Paris 7, Paris, France
Dario Colazzo
LRI, Université Paris-Sud 11, Orsay, France

Kim Nguyễn
LRI, Université Paris-Sud 11, Orsay, France

XML data projection (or pruning) is a natural optimization for main memory query engines: given a query $Q$ over a document $D$, the subtrees of $D$ that are not necessary to evaluate $Q$ are pruned, thus producing a smaller document $D'$; the query $Q$ is then executed on $D'$, hence avoiding to allocate and process nodes that will never be reached by $Q$. In this article, we propose a new approach, based on types, that greatly improves current solutions. Besides providing comparable or greater precision and far lesser pruning overhead, our solution —unlike current approaches— takes into account backward axes, predicates, and can be applied to multiple queries rather than just to single ones. A side contribution is a new type system for XPath able to handle backward axes. The soundness of our approach is formally proved. Furthermore, we prove that the approach is also complete (i.e., yields the best possible type-driven pruning) for a relevant class of queries and Schemas. We further validate our approach using the XMark and XPathMark benchmarks and show that pruning not only improves the main memory query engine's performances (as expected) but also those of state of the art native XML databases.

## 1 Introduction

Main-memory XML query engines are often the primary choice for applications that do not wish or cannot afford to build secondary storage indexes or load a database before query processing. One of the main optimisation techniques recently adopted in this context is XML data projection (or pruning) [27, 13].

The basic idea behind document projection is very simple and powerful at the same time. Given a query $Q$ over a document $D$, sub-trees of $D$ that are not necessary to evaluate $Q$ are pruned, thus yielding a smaller document $D'$. Then $Q$ is executed over $D'$, hence avoiding to allocate and process nodes that will never be reached by navigational specifications in $Q$. This ensures that evaluation over $D'$ is equivalent to and more efficient than the evaluation over $D$.

As shown in [27, 13], XML navigation specifications expressed in queries tend to be very selective, especially in terms of document structure. Therefore, pruning may yield significant improvements both in terms of execution time and in terms of memory usage:

as a matter of facts, for main-memory XML query engines, very large documents can not be queried without pruning.

## 1.1 State of the art

Marian and Siméon [27] propose that the actual data-needs of an XQuery query $Q$ (that is, the part of data that is necessary to the execution of the query) is determined by statically extracting all paths in $Q$. These paths are then applied to $D$ at load time, in a SAX-event based fashion, in order to prune unneeded parts of data. The technique is powerful since: $(i)$ it applies to most of XQuery *core*, $(ii)$ it can be applied to a set of queries over the same document, and $(iii)$ it does not require any *a priori* knowledge of the structure of $D$. However, this technique suffers some limitations. First, the document loader-pruner can manage neither *backward axes* nor path expressions with predicates (sometimes called "qualifiers") which, especially the latter, can contain precious information to optimise pruning. Second, the advantage described in point $(iii)$ becomes a big drawback when "`//`" occur in paths since, in that case, the technique does not behave efficiently in terms of loading time and pruning precision (hence, memory allocation). Indeed, when a `//` is present in a projection path, the pruning process requires to visit all descendants of a node in order to decide whether the node contains a useful descendant. What is worst is that pruning time tends to be quite high and it drastically increases (together with memory consumption) when the number of `//` augments in the pruning path-set. As a matter of fact, in this technique, pruning corresponds to computing a further query whose time and memory occupation may be comparable to those required to compute the original query. In particular, in this technique every occurrence of `//` may yield a full exploration of the tree (e.g., see in [27] the test for the XMark [32] query Q7 which only contains three `//` steps and for which just computing the pruning takes longer than executing the query on the original document). Therefore, pruning execution overhead and its high memory footprint may jeopardise the gains obtained by using the pruned document. Third and finally, as we explain in Section 7, the precision of pruning drastically degrades (down to being nullified) for queries containing the XPath expressions `descendant`::*node*`[`*cond*`]`, which are very useful and used in practice.

Bressan *et al.* [13] introduce a different and quite precise XML pruning technique for a subset of XQuery FLWR expressions. The technique is based on the *a priori* knowledge of a data-guide for $D$. The document $D$ is first matched against an abstract representation of $Q$. Pruning is then performed at run time, it is very precise, and, thanks to the use of some indexes over the data-guide, it ensures good improvements in terms of query execution time. However, the technique is one-query oriented—in the sense that it cannot be applied to multiple queries—, it does not handle XPath predicates, and cannot handle backward axes (recall that the encodings of [31] are defined for XPath, and no extension to XQuery-like languages is known). Also, the approach requires the construction and management of the data-guide and of adequate indexes.

Motivated by efficient XML stream processing, Green *et al.* [23] introduced a framework for discarding sequences of SAX events in an XML data stream. Although their approach allows them to prune an input stream with respect to sets of queries, the language they handle is restricted to forward linear XPath expressions (that is, XPath expressions with only `child` and `descendant` axes and without predicates).

## 1.2 Our contribution

In this article, we present a new pruning approach that is applicable in the presence of typed XML data. This is often the case, as most applications require that data are valid with respect to some external schema (e.g., DTD [19] or XML Schema [37]).

Our technique combines the advantages of the previously mentioned works while relaxing their limitations. Unlike [27, 13, 23], our approach accounts for backward axes, per-forms a fine-grained analysis of predicates, allows (unlike [13]) for dealing with bunches of queries, and (unlike [27]) cannot be jeopardised by pruning overhead. Our solution provides in all cases comparable or greater precision than the other approaches, while it requires always negligible or no pruning overhead. Moreover, contrary to [27, 13], our approach is formally proved to be *sound* (i.e., pruning does not alter the result of queries) and, furthermore, we can also prove it to be *complete* (i.e., it produces the best possible type-driven pruning) for a substantial class of queries and DTDs.

For the sake of presentation we introduce our framework in three steps. In the first step, we consider a simplified version of XPath, we dub XPath$^\ell$, which includes only upward/downward axes and unnested disjunctive predicates. We define for XPath$^\ell$ a static analysis that determines a set of type names, a *type projector*, that is then used to prune the document(s). One of the particular features of this approach is that our pruning algorithm is characterised by a constant (and low) memory consumption and by an execution time linear in the size of the document to prune. More precisely, a pruning based on type projectors is equivalent to a single buffer-less one-pass traversal of the parsed document (it simply discards elements not generated by any of the names in the projector). So if embedded in query processors, pruning can be executed during parsing and/or validation and brings no overhead at all, while if used as an external tool it requires a time always smaller than or equal to the time used to parse the queried document. Soundness and (partial) completeness results for the static analysis are stated.

The second step consists of extending the analysis to the whole XPath (more precisely, to XPath 1.0), that is, we need to show how to deal with missing axes and with general predicates as defined in the XPath specification. This is done by associating to each XPath query $Q$ a XPath$^\ell$ query $P$ that soundly approximates $Q$, in the sense that the projector inferred for $P$ by the static analysis developed at the first step is also a sound projector for $Q$.

The final step of our process is to extend the approach to XQuery (hence, to XPath 2.0). This is obtained in the same way as done in [27], by defining a path extraction algorithm. Our path extraction algorithm improves and extends in several aspects (in particular, in terms of extracted paths' selectivity) the one of [27]. It also computes the XPath$^\ell$ approximation of the extracted paths so that the static analysis of the first step can be directly applied to them.

We prove some important closure properties that guarantee that type projections can always be performed at load time during the validation process, and this without any overhead. In particular for XML documents typed with DTDs or XML Schemas the document can be pruned in streaming.

We gauged and validated our approach by testing it both on the XPathMark [21] and on the XMark [32] benchmarks. The result of this validation confirmed what was expected: thanks to the handling of backward axes and of predicates the precision of our pruning is in general noticeably higher than that of current approaches; the pruning time is linear in the size of the queried document and has a very low memory footprint; the time of the static analysis is always negligible (lower than half a second on the hardware we used for our benchmarks described in Section 9) even for complex queries and DTDs. But benchmarks also brought unexpected (and quite pleasant) results. In particular, they showed that type-based pruning brings benefits that go beyond those of the reduced size of the pruned document: by excluding a whole set of data structures (those whose type names are not included in the type projector), the pruning may drastically reduce the resources that must be allocated at run-time by the query processor. For instance, our benchmarks show that for several XMark and XPathMark queries our pruning yields a document whose size is two thirds of the size of the original document, but the query can then be processed using three times less memory than when processed on the original document. This is a very important gain, especially for DOM-based processors, or memory sensitive processors. Not only our approach is relevant in the case of main memory query engines such as Saxon but it is also shown to be useful for native query engines as efficient as MonetDB [12]. Even in the latter case our experiments demonstrate the relevance of type projection as a complementary optimisation technique. Indeed, this not totally surprising as type projection can be thought of as a way of defining clustering policies in the same line as what was done in the context of object-oriented databases [8, 4, 7]. Clustering and indexing are well-known complementary tools used in the context of query optimisation.

As an aside we want to stress that our technique relies on the definition of a new type system for XPath able to handle backward axes which, alone on its own, constitutes a contribution of this work. In particular the precision of type inference for backward axes goes beyond what is proposed in the XQuery Static Semantic recommendation ([18]).

Finally, we presented a preliminary version of this work at the VLDB 2006 conference [5]. The work in this article, besides including full proofs and having been cleaned up, improves and extends the work in [5] in several important aspects. First and foremost we generalized the definition of type projectors by using as projectors sets of production rules (as opposed to the sets of non-terminals used in [5]) of regular tree grammars (as opposed to the DTDs used in [5]). This generalization was far from being straightforward. In particular, we had to prove the applicability of our technique to the more general framework under consideration (cf. Section 3.2). However the result is worth the effort since the advantages of this generalization are twofold. On the one hand using regular tree grammars allows us to compute type projectors for every kind of XML schema formalism we are aware of as, for instance, DTDs, XMLSchemas, ℂDuce and XDuce types, Relax-Core and TREX schemas. On the other hand, inferring grammar production rules rather than grammar non-terminals allows us to compute context-aware and, thus, more precise projectors. More precisely, the new type projectors introduced in this work can prune a subtree not only based on its tag (as it was done in [5]), but also on any structural condition expressible by a regular tree language. So for instance our pruning process may decide to prune just one of two trees generated by the same non-terminal, because they appear in different contexts (in [5] either both trees were pruned or they were both preserved). Therefore these new projectors are both more general and can perform much finer-grained pruning. Second, although we develop the theory of type projection for a simplified data-model and restricted forms of XPath expressions, we thoroughly detail how to tackle many of the peculiarities of the XML and XPath specifications [34, 35], including the handling of attributes, the presence of absolute axes in XPath predicates or a wide range of predefined XPath functions (all absent in [5]). The path language we formally study extends the one in [5] with top-level unions of paths, predicate conjunctions (“`and`”) and arbitrarily nested predicates (our previous work formally treated only non-nested predicates and resorted to an approximation in the case of nested predicates). Third, we provide an extensive list of experiments showing the overall benefits of type projection for a wide range of queries and query engines. These experiments supersede the early benchmarks realised in [5] and show that despite the advances in XML query technologies in the recent years, our static analysis can significantly improve the performances (both in time and memory consumption) of many different XML query engines.

## 1.3 Plan of the article

The article is organised as follows. Section 2 introduces basic definitions and notations: data model, types, validation. Section 3 presents type projectors, type-based projection, and several theoretical (closure) properties. In Section 4 we define XPath$^\ell$ and its semantics, and formally describe how general XPath predicates can be soundly approximated in it. In Section 5 we present our type projectors inference algorithm for XPath$^\ell$ and state its formal properties. In Section 6 we extend our approach to full XPath and in Section 7 to XQuery. In Section 8 we discuss how to apply our technique to other typing policies as well as to un-typed documents. Section 9 presents our implementation and reports the results of our benchmarks. We finally conclude in Section 10 by presenting the perspectives of this work. Last, for the sake of clarity, all the proofs for the stated results are given in Appendix A.

# 2 Notations

## 2.1 Data Model

For the sake of concision and clarity we present our solution for a simplified version of the XQuery data model where we do not consider node attributes. However, attributes are fully supported in our implementation through a trivial encoding, documented in Section 6. An instance of the XQuery data model can then be generated by the following grammar:

**Definition 2.1 (Data model)**

$$\begin{array}{llll} \textbf{\textit{Tree}} & t & ::= & s_{\mathbf{i}} \mid l_{\mathbf{i}}[f] \\ \textbf{\textit{Forest}} & f & ::= & () \mid f,f \mid t \quad \square \end{array}$$

Essentially, an instance of the XQuery data model is an ordered sequence of labelled ordered *trees* (ranged over by $t$). That is, an ordered *forest* (ranged over by $f$), where each node has a unique *identifier* (ranged over by $\mathbf{i}$) and where $()$ denotes the empty forest. Tree nodes are labelled by *element tags* (ranged over by $l$) while, without loss of generality, we consider only leaves that are text nodes (that is, strings, ranged over by $s$) or empty trees (that is, elements that label the empty forest).

We define a complete partial order $\preceq$ on forests (and thus on trees) by relating a forest with the forests obtained either by adding or by deleting subforests:

**Definition 2.2 (Projection ($\preceq$))** *Given two forests $f$ and $f'$ we say that $f'$ is a* projection *of $f$, noted as $f' \preceq f$, if $f'$ is obtained by replacing some subforests of $f$ by the empty forest. In other terms $\preceq$ is the smallest pre-congruence on forests that contains $() \preceq f$ for all $f$.*$\square$

We also define a notion of good formation, with respect to the data model given in Definition 2.1:

**Definition 2.3 (Good formation)** *A forest is* well formed *if every identifier $\mathbf{i}$ occurs in it at most once. Given a well-formed forest $f$ and an identifier $\mathbf{i}$ occurring in it, we denote by $f@\mathbf{i}$ the unique subtree $t$ of $f$ such that $t = s_{\mathbf{i}}$ or $t = l_{\mathbf{i}}[f']$. The set of identifiers of a forest $f$ is then defined as $\textbf{\textit{Ids}}(f) = \{\mathbf{i} \mid \exists\, t.\ f@\mathbf{i} = t\}$* $\square$

Henceforth we will consider only well-formed forests and confound the notions of a node with that of the identifier of the node.

**Definition 2.4 (Root id)** *Let $t$ be a tree. If $t = s_{\mathbf{i}}$ or $t = l_{\mathbf{i}}[f]$, we define $\textbf{\textit{RootId}}(t) = \mathbf{i}$.*

## 2.2 Types and validation

In this work, we present our approach for an abstract model of types, namely *regular tree grammars*. It is well known that regular tree grammars encompass most of the features of well established and standardized schema specifications such as DTDs, XMLSchemas, RelaxNG definitions, XDuce and ℂDuce's regular expression types. This is for instance documented in [29], from where we borrow the definition of regular tree grammar:

**Definition 2.5 (Regular tree grammar)** *A regular tree grammar is a pair $(\mathscr{S}, E)$ where $\mathscr{S}$ is a set of distinguished names (actually, non-terminal meta-variables) and $E$ is a set of productions rules of the form $\{X_1 \to R_1, \ldots, X_n \to R_n\}$ such that:*

1. *each $R_i$ is either the terminal String—denoting string content—, or the terminal Any—denoting any tree—, or $l[\ r\ ]$ where $l$ ranges over valid element names and $r$ is a regular expression on the non-terminal symbols $X_1, \ldots, X_n$, that is:*

$$\begin{array}{lllll} \textbf{\textit{RegExp}} & r & ::= & \varepsilon & \textit{(empty sequence)} \\ & & \mid & r\,r & \textit{(sequence)} \\ & & \mid & r|r & \textit{(alternation)} \\ & & \mid & r* & \textit{(Kleene star)} \\ & & \mid & X_i & \textit{(non-terminal)} \end{array}$$

*(henceforth, we use $r+$ for $r\ r*$ and $r?$ for $\varepsilon|r$);*

2. *$\mathscr{S} \subseteq \{X_1, \ldots, X_n\}$ is the set of start symbols;*

3. *for any two production rules with the same left hand side $X_i \to l[r]$ and $X_i \to l'[r']$, we have $l \neq l'$;* □

The intuition is that a regular tree grammar describes (i.e., it "types") a set of trees of the data-model. Notice that the left-hand sides of the rules in $E$ do not need to be pairwise distinct. Indeed, production rules such as $X \to R_1, X \to R_2$ are necessary if one wants to encode complex schemas. Furthermore, given a regular tree grammar, it is always possible to equivalently rewrite it so that condition 3 holds: if there are two rules $X_i \to l[r]$ and $X_i \to l[r']$ then they can be merged into a single rule, $X_i \to l[r|r']$.

**Definition 2.6 (Names of a regular expression)** *Given a regular expression $r$ we denote by $\textbf{\textit{Names}}(r)$ the set of non-terminals occurring in it, namely:*

$$\begin{array}{lcl} \textbf{\textit{Names}}(\varepsilon) & = & \varnothing \\ \textbf{\textit{Names}}(r_1\ r_2) & = & \textbf{\textit{Names}}(r_1) \cup \textbf{\textit{Names}}(r_2) \\ \textbf{\textit{Names}}(r_1 \mid r_2) & = & \textbf{\textit{Names}}(r_1) \cup \textbf{\textit{Names}}(r_2) \\ \textbf{\textit{Names}}(r*) & = & \textbf{\textit{Names}}(r) \\ \textbf{\textit{Names}}(X) & = & \{X\} \end{array} \qquad \square$$

By extension, given a set $E = \{X_0 \to R_0, \ldots, X_n \to R_n\}$, we define

$$\textbf{\textit{Names}}(E) = \bigcup_{i \in 0..n} \textbf{\textit{Names}}(R_i)$$

and $\textbf{\textit{Dn}}(E)$ for the set of names defined in $E$ (that is, $\{X_1 \ldots X_n\}$). While for all types $(\mathscr{S}, E)$ we have $\textbf{\textit{Names}}(E) = \textbf{\textit{Dn}}(E)$, we handle incomplete sets of rules during the formalisation of the algorithms, whence the need for both notations. We also say that $r$ is a regular expression over $(\mathscr{S}, E)$, if $r$ is a regular expression over names in $\textbf{\textit{Dn}}(E)$. We will denote by $\mathfrak{L}(r)$ the language recognized by the regular expression $r$. We will use $W, X, Y, Z$ to range over *names*. We use Greek letters to range over sets of rules. As $(\mathscr{S}, E)$ represents a regular tree grammar we shall use $\pi$ to stress that the set of rules is a *type projector* [cf. Definition 3.1] and $\kappa$ and $\tau$ to stress that the set is used as a context or as a type, respectively [cf. Section 5.1]. Last, we shall use $S$ to range over sets of (node) identifiers.

We illustrate the syntax of regular tree grammars with the following example:

**Example 2.7 (A regular tree grammar for the bibliography DTD)** The well known bibliography DTD (taken from the XML Query use cases [15]) can be written as a regular tree grammar $(\{X\}, E)$, with unique start symbol $X$ and the following set $E$ of rules:

$$\begin{array}{rcl} X & \to & \texttt{bib}[Book*] \\ Book & \to & \texttt{book}[Title, (Author+ \mid Editor+), Publ] \\ Title & \to & \texttt{title}[String] \\ Author & \to & \texttt{author}[String] \\ Editor & \to & \texttt{editor}[String] \\ Publ & \to & \texttt{publisher}[String] \end{array}$$

This regular tree grammar "types" all XML documents (i.e., trees of the data model) that are rooted in a `bib` element, that contains a possibly empty list of `book` elements, each one containing a list starting with a `title` element containing a string, followed by a non-empty homogeneous list formed either by `author` elements or `editor` elements, and ended by a publisher element.

The concept of typing an XML document by a regular tree grammar is formalized by the notion of *validity* defined as follows:

**Definition 2.8 (Valid Trees)** *A tree t is* valid *with respect to a type* $(\mathscr{S}, E)$*, if there exists a mapping (interpretation)* $\mathfrak{I}$ *from* $\textbf{\textit{Ids}}(t)$ *to* $\textbf{\textit{Names}}(E)$ *such that:*

1. $\mathfrak{I}(\textbf{\textit{RootId}}(t)) \in \mathscr{S}$
2. *for each* $\mathbf{i}$ *in* $\textbf{\textit{Ids}}(t)$*, if* $t@\mathbf{i} = s_{\mathbf{i}}$ *then either* $\mathfrak{I}(\mathbf{i}) \to Any \in E$ *or* $\mathfrak{I}(\mathbf{i}) \to String \in E$
3. *for each* $\mathbf{i}$ *in* $\textbf{\textit{Ids}}(t)$*, if* $t@\mathbf{i} = l_{\mathbf{i}}[t_1, ..., t_n]$*, then either we have* $\mathfrak{I}(\mathbf{i}) \to Any \in E$ *or we have* $\mathfrak{I}(\mathbf{i}) \to l[r] \in E$ *and* $\mathfrak{I}(\textbf{\textit{RootId}}(t_1)), \ldots, \mathfrak{I}(\textbf{\textit{RootId}}(t_n)) \in \mathfrak{L}(r)$*.*

*In this case we say that t is* $\mathfrak{I}$*-valid with respect to* $(\mathscr{S}, E)$ *and write* $t \in_{\mathfrak{I}} (\mathscr{S}, E)$ *to indicate it.* □

For instance the following tree (in which we omit the node identifiers)

```
bib[
  book[
     title["Divina Commedia"],
     author["Dante"],
     publisher["Ludovico Dolce"]
  ]
]
```

is valid with respect to the type $(\{X\}, E)$ defined in Example 2.7. There exist various techniques and algorithms to validate XML trees against regular tree grammars (for instance, by using tree automata: cf. Algorithm 4.4 in [29]). Note however that due to our use of regular tree grammars, the interpretation $\mathfrak{I}$ might not be unique and that a validating algorithm will generate—for a document $t$ and a type $(\mathscr{S}, E)$—*one* possible interpretation such that $t$ is $\mathfrak{I}$-valid with respect to $(\mathscr{S}, E)$.

Given a tree $t$ valid with respect to a type $(\mathscr{S}, E)$, we can use subsets of $E$ to project that tree. Essentially, from the rules in $E$ we compute another set of "simpler" rules which denotes only the nodes to be kept. In order to define formally this notion we need to define the reachability relation $\Rightarrow_E$, that we introduce below together with several other definitions that we use later in the article.

**Definition 2.9 (Forward Reachability)** *Given a type* $(\mathscr{S}, E)$ *and* $Z \in \textbf{\textit{Dn}}(E)$*, we write* $Z \Rightarrow_E Y$ *if and only if* $Z \to R \in E$ *and* $Y \in \textbf{\textit{Names}}(R)$*. We use* $\Rightarrow_E^+$ *and* $\Rightarrow_E^*$ *to denote respectively the transitive closure and the transitive and reflexive closure of* $\Rightarrow_E$*.* □

Strings of names are called *chains* and ranged over by $c$, $c_i$, $c'$,... In particular we use $\textbf{\textit{Chains}}_{(X,E)}(Y)$ to denote the set of all chains rooted at $Y$, and defined as $\{Y X_1 \ldots X_n \mid Y \Rightarrow_E X_1 \Rightarrow_E \ldots \Rightarrow_E X_n, n \geq 0\}$. We use $\textbf{\textit{Names}}(c)$ to denote the set of all names occurring in a chain $c$.

At the beginning of the section we defined the projection of a forest as a forest obtained by replacing some subforests by the empty tree. Here we define an analogous concept for types, called *erasure* according to which a type is obtained from another by replacing some non-terminals by the empty regular expression.

**Definition 2.10 (Erasure of a regular expression)** *Let r be a regular expression and N a set of names. We define the erasure of r with respect to N and we note* $r|_N$ *the regular expression inductively defined as:*

$$\begin{array}{rcll} \varepsilon|_N &=& \varepsilon & \\ (r_1\, r_2)|_N &=& r_1|_N\, r_2|_N & \\ (r_1 \mid r_2)|_N &=& r_1|_N \mid r_2|_N & \\ (r*)|_N &=& (r|_N)* & \\ X|_N &=& X & \text{if } X \in N \\ X|_N &=& \varepsilon & \text{if } X \notin N \ \square \end{array}$$

We generalize this notion to production rules of a grammar:

**Definition 2.11 (Erasure of a rule)** *Let $X \to R$ be a production rule, and $N$ a set of names. We define the* erasure *of $X \to R$ with respect to $N$, noted $(X \to R)|_N$, as:*

$$\begin{array}{rcl} (X \to l[\, r\,])|_N &=& X \to l[\, r|_N\,] \\ (X \to String)|_N &=& X \to String \\ (X \to Any)|_N &=& X \to Any \qquad \square \end{array}$$

We recall that *String* and *Any* are special *terminals* denoting string and any content, respectively. We can finally define the erasure of a grammar:

**Definition 2.12 (Erasure of a tree grammar)** *Let $(\mathscr{S},E)$ and $(\mathscr{S}',E')$ be tree grammars. We say that $(\mathscr{S}',E')$ is an* erasure *of $(\mathscr{S},E)$, noted $(\mathscr{S}',E') <: (\mathscr{S},E)$, if and only if all the following conditions hold*

1. *$\mathscr{S}' \subseteq \mathscr{S}$;*
2. *if $X \to String \in E'$, then $X \to String \in E$;*
3. *if $X \to Any \in E'$, then $X \to Any \in E$;*
4. *for all rules $X \to l[\, r'\,] \in E'$, there exists a rule $X \to l[\, r\,] \in E$ such that $r' = r|_N$ for some $N \subseteq \textbf{Names}(r)$.* □

In summary, an erasure of a type grammar erases some rules and some non-terminals in the regular expressions.

Finally, we conclude this section by recalling few definitions taken from [29] that will be useful for establishing further results.

**Definition 2.13 (Competing non-terminals)** *Let $(\mathscr{S},E)$ be a tree grammar. Let $A,B \in \textbf{Names}(E)$ be two non-terminals such that $A \neq B$. $A$ and $B$ are* competing *if and only if there exist $A \to l[\, r\,] \in E$ and $B \to l'[\, r'\,] \in E$ such that $l = l'$.* □

The definitions that are actually interesting are those of *local* and *single-type tree* grammars, which can by defined in terms of competing non-terminals:

**Definition 2.14 (local tree grammar)** *A regular tree grammar $(\mathscr{S},E)$ is a* local *tree grammar if and only if:*

- $|\mathscr{S}| \leq 1$
- *$E$ does not contain any competing non-terminals*
- *For all $Y \in \textbf{Names}(E)$ there is exactly one rule in $E$ whose left-hand-side is $Y$.*

**Definition 2.15 (single-type tree grammar)** *A regular tree grammar $(\mathscr{S},E)$ is a* single-type tree *grammar if and only if:*

1. *For all $X \to l[\, r\,] \in E$, if $A$, $B$ in $\textbf{Names}(r)$ and $A \neq B$, then $A$ and $B$ are not competing and*
2. *no pair of distinct non-terminals in $\mathscr{S}$ is competing.* □

The interest of these two definitions is that—as shown in [29]—they characterize the structural constraints that can be expressed by the two most widespread schema formalisms, namely DTDs (which roughly correspond to local tree grammars) and XML Schemas (which are, essentially, single-type tree grammars).

# 3 Type projectors

In this section we shall first precisely define what type projectors are and then establish some useful closure results on type projectors.

## 3.1 Definition

**Definition 3.1 (Type Projector)** *Given a type* $(\mathscr{S}, E)$*, a (possibly empty) set of rules* $\pi \subseteq E$ *is a* type projector *if and only if* $(\mathscr{S} \cap \textbf{\textit{Names}}(\pi), \pi)$ *is a regular tree grammar erasure of* $(\mathscr{S}, E)$*.*

A type projector is thus a set of rules obtained from the type $(\mathscr{S}, E)$ by erasing some rules and some non-terminals in the remaining rules.

A type projector for a given type describes a particular pruning for XML documents of that type, that is, a *type driven projection*:

**Definition 3.2 (Type Driven Projections)** *Let* $\pi$ *be a type projector for* $(\mathscr{S}, E)$ *and* $t$ *a forest such that* $t \in_{\mathfrak{I}} (\mathscr{S}, E)$*. The* $\pi$*-projection of* $t$*, noted as* $t \backslash_{\mathfrak{I}} \pi$*, is defined as follows:*

$$
\begin{array}{rcll}
()\backslash_{\mathfrak{I}}\pi & = & () & \\
s_{\mathbf{i}}\backslash_{\mathfrak{I}}\pi & = & s_{\mathbf{i}} & \textit{if } \mathfrak{I}(\mathbf{i}) \to \textit{String} \in \pi \textit{ or } \mathfrak{I}(\mathbf{i}) \to \textit{Any} \in \pi \\
s_{\mathbf{i}}\backslash_{\mathfrak{I}}\pi & = & () & \textit{if } \mathfrak{I}(\mathbf{i}) \to \textit{String} \notin \pi \textit{ and } \mathfrak{I}(\mathbf{i}) \to \textit{Any} \notin \pi \\
l_{\mathbf{i}}[f]\backslash_{\mathfrak{I}}\pi & = & l_{\mathbf{i}}[f] & \textit{if } \mathfrak{I}(\mathbf{i}) \to \textit{Any} \in \pi \\
l_{\mathbf{i}}[f]\backslash_{\mathfrak{I}}\pi & = & l_{\mathbf{i}}[f\backslash_{\mathfrak{I}}\pi] & \textit{if } \mathfrak{I}(\mathbf{i}) \to l[r] \in \pi \textit{ and } \mathfrak{I}(\mathbf{i}) \to \textit{Any} \notin \pi \\
l_{\mathbf{i}}[f]\backslash_{\mathfrak{I}}\pi & = & () & \textit{if } \mathfrak{I}(\mathbf{i}) \to l[r] \notin \pi \textit{ and } \mathfrak{I}(\mathbf{i}) \to \textit{Any} \notin \pi \\
(f, f')\backslash_{\mathfrak{I}}\pi & = & (f\backslash_{\mathfrak{I}}\pi), (f'\backslash_{\mathfrak{I}}\pi) &
\end{array}
$$

□

In words, pruning erases (by replacing it by an empty forest) every node that cannot be derived by a rule in $\pi$.

**Lemma 3.3** *Let* $\pi$ *be a type projector for* $(\mathscr{S}, E)$*. Then for every tree* $t \in_{\mathfrak{I}} (\mathscr{S}, E)$ *it holds* $(t \backslash_{\mathfrak{I}} \pi) \preceq t$*.*

As the knowledgeable reader might have already noticed, validation (as in Definition 2.8) and type-driven projection are quite similar. Given a tree $t$ and a type $(\mathscr{S}, E)$, a validation algorithm builds an interpretation $\mathfrak{I}$ of $t$ with respect to that type. More precisely, the algorithm associates to each node of $t$ a non terminal of $E$. If it cannot find at least one, validation fails and the tree is not valid with respect to $(\mathscr{S}, E)$. A type-driven projecting algorithm works *exactly in the same way* but when a node cannot be associated with a name it is simply discarded together with the associated subtree. Projecting a document can be seen as an instance of validation. This observation is precious to determine the complexity of type-driven projection, given a particular type projector $\pi$. If $\pi$ is a local tree grammar or a single-type tree grammar (that is, a DTD or an XML-Schema, see [29]) then projection can be performed in streaming. On the contrary, if $\pi$ ends-up being an general tree grammar, then projection might require in the worst case to keep the whole tree in memory (see our remark at the end of Section 3.2, for how to use type projection in this particular setting).

## 3.2 Closure properties

The fact that *if* a type projector is a DTD or an XMLSchema, then type-driven projection can be done efficiently is already a good thing. However, we can show a stronger result: a type projector inherits the properties of the type it was deduced from. This is important since in practice if someone chooses to use DTDs or XML-Schemas to specify their documents, the projection process should not be more expensive than the validation process.

Indeed, a nice property of the *erasure* of a type is that it preserves both the local tree and single type property. In other words, the erasure of a DTD remains a DTD and the erasure of an XML-Schema remains an XML-Schema. This is stated by the two following lemmas.

**Lemma 3.4 (Erasure preserves locality)** *Let* $(\mathscr{S},E)$ *be a local tree grammar and* $(\mathscr{S}',E')$ *a regular tree grammar. If* $(\mathscr{S}',E') <: (\mathscr{S},E)$ *then* $(\mathscr{S}',E')$ *is a local tree grammar.*

**Lemma 3.5 (Erasure preserves single-typedness)** *Let* $(\mathscr{S},E)$ *be a single-type tree grammar and* $(\mathscr{S}',E')$ *a regular tree grammar. If* $(\mathscr{S}',E') <: (\mathscr{S},E)$ *then* $(\mathscr{S}',E')$ *is a single-type tree grammar.*

Last but not least, we show that if two projectors coming from the same type enjoy the local (resp. single-type) property, then their union is also local (resp. single-type). This property of type projectors is instrumental to our approach. Indeed, given a set of paths, we will compute a type projector for it by taking the union of all the type projectors of the individual paths. However, if taking the union of type projectors caused the loss of local or single-type properties, the interest of extending our approach to sets of paths (and thus to XQuery or to bunches of queries) would be quite limited.

The key observation here is that, while in general local and single-type tree grammars are not closed under union, two type-projectors that *come from the same type* share a common structure and therefore are not completely independent one from the other. In particular we can show that the union of two type projectors for the same type cannot introduce competing non-terminals in the resulting type projector. In terms of term-rewrite systems, we can say that the union of two type projectors does not introduce a critical pair (of non-terminals).

**Lemma 3.6 (Union closure of local type projectors)** *Let* $(\mathscr{S},E)$ *be a local tree grammar. Let* $(\mathscr{S}_1,E_1)$ *and* $(\mathscr{S}_2,E_2)$ *be two tree grammars such that* $(\mathscr{S}_1,E_1) <: (\mathscr{S},E)$ *and* $(\mathscr{S}_2,E_2) <: (\mathscr{S},E)$*. Then* $(\mathscr{S}_1 \cup \mathscr{S}_2, E_1 \cup E_2)$ *is a local tree grammar.*

**Lemma 3.7 (Union closure of single-type type projectors)** *Let* $(\mathscr{S},E)$ *be a single-type tree grammar. Let* $(\mathscr{S}_1,E_1)$ *and* $(\mathscr{S}_2,E_2)$ *be two tree grammars such that* $(\mathscr{S}_1,E_1) <: (\mathscr{S},E)$ *and* $(\mathscr{S}_2,E_2) <: (\mathscr{S},E)$*. Then* $(\mathscr{S}_1 \cup \mathscr{S}_2, E_1 \cup E_2)$ *is a single-type tree grammar.*

To conclude this presentation of the formal properties of type-projectors we could note that a third category of deterministic regular tree grammars, namely *restrained-competition tree grammars* (see [29]), is not closed under erasure. Therefore, for this kind of schemas (and associated type-projectors) pruning might require a full buffering of the input document. However this is only of mild importance since, to the best of our knowledge, no well-known schema specification relies on it. All the other schema specifications that we are aware of (XDuce and ℂDuce regular expression types, TREX, Relax Core,...) possess the full expressive power of regular tree languages which, as it is well-known, are closed under erasure and union (see for instance [16]). This means that type driven projection proposed here can be applied to these kinds of schemas, as well. However, projection remains as expensive as validation which, for these particular schemas, implies that the whole document might need to be loaded into memory to actually decide which subtrees must be pruned. Practical solutions to this problem are discussed in Section 8.2

# 4 XPath$^\ell$

In XPath, queries are expressed by defining a path of steps separated by "/". For instance,

$$Q = \texttt{/descendant::}author\texttt{/child::text[self::node} = \text{"}Dante\text{"}\texttt{]/}$$
$$\texttt{parent::}book\texttt{/child::}title$$

is the query that returns all titles of books whose author is "*Dante*". First, the navigational part instructs to descend to all text nodes whose parent is an author (by following the path /descendant::*author*/child::*text*), then the predicate selects those nodes that are the string "*Dante*" (with the test self::node ="*Dante*"), and finally the navigation ascends to the *book* element and descends to the *title*.

The inference rules we define in Section 5 do not work directly on queries such as $Q$. The rules are defined for a subset of XPath that we dub XPath$^\ell$ and introduce in this section. XPath$^\ell$ (for XPath $\ell$*ight*) includes forward and backward axes and a special kind of predicates. In order to statically analyse $Q$ (or any other XPath query that is not in XPath$^\ell$), we will find a XPath$^\ell$ query that approximates $Q$ soundly with respect to the pruning inferred (Section 6), and use it to deduce the pruning for $Q$. Of course, these approximations, as well as those we introduce later on, will only be used to determine the pruning: the pruned document will be queried by the original query. Therefore we are going to proceed as follows. In this section we define XPath$^\ell$, which is roughly equivalent to the structural subset of positive XPath Core, without absolute paths. Then in Section 5, we introduce our type and type-projector inference algorithms, which work on XPath$^\ell$ queries. To complete the treatment of XPath we show in Section 6 how to compute a sound approximation of a query $Q$ with respect to type projection. In other words, given a (full) XPath query $Q$, we will compute an XPath$^\ell$ query $Q'$ such that the type projector inferred from $Q'$ preserves the semantics of $Q$.

Let us start with defining XPath$^\ell$ paths and their semantics. From now on, "path" refers to an XPath$^\ell$ query as defined hereafter unless otherwise specified.

**Definition 4.1 (XPath$^\ell$ path)** *An XPath$^\ell$ path is a term inductively generated by the following grammar:*

$$\begin{array}{rcl}
Path & ::= & Step \mid Path/Path \mid Path \,\|\, Path \\
Step & ::= & Axis::Test \mid Axis::Test[Cond] \\
Axis & ::= & \texttt{self} \mid \texttt{child} \mid \texttt{descendant} \mid \texttt{parent} \mid \texttt{ancestor} \\
Test & ::= & tag \mid \texttt{node} \mid \texttt{text} \\
Cond & ::= & Cond \; \textit{or} \; Cond \mid Cond \; \textit{and} \; Cond \mid Path
\end{array}$$

*where tag is a meta-variable ranging over element tags.* □

As customary, "and" takes precedence over "or" and the path delimiter "/" takes precedence over the top-level union "‖". We will also use the (possibly indexed) meta-variables $P$ and $C$ to range over paths and conditions, respectively.

The formal semantics of paths is inductively defined on the productions of Definition 4.1. First, we formalise *Test* filtering as the set of nodes that satisfy a given test. Then *Axis* selection as the set of nodes reachable from some context nodes by following some Axis. Finally, we combine these notions to define the semantics of paths. The definitions comply with the semantics of XPath 1.0 (see [35]).

**Definition 4.2 (Node test semantics)** *Given a tree $t$ and a set of nodes $S \subseteq \textbf{\textit{Ids}}(t)$ we define:*

$$\begin{array}{rcl}
S::_t l & = & S \cap \{\mathbf{i} \in \textbf{\textit{Ids}}(t) \mid t@\mathbf{i} = l_{\mathbf{i}}[f]\} \\
S::_t\texttt{node} & = & S \\
S::_t\texttt{text} & = & S \cap \{\mathbf{i} \in \textbf{\textit{Ids}}(t) \mid \exists s,\ t@\mathbf{i} = s_{\mathbf{i}}\}
\end{array} \quad \square$$

**Definition 4.3 (Axes selection)** *Given a tree $t$ and a set of nodes $S \subseteq \textbf{\textit{Ids}}(t)$ (called context nodes), we define $[\![Axis]\!]_t(S)$ as the set of nodes obtained by applying Step to each node in $S$:*

$$\begin{array}{rcl}
[\![\texttt{self}]\!]_t(S) & = & S \\
[\![\texttt{child}]\!]_t(S) & = & \bigcup_{\mathbf{i} \in S}\{\mathbf{i}' \mid (\mathbf{i},\mathbf{i}') \in \textbf{\textit{Edg}}(t)\} \\
[\![\texttt{parent}]\!]_t(S) & = & \bigcup_{\mathbf{i} \in S}\{\mathbf{i}' \mid (\mathbf{i}',\mathbf{i}) \in \textbf{\textit{Edg}}(t)\} \\
[\![\texttt{descendant}]\!]_t(S) & = & \bigcup_{\mathbf{i} \in S}\{\mathbf{i}' \mid (\mathbf{i},\mathbf{i}') \in \textbf{\textit{Edg}}(t)^+\} \\
[\![\texttt{ancestor}]\!]_t(S) & = & \bigcup_{\mathbf{i} \in S}\{\mathbf{i}' \mid (\mathbf{i}',\mathbf{i}) \in \textbf{\textit{Edg}}(t)^+\}
\end{array}$$

*where* $\boldsymbol{Edg}(t)$ *is the* edge *relation of* $t$*, that is*

$$\boldsymbol{Edg}(t) = \{(\mathbf{i},\mathbf{i}') \mid t@\mathbf{i} = l_{\mathbf{i}}[f,t',f'] \wedge \boldsymbol{RootId}(t') = \mathbf{i}'\}$$

*and* $\boldsymbol{Edg}(t)^+$ *denotes its transitive closure.* □

Since predicates may contain paths and conversely, path and predicate semantics are mutually defined.

**Definition 4.4 (XPath$^\ell$ semantics)** *Given* $t$*, a set* $S \subset \boldsymbol{Ids}(t)$ *and a path* $P$*, we define the evaluation of path* $P$ *over the set of context nodes* $S$ *as the function* $[\![P]\!]_t(S)$ *defined as:*

$$\begin{array}{rcl}
[\![Axis::Test]\!]_t(S) & = & ([\![Axis]\!]_t(S))::_t Test \\
[\![Axis::Test[C]]\!]_t(S) & = & ([\![Axis]\!]_t(S))::_t Test \cap \{\mathbf{i} \in S \mid \boldsymbol{Check}_t[C](\mathbf{i})\} \\
[\![Path_1/Path_2]\!]_t(S) & = & [\![Path_2]\!]_t([\![Path_1]\!]_t(S)) \\
[\![Path_1 \,|\, Path_2]\!]_t(S) & = & [\![Path_2]\!]_t(S) \cup [\![Path_1]\!]_t(S)
\end{array}$$

*where* $\boldsymbol{Check}_[_](_)$ *is the Boolean function defined as:*

$$\begin{array}{rcl}
\boldsymbol{Check}_t[Path](\mathbf{i}) & = & [\![Path]\!]_t(\{\mathbf{i}\}) \neq \varnothing \\
\boldsymbol{Check}_t[C_1 \; or \; C_2](\mathbf{i}) & = & \boldsymbol{Check}_t[C_1](\mathbf{i}) \vee \boldsymbol{Check}_t[C_2](\mathbf{i}) \\
\boldsymbol{Check}_t[C_1 \; and \; C_2](\mathbf{i}) & = & \boldsymbol{Check}_t[C_1](\mathbf{i}) \wedge \boldsymbol{Check}_t[C_2](\mathbf{i}) \;\; \square
\end{array}$$

It is easy to see that the last definition is well founded since terms are inductively generated by the productions of the grammar in Definition 4.1.

Although the paths in XPath$^\ell$ are quite simple, the definition of their static analysis can result quite complex: the simultaneous presence in a single step of axes, tests, and predicates can cause a case explosion in the definition of the analysis. This is not a problem for a static analyzer, but it is a problem for a human reader. Fortunately, for the human reader, XPath$^\ell$ paths can be further simplified and transformed into equivalent normal forms in which all non trivial axes, tests and predicates are distributed over different steps. The idea is then to normalize paths before passing them to the static analyzer so that the definition of the latter can result much simpler. The normal forms that will be analyzed by the static analysis of Section 5 are defined as follows

**Definition 4.5 (Single step normal form)** *Let* $P$ *be an XPath$^\ell$ query. The* single step normal form *of* $P$*, noted* $\boldsymbol{Snf}(P)$*, is defined as:*

$$\begin{array}{rcl}
\boldsymbol{Snf}(Axis :: \texttt{node}) & = & Axis :: \texttt{node} \\
\boldsymbol{Snf}(\texttt{self} :: Test) & = & \texttt{self} :: Test \\
\boldsymbol{Snf}(\texttt{self}::\texttt{node[}C\texttt{]}) & = & \texttt{self}::\texttt{node[}\boldsymbol{Dnf}(C)\texttt{]} \\
\boldsymbol{Snf}(Axis :: Test) & = & Axis :: \texttt{node/self}::Test \quad (\text{if } Axis \neq \texttt{self} \wedge Test \neq \texttt{node}) \\
\boldsymbol{Snf}(Axis :: Test\texttt{[}C\texttt{]}) & = & Axis :: \texttt{node/self}::Test\texttt{/self}::\texttt{node[}\boldsymbol{Dnf}(C)\texttt{]} \\
 & & \qquad (\text{if } Axis \neq \texttt{self} \wedge Test \neq \texttt{node}) \\
\boldsymbol{Snf}(P_1/P_2) & = & \boldsymbol{Snf}(P_1)/\boldsymbol{Snf}(P_2)
\end{array}$$

*where* $\boldsymbol{Dnf}(C)$ *is a disjunctive normal form of the Boolean proposition* $C$ *(whose atoms are paths).* □

It is clear from this definition that $P$ and $\boldsymbol{Snf}(P)$ have the same semantics. Indeed, if we have a step

$$Axis :: Test\texttt{[}C\texttt{]}$$

then its single step normal form

$$Axis :: \texttt{node/self} :: Test/\texttt{self} :: \texttt{node}[\,\textbf{\textit{Dnf}}(C)\,]$$

only makes the order of node selection more explicit[1]. For a given set of context nodes $S$, we first select all nodes that can be reached by the *Axis*. Then we keep only nodes that match the *Test*. Finally we further refine the result by filtering the nodes that satisfy the predicate $C$, put in disjunctive normal form. The disjunctive normal form of our predicates is obtained by distributing the "`or`" over the "`and`" yielding a formula of the form $\texttt{Or}_i\,\texttt{And}_j\,P_{ij}$ (where $P_{ij}$ are paths). Although this may yield an exponential blow-up of the formula, remember that we introduce this simplification only to provide a concise and human readable presentation of the static type inference algorithms. An actual implementation can work directly on the abstract syntax tree of the formula without resorting to this transformation.

# 5 Static Analysis

In this section we define deduction rules to statically infer from a XPath$^\ell$ path $P$ and a type $(\mathscr{S},E)$ a type-projector for any input document valid with respect to $(\mathscr{S},E)$. We show that the analysis is sound, and that it enjoys completeness for a large class of queries when $E$ is a $*$-guarded and non-recursive local tree grammar (see Definition 5.7 later on). Soundness means that executing the query on the original document and on the document pruned by the inferred projector yields the same result. Completeness means that the analysis infers the best correct projector, that is, that if we take a type projector smaller (i.e., more selective) than the inferred one, then there exists a document validating $(\mathscr{S},E)$ for which the result of the two executions is not the same. When the conditions on schemas or on queries are relaxed, then the analysis is still sound but it may be not complete. Nevertheless, as we will formally illustrate, it is still very precise.

In order to define our static type-projector inference algorithm we proceed in two steps.

1. Given a path $P$ and a regular expression grammar $(\mathscr{S},E)$ the rough idea is to use a type system to associate $P$ with the set of all trees that may appear in the result of applying $P$ to a document validating $(\mathscr{S},E)$. In order to achieve a great precision, we then "type" $P$ by the set of all rules of $E$ that *validate* any tree in the result.[2] This is done in Section 5.1.

2. Next, we use the type system defined in the previous point to define inference of type projectors. In particular we use the cases in which the previous type system returns an empty set of rules to determine the points in which pruning must be performed. This is done in Section 5.2.

## 5.1 Type inference

Given a path *Path* and a schema $(\mathscr{S},E)$ we want to find a subset of rules in $E$ that can generate all the trees that can occur in the result of *Path* when applied to a tree validating $(\mathscr{S},E)$. Formally, we want to infer a set $\tau \subseteq E$ such that

$$\forall t \in_{\mathfrak{I}} (\mathscr{S},E),\;\; \mathfrak{I}([\![Path]\!]_t(\textbf{\textit{RootId}}(t))) \subseteq \textbf{\textit{Dn}}(\tau) \tag{1}$$

The equation above states the soundness of the analysis. In words it says that if we take any tree $t$ valid for $(\mathscr{S},E)$ and we apply the path *Path* to it, then the type $\tau$ inferred in the type systems defines every symbol interpreting a node in the result. As usual, soundness alone is not interesting since there always are sets that trivially satisfy it (notably, the set of all rules in $E$). What we aim at is an analysis that is as selective as possible, that is, an analysis that is precise enough to guarantee, on a large class of types and for a large class of queries, that whenever the path semantics is empty over all possible instances of the input type, then the inferred type $\tau$ is empty, as well:

$$\forall t \in_{\mathfrak{I}} (\mathscr{S},E),\ \ \mathfrak{I}([\![Path]\!]_t(\textbf{\textit{RootId}}(t)))=\varnothing \implies \tau=\varnothing \tag{2}$$

[1] As an aside, note that this kind of equivalence does not hold for full XPath because of the `position()` function. Indeed, `descendant`$::a$`[position() = 1]` and `descendant::node/self`$::a$`[position() = 1]` do not return, in general, the same result. The former returns the first "a"-node in pre-order while the latter returns all the "a"-nodes of the document.

[2] This yields a finer-grained analysis since different rules may generate the same tree but in different contexts.

(the converse is a consequence of (1)). In other terms we want that if there does not exist any instance of the type that matches the path, then the path is typed by the empty set.

The precision described by (2) will then be used during the inference of type-projectors to discard elements that are useless in the evaluation of *Path*, that is, all the sub-trees of the original document that cannot be matched by *Path*.

We start by inferring types for single-step paths without predicates.

**Definition 5.1 (Unconditional Single Step Typing)** *The type of an unconditional single-step query Axis::Test for the schema* $(\mathscr{S},E)$ *is given by:*

$$\mathbf{T}_E(\mathbf{A}_E(\mathscr{S},Axis),Test)$$

*where axes are typed as:*

$$\begin{array}{rcl}
\mathbf{A}_E(\tau,\texttt{ancestor}) & = & \bigcup_{Y\in \textbf{\textit{Dn}}(\tau)}\{Z\to R\in E \mid Z\Rightarrow_E^+ Y\}\\
\mathbf{A}_E(\tau,\texttt{child}) & = & \bigcup_{Y\in \textbf{\textit{Dn}}(\tau)}\{Z\to R\in E \mid Y\Rightarrow_E Z\}\\
\mathbf{A}_E(\tau,\texttt{parent}) & = & \bigcup_{Y\in \textbf{\textit{Dn}}(\tau)}\{Z\to R\in E \mid Z\Rightarrow_E Y\}\\
\mathbf{A}_E(\tau,\texttt{descendant}) & = & \bigcup_{Y\in \textbf{\textit{Dn}}(\tau)}\{Z\to R\in E \mid Y\Rightarrow_E^+ Z\}\\
\mathbf{A}_E(\tau,\texttt{self}) & = & \tau
\end{array}$$

*and tests are typed as:*

$$\begin{array}{rcl}
\mathbf{T}_E(\tau,\texttt{node}) & = & \tau\\
\mathbf{T}_E(\tau,a) & = & \{Y\to R \mid Y\in \textbf{\textit{Dn}}(\tau), R{=}a[R'] \text{ or } R{=}Any\}\\
\mathbf{T}_E(\tau,\texttt{text}) & = & \{Y\to R \mid Y\in \textbf{\textit{Dn}}(\tau), R{=}String \text{ or } R{=}Any\}
\end{array}$$

□

This definition introduces two typing operators, one for axes, $\mathbf{A}_(_,_)$, and one for tests, $\mathbf{T}_(_,_)$. Firstly, $\mathbf{A}_E(\tau,Axis)$ returns all the rules that can be reached from names in $\tau$ following *Axis*. If *Axis* is `self`, `child` or `descendant`, our definition coincide with the static semantics of XQuery and XPath, as defined by Draper *et al.* in [18]. However, Draper *et al.*'s static semantics is much less precise than ours in case of backward axis. Translated in our formalism, the type of `parent` and `ancestor` for any $\tau$ would be $\{X\to Any\}$ for some name $X$.[3]

Secondly, $\mathbf{T}_E(\tau,test)$ restrict the rules in $\tau$ to only rules which type elements compatible with *test*.

The soundness of this definition, that is, the property stated by Formula (1) is given by the following lemma.

**Lemma 5.2** *Let t be a tree* $\mathfrak{I}$*-valid with respect to the schema* $(\mathscr{S},E)$*. For every* $S\subseteq \textbf{\textit{Ids}}(t)$ *and type* $\tau$*, if* $\mathfrak{I}(S)\subseteq \textbf{\textit{Dn}}(\tau)$*, then*

1. $\mathfrak{I}([\![Axis]\!]_t(S))\subseteq \textbf{\textit{Dn}}(\mathbf{A}_E(\tau,Axis))$

[3] More precisely `parent`::*test* and `ancestor`::*test* return the union type `element()|document()` independently of *test*; where `element()` is the type of any element node and `document()` is the type of the document node which we don't consider in our data-model.

2. $\Im(S::_t Test) \subseteq \textbf{\textit{Dn}}(\mathbf{T}_E(\tau, Test))$

It is easy to check that the property stated by Formula (1) is a direct consequence of Definition 4.4 and the composition of the two properties of the lemma above.

The presence of upward axes makes the typing of composed paths much more difficult. To ensure precision, that is the property stated by Formula (2), we have to be careful in dealing with types in which an element may occur in the content of different elements. The naive solution consisting of inferring a type for composed paths by composing the functions we just defined for single steps, works only in the absence of upward axes. This can be illustrated by an example. Consider the following grammar rooted at $X$:

$$X \to a[Y],\ X \to b[Z],\ Y \to c[\,],\ Z \to d[\,]$$

and observe that $X$ yields two possible definitions. Now consider the path

$$\texttt{self}::a/\texttt{child}::c/\texttt{parent}::\texttt{node}$$

applied to documents of the above type, then the precise type that this path should have is $\{X \to a[Y]\}$. However if we naively iterate Definition 5.1, we obtain at the first step $\{X \to a[Y]\}$, onto which we apply `child::c`, which yields $\{Y \to c[\,]\}$ to which we finally apply `parent::node` which gives us $\{X \to a[Y],\ X \to b[Z]\}$, which is sound but imprecise. This is due to the fact that the single step typing blindly selects all rules associated with a name which can generate $Y$, here all the rules associated with $X$.

To solve this problem we introduce particular sets of rules, called *contexts*, to be updated at each step and containing rules already encountered in previous steps. We then use them to refine type inference for upward axes. In the previous example, when typing the first two steps we build a *context*

$$\{X \to a[Y],\ Y \to c[\,]\}$$

indicating that for the moment the two rules are the only ones visited by the traversal. Then, we use Definition 5.1 to type `parent::node` thus obtaining $\{X \to a[Y],\ X \to b[Z]\}$, as before, but this time we intersect it with the context thus obtaining the precise answer $\{X \to a[Y]\}$. We now formalize this idea:

**Definition 5.3 (Type inference)** *Let $(\mathscr{S}, E)$ be a type and P an XPath$^\ell$ query in single step normal form. Let $\tau$ and $\kappa$ be two sets of rules of E. If the deduction system in Figure 1 deduces for a path P the judgment,*

$$(\tau, \kappa) \vdash_E P : (\tau', \kappa')$$

*then we say that P has type $(\textbf{\textit{Dn}}(\tau'), \tau')$.* □

The idea underlying the judgments of the definition is that if the system proves $(\tau, \kappa) \vdash_E P : (\tau', \kappa')$, then from an input set of rules $\tau$ and an input context $\kappa$ the application of $P$ returns an output set of rules $\tau'$ and an updated context $\kappa'$. In other terms $\tau$ is (the production part of) a type that approximates the current nodes, $\kappa$ is the context that was visited to type them, $\tau'$ is (the production part of) a type that approximates the set of nodes reachable from the current ones by following $P$, and $\kappa'$ is the additional context visited to reach them. In Figure 1 environments—that is pairs of sets of rules—are ranged over by $\Sigma$ for concision. $\Sigma$ being a pair $(\tau, \kappa)$, we use $\Sigma_{\mathsf{typ}}$ to denote its first projection (i.e., the "type" component $\tau$) and $\Sigma_{\mathsf{ctx}}$ to denote the second projection (i.e., the "context" component $\kappa$).

**Definition 5.4 (Environment well-formedness)** *Let $(\tau, \kappa)$ be an environment and E a set of rules. If $\tau \subseteq E$ and $\kappa \subseteq \tau \cup \mathbf{A}_E(\tau, \texttt{ancestor})$, then we say that $(\tau, \kappa)$ is* well formed *with respect to E.* □

In other words, a context is well-formed if it contains only rules from which the names in $\textbf{\textit{Dn}}(\tau)$ are reachable. We say that a judgment $\Sigma \vdash_E P : \Sigma'$ is well formed if both $\Sigma$ and $\Sigma'$

Primitive Single Step

$$\textbf{(down-axis)}\ \frac{Axis \in \{\texttt{self}, \texttt{child}, \texttt{descendant}\}}{\Sigma \vdash_E Axis :: \texttt{node} : (\mathbf{A}_E(\Sigma_{\mathsf{typ}}, Axis)\ ,\ \Sigma_{\mathsf{ctx}} \cup \mathbf{A}_E(\Sigma_{\mathsf{typ}}, Axis))}$$

$$\textbf{(up-axis)}\ \frac{Axis \in \{\texttt{parent}, \texttt{ancestor}\}}{\Sigma \vdash_E Axis :: \texttt{node} : (\mathbf{A}_E(\Sigma_{\mathsf{typ}}, Axis)) \cap \Sigma_\kappa\ ,\ \mathbf{A}_E(\Sigma_{\mathsf{ctx}}, Axis) \cap \Sigma_{\mathsf{ctx}}}$$

$$\textbf{(test)}\ \frac{Test \neq \texttt{node}}{\Sigma \vdash_E \texttt{self} :: Test : (\tau, (\Sigma_{\mathsf{ctx}} \cap \mathbf{A}_E(\tau, \texttt{ancestor})) \cup \tau)}\ (*)$$

(*) where $\tau = T_E(\Sigma_{\mathsf{typ}}, Test)$

$$\textbf{(predicate)}\ \frac{(\forall X_i \to R_i\ \in \Sigma_{\mathsf{typ}}) \qquad (\{X_i \to R_i\}, \Sigma_{\mathsf{ctx}}) \vdash_E P_{jk} : \Sigma^{ijk}}{\Sigma \vdash_E \texttt{self} :: \texttt{node}[\underset{j}{\texttt{Or}}\ \underset{k}{\texttt{And}} P_{jk}] : (\ \tau\ ,\ (\Sigma_{\mathsf{ctx}} \cap \mathbf{A}_E(\tau, \texttt{ancestor})) \cup \tau)}\ (**)$$

(**) where $\tau = \{X_i \to R_i \mid \bigcup_i \bigcap_j \Sigma^{ijk}_{\mathsf{typ}} \neq \varnothing\}$

Composed paths

$$\textbf{(sequence)}\ \frac{\Sigma \vdash_E Step : \Sigma'' \qquad \Sigma'' \vdash_E Path : \Sigma'}{\Sigma \vdash_E Step/Path : \Sigma'}$$

$$\textbf{(union)}\ \frac{\Sigma \vdash_E Path_1 : (\tau_1, \kappa_1) \qquad \Sigma \vdash_E Path_2 : (\tau_2, \kappa_2)}{\Sigma \vdash_E Path_1 | Path_2 : (\tau_1 \cup \tau_2, \kappa_1 \cup \kappa_2)}$$

Figure 1: Inference rules for XPath$^\ell$ queries

are well formed with respect to $E$. We can remark that the rules in Figure 1 are syntax directed —at most one rule apply for a given judgment— and they preserve context well-formedness.

The rules are relatively simple to understand. The first two rules implement our main idea: when we follow an axis *Axis*, we compute the type by $\mathbf{A}_E(\Sigma_{\mathsf{typ}}, Axis)$; if the axis is a downward one, then we add this type to the current context, otherwise if the axis is an upward one, then we intersect it with the current context (both for the type part and for the context part). The rule for **(test)** is slightly more difficult since it discards from the current set of rules those that do not satisfy the test: the type is computed by $\mathbf{T}_E(\Sigma_{\mathsf{typ}}, Test)$, while the context is obtained by removing all the rules that were in there just because they generated one of the discarded nodes; to do so it generates (the type of) all ancestors of the nodes satisfying the test, and intersects them with the current context. The fourth rule, **(predicate)**, is the most difficult one. Recall that we work with single step normal forms and, therefore, that the predicates are Boolean formulas over paths in disjunctive normal form; the type $\tau$ is obtained by discarding from $\Sigma_{\mathsf{typ}}$ all rules for which the predicate never holds; thus for each $X_i \to R_i$ in $\Sigma_{\mathsf{typ}}$ we compute the type of all the paths $P_{jk}$ in the predicate, and keep in $\tau$ only rules for which at least one path may yield a non-empty result; the context is then computed as in the deduction rule **(test)**, by discarding from the context all rules that generated only rules discarded from $\Sigma_{\mathsf{typ}}$. The deduction rule **(sequence)** chains the result of one step to the following one. Lastly, the rule **(union)** handles the top-level union operator "$|$".

Let us illustrate how the algorithm works on an example. Consider the grammar with rules

$$\{A \to a[B|C|E], B \to b[D], C \to c[\,], D \to d[], E \to b[\,]\}$$

and rooted in $A$, and the path

$$\texttt{child}::\texttt{node}/\texttt{self}::b/\texttt{self}::\texttt{node}[\texttt{child}::\texttt{node}/\texttt{self}::d]$$

Notice that the path above is nothing but the single step normal form of

$$\texttt{child}::b[\texttt{child}::d]$$

We start from an initial environment

$$\Sigma = (\{A \to a[B|C|E]\}, \{A \to a[B|C|E]\})$$

in which both the context and the type component contain all the rules whose left hand side is a root of the grammar (in this case we have just one rule). The first step is typed with the **(down-axis)** rule, giving the result $\Sigma^1$ where

$$\Sigma^1_{\mathsf{typ}} = \{B \to b[D], C \to c[\,], E \to b[\,]\}$$

and

$$\Sigma^1_{\mathsf{ctx}} = \{A \to a[B|C|E], B \to b[D], C \to c[\,], E \to b[\,]\}$$

The second step is typed by applying the rule **(test)**, which returns $\Sigma^2$:

$$\Sigma^2_{\mathsf{typ}} = \{B \to b[D], E \to b[\,]\}$$

and more interestingly, the context

$$\Sigma^2_{\mathsf{ctx}} = \{A \to a[B|C|E], B \to b[D], E \to b[\,]\}$$

Indeed, the intersection of $\Sigma^1_{\mathsf{ctx}}$ with the name generated by the ancestors of $B$, namely $A$ yields exactly $\{A \to a[B|C|E]\}$ to which we add the result of the current step:

$$\{B \to b[D], E \to b[\,]\}$$

As we said, this intersection ensures that we only keep in the context rules from which we can derive the current type. In this example, the rules for $C$ which was introduced by the wildcard step `child::node` is removed by the typing of the more restrictive step `self::`$b$. The third step is typed by the **(predicate)** rule. Intuitively, this rule types independently the path `child::node/self::`$d$ and keeps in the result only the input rules for which the path yields a non-empty result which, in this case, is the rule for $B$:

$$\Sigma^3_{\mathsf{typ}} = \{B \to b[D]\}$$

As before, the context is purged from rules that do not generate the current type:

$$\Sigma^3_{\mathsf{ctx}} = \{A \to a[B|C|E], B \to b[D]\}$$

Before proving the main theorems of type inference, namely soundness and completeness, let us first show that the inference rules of Figure 1 form indeed an algorithm.

**Lemma 5.5 (Termination of type inference)** *Let $(\mathscr{S}, E)$ be a type, $P$ a path, and $\Sigma$ and $\Sigma'$ two environments. If there is a derivation for the judgment $\Sigma \vdash_E P : \Sigma'$, then this derivation is unique and finite.*

We can now proceed to prove the soundness of the type system.

**Theorem 5.6 (Soundness of type inference)** *Let $(\mathscr{S}, E)$ be a type and $P$ a path. Let $E_0 = \{X \to R \mid X \to R \in E, X \in \mathscr{S}\}$. If $(E_0, E_0) \vdash_E P : (\tau, \kappa)$ then:*

$$\boldsymbol{Dn}(\tau) \supseteq \bigcup_{t \in \mathfrak{I}(\mathscr{S},E)} \mathfrak{I}([\![P]\!]_t(\boldsymbol{RootId}(t)))$$

The type system is sound. It is also complete for a particular class of schemas, namely local tree grammars that are $*$-guarded, non-recursive, and parent-unambiguous. Intuitively, a type is $*$-guarded when every union occurring in its productions is guarded by $*$ (or by +), it is non recursive if the depth of all documents validating it is bounded, while it is parent-unambiguous if no rule types both the parent and a strict ancestor of the parent of another name. Formally, we have the following definition:

**Definition 5.7** *Let $(\mathscr{S}, E)$ be a local tree grammar.*

1. *$E$ is* $*$-guarded *if for each $Y \to l[r]$ in $E$, the regular expression is a product $r = r_1 \cdots r_n$ and whenever $r_i$ contains a union, then $r_i = (r')*$;*
2. *$E$ is* non-recursive *if it is never the case that $Y \Rightarrow_E^+ Y$, for any name $Y \in \boldsymbol{Names}(E)$;*
3. *$E$ is* parent-unambiguous *if for all chains $c$ and names $Y, Z$ such that $cYZ \in \boldsymbol{Chains}_{(\mathscr{S},E)}(X)$ the implication*

   $$cYc'Z \in \boldsymbol{Chains}_{(\mathscr{S},E)}(X) \quad \Longrightarrow \quad c' = \varepsilon$$

   *holds ($\varepsilon$ denotes the empty chain).* □

Non-recursiveness and $*$-guardedness are properties enjoyed by a large number of commonly used DTDs. As an example, the reader can consider the DTDs of the XML Query Use Cases [15]: among the ten DTDs defined in the Use Cases, seven are both non-recursive and $*$-guarded, one is only $*$-guarded, one is only non-recursive, and just one does not satisfy either property. Furthermore our personal experience is that most of the DTDs available on the web are $*$-guarded. Concerning the parent-unambiguous property, although DTDs satisfying this property are less frequent (five on the ten DTDs in [15]), its absence is in practice not very problematic since, as we will see, only the presence of the `parent` axis may hinder completeness in that case.

Before proving the completeness of type inference, we illustrate on simple examples what happens when one of the conditions is not fulfilled. For $*$-guardedness, consider the grammar

$$X \to a[\,B|C\,],\ B \to b[\,],\ C \to c[\,]$$

rooted in $X$, together with the path:

$$\texttt{child::node/self::}b\texttt{/parent::node/child::node}$$

For the first two steps, our algorithm would determine the exact type and context:

$$\Sigma^2 = (\{B \to b[\,]\}, \{X \to a[\,B|C\,], B \to b[\,]\})$$

For the `parent` step, the type and context are:

$$\Sigma^3 = (\{X \to a[B|C\,]\}, \{X \to a[\,B|C\,]\})$$

which are also exact. However, the last step induces the final type:

$$\Sigma^4_{\mathsf{typ}} = \{B \to b[\,], C \to c[\,]\}$$

and the context:

$$\Sigma_{\mathsf{ctx}}^4 = \{X \rightarrow a[\ B|C\ ], B \rightarrow b[\ ], C \rightarrow c[\ ]\}$$

This is not exact because a document matching the first part, $\texttt{child}::\texttt{node/self}::b$ does not have any "*c*" tag and therefore the rule $C$ in the output type is superfluous: this query will never return a node with type $C$ for a document of the considered type. Note that the condition that unions in regular expressions must be guarded must also hold for rules, namely that there must not be two rules $Y \rightarrow l[r_1]$ and $Y \rightarrow l'[r_2]$ in the input type. Indeed these two rules behave like an un-guarded union and therefore jeopardize completeness. Local tree grammars forbid such rules and are thus an essential condition of the input type for completeness to hold.

The recursiveness of the schema also interacts with the `parent` axis in a way that prevents completeness of type inference. Consider the grammar:

$$\{A \rightarrow a[\ B\ ],\ B \rightarrow b[\ B?\ ]\}$$

and the path expression:

$$\texttt{child}::\texttt{node/self}::b/\texttt{child}::\texttt{node/self}::b/\texttt{parent}::\texttt{node}$$

Our type inference algorithm deduces on the second $\texttt{self}::b$ step that the output type is be $\{B \rightarrow b[B?]\}$. However, the last step, $\texttt{parent}::\texttt{node}$ is typed with a type

$$\{A \rightarrow a[\ B\ ],\ B \rightarrow b[\ B?\ ]\}$$

this is because in the grammar, $A$ is a name reachable from $B$ with a parent axis. However, consider any document valid with respect to this grammar. Either it has only one $b$ element, in which case the result is empty, since we try to match two levels of $b$'s with the query. Or it has at least two $b$'s and then the output is always a $b$ node (the topmost one). Therefore, an $a$ node is never part of the result, while the type $A$ is returned by our algorithm.

Lastly, with the following parent-ambiguous grammar:

$$\{A \rightarrow a[\ B \mid C\ ], B \rightarrow b[\ ], C \rightarrow c[B]\}$$

the algorithm fails to type *exactly* (but the output type is still sound) the query:

$$\texttt{child}::\texttt{node/self}::c/\texttt{child}::\texttt{node/self}::b/\texttt{parent}::\texttt{node}$$

By a similar reasoning, we can see that the algorithm returns the rules

$$\{A \rightarrow a[B|C], C \rightarrow c[B]\}$$

while only nodes with tag $c$ can be returned by this query.

Intuitively, the reason why completeness does not hold in the three previous examples is that there are chains in the grammar that may not reflect actual paths in a document. For instance in the last example, in a document "`a[ b[] ]`", the chain "$ACB$" has no interpretation (since there are no `c`-nodes). In this case, there exists a *valid* document which does not contain all the paths described by the possible chains in its type. Therefore, the type inference algorithm will use chains and rules which are actually not part of the interpretation of some documents of the type at issue. Fortunately, if a local tree grammar is $*$-guarded, non-recursive, and parent-unanbiguous, then there always exists a document in which *all* the chains in the grammar are instanciated by some path. We call such a document a *witness* of the grammar. We prove the existance of such a witness before stating the completeness theorem.

**Lemma 5.8 (Witness of a grammar)** *Let $(\mathscr{S}, E)$ be a non-recursive, $*$-guarded, parent-unambiguous local tree grammar. There exists a document $t$, $\mathfrak{I}$-valid with respect to $(\mathscr{S}, E)$ such that:*

$$\forall X \in \boldsymbol{Dn}(E), \exists \mathbf{i} \in \boldsymbol{Ids}(t) \text{ such that } \mathfrak{I}(\mathbf{i}) = X$$

*we call such a document a* witness *of the schema $(\mathscr{S}, E)$.*

**Corollary 5.9** *Let $(\{X\},E)$ be a non-recursive,$*$-guarded, parent-unambiguous local tree grammar and t be its witness. Let $\{Y_1 \ldots, Y_n\} \subseteq \boldsymbol{Dn}(E)$. If $Y_1 \Rightarrow_E \ldots \Rightarrow_E Y_n$, then there exists $\{\mathbf{i}_1, \ldots, \mathbf{i}_n\} \subseteq \boldsymbol{Ids}(t)$ such that*

$$\forall i \in \{2 \ldots n\}, ((\mathbf{i}_{i-1}, \mathbf{i}_i) \in \mathscr{E}(t)) \wedge \mathfrak{I}(id_{i-1}) = Y_{i-1} \wedge \mathfrak{I}(\mathbf{i}_i) = Y_i$$

We are now equipped to state (and prove) the completeness theorem:

**Theorem 5.10 (Completeness of type inference)** *Let $(\mathscr{S}, E)$ be a $*$-guarded non-recursive and parent unambiguous local tree grammar, and P a path. Let*

$$E_0 = \{X \to R \mid X \to R \in E,\ X \in \mathscr{S}\}.$$

*If $(E_0, E_0) \vdash_E P : (\tau, \kappa)$ then:*

$$\boldsymbol{Dn}(\tau) \subseteq \bigcup_{t \in_{\mathfrak{I}} E} \mathfrak{I}(\llbracket P \rrbracket_t(\boldsymbol{RootId}(t)))$$

One of the main reasons why completeness does not hold in general is because the intersections operated by the type rule for `parent` are not powerful enough to guarantee precision for recursive or parent-ambiguous grammar. In a nutshell, this happens because in the presence of parent-ambiguous grammar the type analysis may produce contexts containing false parent types (with respect the current type $\tau$). This suggests that in order to be extremely precise, instead of sets of rules, contexts should rather be sets of *chains* of names, computed and opportunely managed by the type analysis. However $(i)$ managing sets of chains instead of simple sets of rules dramatically complicates the treatment, due to the interaction of recursive axes like `descendant` and recursive grammars, $(ii)$ the problem may arise only for queries that use parent axis and the concomitance of parent-ambiguity make the event rare in practice, $(iii)$ the loss of precision looks in most cases negligible, $(iv)$ even though it would be possible to obtain more precise results for a larger class of grammars, it is well known that exact type-inference for XPath routinely escapes regular tree languages and therefore all existing formalisms to type XML: at some point, an approximation in the type inference process is necessary to remain in the realm of regular types. Therefore we considered that such a small gain (remember that completeness is just some icing on the cake since, while it helps to gauge the precision of the approach, its absence does not hinder its application) did not justify the dramatic increase in complexity needed to relax the condition on the type for completeness to hold.

Of course, the completeness theorem is only stated for XPath$^\ell$ queries and does not account for full XPath queries. Yet it illustrates how precise our type system is in the best case. We will show on various example that on less favorable cases for schemas or for XPath$^\ell$ queries which need to be approximated, the type inference still remains very precise.

## 5.2 Type-Projection inference

In this section we use the type inference of the previous section to infer type-projectors. Once more, naive solutions do not work. For instance, for simple paths $Step_1/\ldots/Step_n$, we may consider as type projector with respect to $(\mathscr{S}, E)$ the set

$$\bigcup_{i=1 \ldots n} \tau_i \cup \{X_i \to R_i \mid X_i \in \mathscr{S}\}$$

where for $i = 1 \ldots n$:

$$\Sigma \vdash_E Step_1/\ldots/Step_i : (\tau_i, -)$$

with $\Sigma = \{X_i \to R_i \mid X_i \in \mathscr{S}\}, \{X_i \to R_i \mid X_i \in \mathscr{S}\}$ (we use "$-$" as a placeholder for uninteresting parameters). This definition is sound but not precise at all, as can be seen by considering the query $\texttt{descendant} :: \texttt{node}/Path$: the use of the above union yields a set containing $\tau_1$ defined as

$$\Sigma \vdash_E \texttt{descendant} :: \texttt{node} : (\tau_1, -)$$

that is, all descendants of the root start symbols in $\mathscr{S}$ (no pruning is performed). Instead, we would like to discard, at least, all rules that are descendants of $\mathscr{S}$ but that are not ancestors of a node matching *Path*. These are the rules
$Y \to R \in T_E(\mathbf{A}_E(\mathscr{S}, \texttt{descendant}), \texttt{node})$ such that

$$(\{Y \to R\}, \kappa) \vdash_E \texttt{descendant} :: \texttt{node}/Path : (\varnothing, -)$$

for some appropriate context $\kappa$. A similar reasoning applies to `ancestor`.

As for the type inference, we define type-projector inference by a judgment and associated inference rules:

**Definition 5.11 (Type-projector inference)** *Let $(\mathscr{S}, E)$ be a type and $P$ an XPath$^\ell$ query in simple step normal form, and $\tau$ and $\kappa$ be subsets of $E$. If the deduction system in Figure 2 proves the judgment*

$$(\tau, \kappa) \Vdash_E P : \pi$$

*then the* type-projector *induced by $\pi$ is the grammar:*

$$(\mathscr{S} \cap \boldsymbol{Dn}(\pi), \{(X \to R)|_{\boldsymbol{Dn}(\pi)} \mid X \to R \in \pi\})$$

.

□

Obtaining a type projector from a set of rules returned by the judgment is straightforward. In essence, the derivation collects in $\pi$ the rules of $E$ that are sufficient to answer the query. Since in general not all rules in $E$ are kept, then the rule in $\pi$ may use names that are not defined in $\pi$. Therefore, the erasure operation (defined in Definition 2.10) simply removes references to names not defined by any rule in $\pi$ (the definition of $R|_{\mathscr{S}}$ is straightforward: it is $R$ where every occurrence of a name in $\mathscr{S}$ is replaced by $\varepsilon$).

The rules in Figure 2 reflect the intuition we gave earlier. At each step, we execute the type inference algorithm on the current set of rules and accumulate only those for which the resulting type is not empty. Informally, each rule preserves the following properties:

**well-formedness**: if a rule $Y \to R$ is added to the type projector $\pi$ then there must be a rule $X \to R' \in \pi$ such that $Y \in \boldsymbol{Names}(R')$.

**precision**: given a path $P$ and a rule $Y \to R$. If $(\{Y \to R\}, -) \vdash_E P : (\varnothing, -)$ then $Y \to R$ must not be added to the projector.

Let us explain how the different rules preserve these properties. The easiest case is the one of a query consisting of a single step, handled by the Rule **(p-step)**. In this rule, we just apply the type inference algorithm to determine the output type of the results. The resulting projector is the set of rules in the results to which we add their upward context $\kappa$, that is the rules linking the results to a start symbol. The rules **(p-union)** and **(p-iterate)** are only inductive cases which allows us to handle respectively top-level union and projectors applied to a set of rules. In particular, **(p-iterate)** splits the checking of a path over all the possible rules specified in the type component of the environment (each one identifies a different set of current nodes). This allows us to define the so-called "Path Rules" in much a simpler way since they can be written for environments in which the type component is just a singleton. The Path Rules actually perform the projection and they all follow the same scheme. The Rule **(p-test)** handles a simple node test. If the type inference returns some *non-empty* type $\Sigma_{\mathsf{typ}}$ for the step, then we can compute the projector for the continuation $P$ and add its result to the rule for the current node. The Rule **(p-predicate)** is similar: the type $\Sigma_{\mathsf{typ}}$ returned by the type inference is the set of nodes for which the predicate is satisfied. We then recursively compute the projector for the continuation $P$ as well as for the paths $P_{ij}$ occurring in the predicate. In the end, we return the union of all the computed projectors to which we add the rule for the current node. Again we only do this if the type inference returned a non-empty type. The following rules handle the actual navigation. They are split in two sets, one for the `parent` and `child` axes another for their recursive variant, `ancestor` and `descendant`. Since they are the most delicate rules let us explain them in details. The two cases are similar. In the Rule **(p-single)**, the algorithm first retrieves all the rules matching the axis (`child` or `parent`). These rules are collected in $\tau$ and the analysis yields a current context $\kappa'$. Then, by using $n$ calls to the type inference algorithm ($n$ being the number of rules in $\tau$), it collects among $\tau$ only the rules which are a suitable starting point for the rest of the path, that is all the rules yielding a non-empty result type when typed against $P$. These rules are collected in $\tau'$ which, as it can be easily seen, a subset of $\tau$. Finally, $\tau'$ and $\kappa'$ are used as the environment to infer the projection with the rest of the path. The **(p-many)** rule handles the recursive axes, `descendant` and `ancestor`. The rule is almost the same as Rule **(p-single)** with the exception that it does not test whether the continuation $P$ yields a non empty result on the node but on a `descendant` (or `ancestor`) of the node, to ensure that we put not only the correct rules in the projector but also the rules leading to them, and therefore that we maintain well-formedness. If for any of these rules one of the side conditions does not hold, then the rule **(p-erase)** is applied and returns an empty projector for the current path.

Before proving the formal properties of the type-projection inference, we illustrate its behavior by unrolling it on an example. Consider the grammar:

$$\{A \to a[(B|C)*], B \to b[D], C \to c[E], D \to d[E], E \to e[\,]\}$$

with start symbol $A$ and the path $P$:

$$\texttt{descendant}::\texttt{node}/\texttt{self}::e/\texttt{ancestor}::\texttt{node}/\texttt{self}::b$$

which is the single step normal form of

$$//e/\texttt{ancestor}::b$$

To ease the reading, we identify every rule with the non-terminal it defines. Therefore in what follows when we write, say, $A$ in types or contexts, we actually mean $A \to a[(B|C)*]$. The algorithm computes the type projector for $P$ as follows. The initial environment is $(\{A\},\{A\})$. We apply the rule **(p-many)** for the first step. The first premise computes the type of `descendant::node` applied to $A$, which returns the type and context (these instantiate the $(\tau',\kappa')$ of the rule):

$$(\{B,C,D,E\},\{A,B,C,D,E\})$$

Then the second premise filters out the unwanted names and keeps only those for which the whole path may succeed. This gives us an intermediary type: $\{B,D,E\}$ (and unchanged context) onto which we can compute the projector for the path:

$$\texttt{child}::\texttt{node}/\texttt{self}::e/\texttt{ancestor}::\texttt{node}/\texttt{self}::b$$

the final result for this rule will be the projector for the above path to which we add $\{A\}$ *(i)*. At this point, since the input type contains many rules, we can apply the rule **(p-iterate)** which will apply the continuation path on $\{B\}$, $\{C\}$ and $\{D\}$. It is easy to see that on $\{B\}$ the side condition for the rule **(p-single)** is not fulfilled, since the type inference returns empty. The rule **(p-erase)** applies and returns an empty projector. The projection continues with only $\{C\}$ and $\{D\}$ left (the context is unchanged until now, $\{A,B,C,D,E\}$). First let's consider the derivation for $\{D\}$. The current step is $\texttt{child}::\texttt{node}$ which was introduced by the previous **(p-many)** rule. On this step, we apply the rule **(p-single)**. This rule adds $\{D\}$ *(ii)* in the final projector and continues by computing a projector from $\{E\}$ using the path:

$$\texttt{self}::e/\texttt{ancestor}::\texttt{node}/\texttt{self}::b$$

When we apply the same rule to $\{C\}$ however, while the first premise returns a non empty type, the second one returns an empty result, since from a node with type $C$ the path

$$\texttt{child}::\texttt{node}/\texttt{self}::e/\texttt{ancestor}::\texttt{node}/\texttt{self}::b$$

yields an empty result. Thus the rule is not applied and the result of the projector for the remaining path for the node type $\{C\}$ is the empty projector. We continue with our only set, $\{E\}$. We compute the projector for $\texttt{self}::e$ which adds $\{E\}$ to the final projector *(iii)* and computes the projector for the path:

$$\texttt{ancestor}::\texttt{node}/\texttt{self}::b$$

It is easy to to see that these will return $\{B\}$ as a projector *(iv)*. If we summarize, we obtain from (*i, ii, iii,* and *iv*) the set of rules

$$\pi = \{A \to a[(B|C)*], B \to b[D], D \to d[E], E \to e[\,]\}$$

The actual type projector is:

$$(\mathscr{S} \cap \textbf{\textit{Dn}}(\pi), \{(X \to R)|_{\textbf{\textit{Dn}}(\pi)} \mid X \to R \in \pi\})$$

that is:

$$(\{A\}, \{A \to a[B*], B \to b[D], D \to d[E], E \to e[\,]\})$$

This example shows how the two properties of precision and well-formedness are preserved:

**well-formedness**: what we obtained at the end is a valid type without unneeded rules.

**precision**: although the query references $e$ nodes explicitly, we do not naively keep all the $e$ nodes but only those that are useful to compute the query, namely those occurring *below* a $b$ node.

We can now present the formal properties of type-projection inference

**Lemma 5.12 (Termination of type-projector inference)** *Let $(\mathscr{S}, E)$ be a type, $P$ a path, and $\Sigma$ and $\Sigma'$ environments. The judgment $\Sigma \Vdash_E P : \Sigma'$ has a unique and finite derivation.*

The lemma above states that the rules in Figure 2 describe a terminating algorithm. We show now that they compute a type-projector by formalizing the "well-formedness" property that we outlined above. The intuition is that when the output type for a step is computed (e.g., in the first premise of the rule **(p-predicate)**), then the *context* corresponding to this computation is kept and passed as a parameter for the inference of the remainder of the path. On the last step, (rule **(p-step)**) the context is added to the type projector. There, it ensures that whenever a rule $Y \to R$ is added to the type-projector, all the rules needed to derive $Y \to R$ from the start symbols are added to the type-projector as well. This is what we formally state by the following lemma:

**Lemma 5.13 (Well-formedness of type-projector inference)** *Let $(\mathscr{S}, E)$ be a type, $\tau$, $\tau'$, and $\kappa$ sets of rules, and $P$ a path. If $(\tau, \kappa) \Vdash_E P : \tau'$, then $(\tau, \kappa) \vdash_E P : (\tau'', \kappa'')$ implies $\kappa'' \subseteq \tau'$.*

We can now state the soundness of type-projection inference:

**Theorem 5.14 (Soundness of type-projector inference)** *Let $(\mathscr{S}, E)$ be a type and P an XPath$^\ell$ query. Let S be the set of rules: $S = \{X \to R \mid X \in \mathscr{S}\}$. If*

$$(S,S) \Vdash_E P : \tau$$

*then $\tau$ is a type-projector for $(\mathscr{S}, E)$ and for every $t \in_{\mathfrak{I}} (\mathscr{S}, E)$ we have:*

$$[\![P]\!]_{t \backslash_{\mathfrak{I}} \tau}(\textbf{\textit{RootId}}(t)) = [\![P]\!]_t(\textbf{\textit{RootId}}(t))$$

In words, if $\tau$ is the projector inferred for a query $P$ and a grammar $(\mathscr{S}, E)$, then for every tree $t$ validating the grammar, the result of executing $P$ on $t$ or on its pruned version $t \backslash_{\mathfrak{I}} \tau$ is the same.

Completeness requires not only completeness of the type system ($*$-guarded, non-recursive, and parent-unambiguous DTDs), but also the following condition on queries:

**Definition 5.15** *An XPath query Q is* strongly-specified *if:*

*i.* *its predicates do not use backward axes,*

*ii.* *along Q and along each path in the predicates of Q there are no two consecutive (possibly conditional) steps whose Test part is* `node`

*iii.* *each predicate in Q contains at most one path and this does not terminate by a step whose Test is* `node`. □

For instance, among the following queries, only the first two are strongly-specified:

– `descendant::node/self::`$a$`/ancestor::node`
– `descendant::node[child::`$b$`]/self::`$a$`/parent::node`
– `descendant::node/ancestor::node/self::`$a$
– `descendant::node[child::`$b$`/child::node]/self::`$a$
– `child::`$a$`[descendant::node/parent::`$b$`/child::`$c$`]`

In the third query, there are two consecutive steps with a "`node`" test, which violates condition ($ii$). In the fourth query the predicate contains a path ending with "`node`"—failing to satisfy condition ($iii$)—and for the last query, the predicate contains backward axes, which violates condition ($i$).

Once more, we are in presence of a very common class of queries: for instance, almost all paths in the XMark and XPathMark benchmarks are strongly specified.

If all the conditions are met, then we can show that our algorithm is complete, in the sense that it infers the best possible sound projector. In words, if we remove any rule (and its consequences) from a projector inferred for a path $P$ and a grammar $(\mathscr{S}, E)$, then we obtain a projector for which there exists a tree $t$ validating the grammar for which the execution on $t$ and on its pruned version yield different results. Formally:

**Theorem 5.16 (Completeness of projector inference)** *Let $(\mathscr{S}, E)$ be a $*$-guarded, non-recursive, and parent-unambiguous local tree grammar, and P a strongly-specified XPath$^\ell$ path. Let S be the set of rules: $S = \{X \to R \mid X \in \mathscr{S}\}$. If*

$$(S,S) \Vdash_E P : \tau$$

*then there exists $t \in_{\mathfrak{I}} (\mathscr{S}, E)$ such that for each $Y \to R \in \tau$, if $\pi = \tau \setminus (\{Y \to R\} \cup \mathbf{A}_E(\{Y \to R\}, \texttt{descendant}))$, then:*

$$[\![P]\!]_{t \backslash_{\mathfrak{I}} \pi}(\textbf{\textit{RootId}}(t)) \neq [\![P]\!]_t(\textbf{\textit{RootId}}(t))$$

The fact that completeness may not hold for not $*$-guarded, non-recursive, or parent-ambiguous local tree grammar, is a consequence of the analogous property of the type system. To see that also strong-specification is a necessary condition consider documents valid with respect to the following grammar rooted at $X$:

$$\{X \to a[Y,W],\ W \to c[\,],Y \to b[Z],\ Z \to d[\,]\}$$

If we query a document of that type with the following non strongly-specified query (it does not satisfy $(iii)$)

$$\texttt{self}::a\texttt{[child}::\texttt{node]},$$

then $\{X,Y\}$ is an optimal projector for this query (once more, we use a name to denote the rule that defines it), but the presence of the condition `child::node` forces the system to include also $W$ in the inferred projector, thus breaking completeness. A similar reasoning applies for $\texttt{self}::a[\texttt{child}::b \texttt{ or } \texttt{child}::c]$, which does not satisfy condition $(iii)$ because of the presence of multiple path in the predicate. Concerning the presence of backward axes in predicates, consider the query

$$\texttt{self}::a\texttt{[descendant}::\texttt{node/ancestor}::a\texttt{]}$$

which does not satisfy condition $(i)$. An optimal projector for this query on the same grammar is $\{X,Y\}$. However, since the `ancestor` condition is true for all descendants of $a$ nodes, $\{W,Z\}$ is included in the projector as well. Finally, with a similar reasoning on the same grammar, it is clear that the query

$$\texttt{descendant}::\texttt{node/ancestor}::\texttt{node/self}::a$$

for which condition $(ii)$ does not hold, jeopardises completeness. The first step selects all the rules in the grammar that can be derived from the start symbol (that is, all the rules). None of these rule are discarded by the projector-inference since for none of them the output type of

$$\texttt{ancestor}::\texttt{node/self}::a$$

is empty. The point here is that for the given grammar, there is no need to keep all the nodes, but only one child of the root. Indeed, having one element below the root guaranties that the sequence `descendant::node`, `ancestor::node` is not empty and therefore that the root can be selected.

Of course, it is possible to state completeness for other classes of queries but, once more, this seems a satisfactory compromise between simplicity and generality.

# 6 Extension to full XPath

The formal developments of the previous section only deal with the XPath$^\ell$ language. This language allows one to express *structural queries*, that is, queries whose predicates contain only conjunctions or disjunctions of paths. In this section we show how to translate a (full) XPath query into a (set of) XPath$^\ell$ queries and perform type-projection inference for the latter that is sound for the former. In other terms, we show that our translation is a sound approximation with respect to type-projection. Finally, we also show how to encode the XPath axes not present in XPath$^\ell$ and how to extend our theoretical framework to handle most XML and XPath peculiarities (attributes, absolute paths,...)

## 6.1 Handling XPath predicates

We extend Definition 4.1 to XPath 1.0 paths ([35]):

**Definition 6.1** *A path is a finite production of the following grammar:*

$$
\begin{array}{lcl}
Path & ::= & Step \mid Path/Path \mid Path \,\|\, Path \\
Step & ::= & Axis::Test \mid Axis::Test[Cond] \\
Axis & ::= & \texttt{self} \mid \texttt{child} \mid \texttt{descendant} \mid \texttt{parent} \mid \texttt{ancestor} \\
Test & ::= & tag \mid \texttt{node} \mid \texttt{text} \\
Cond & ::= & Cond \ \texttt{or} \ Cond \mid Cond \ \texttt{and} \ Cond \mid Expr \\
Expr & ::= & Expr\ cmp\ Expr \mid Arith \\
Arith & ::= & Arith\ op\ Arith \mid Atom \\
Atom & ::= & f(Expr,\ldots,Expr) \mid Path \mid v
\end{array}
$$

*where:*

*tag* *ranges over element tags*

*cmp* $\in \{$=, !=, <=, <, >, >=$\}$

*op* $\in \{$+, -, *, div, mod$\}$

*f* *ranges over a set of built-in functions of the Core Function Library.*

*v* *ranges over values: strings, sequences, integers,. . .*

□

We wish to provide a safe translation from an XPath query $Q$ to an XPath$^\ell$ query $P$ that approximates $Q$ and use it to infer a type projector. By safe we mean that the type-projector inferred for $P$ must not change the semantics of $Q$.

What exactly is an approximating query in this context? A naive approach to define query approximation is to consider inclusion of the results. According to it the query $P$ translation of $Q$ should always select more nodes than $Q$. However this works only as long as we do not use non-structural conditions (that is, predicates that make a query be non-structural). This is clear for example when we use the negation function `not`. Consider the query

$$\texttt{descendant}::a\texttt{[not(child}::b\texttt{)]}$$

For all documents, the query `descendant`::$a$ returns more results than the query above. However, a projector inferred for `descendant`::$a$ would discard $b$ nodes not occurring before an $a$ node, and therefore possibly also some $b$ nodes children of an $a$ node. In this way it would change the result of the original query. What the approximating query needs to reflect cannot be defined in terms of inclusion of results but rather in terms of *data-need*. We must ensure that the approximation traverses at least the same nodes as the original one to ensure that the former will not be pruned. However, we want the approximation also to be as precise as possible. For instance “`descendant::node`” is a sound approximation for any XPath query but the projector we infer from it is utterly imprecise: it performs no pruning.

As the reader will have understood, the tricky part is to approximate non-structural conditions. We do it as follows:

**Definition 6.2 (Approximation of a path)** *Let $P$ and $S$ respectively denote a path and a set of paths of XPath$^\ell$. Let $P/S$ denote the set of XPath$^\ell$ paths defined as $\bigcup_{P'\in S}\{P/P'\}$. Given an XPath expression $Q$, its* approximation $\mathbf{P}(Q)$ *is the set of XPath$^\ell$ paths defined as:*

$$
\begin{array}{rcl}
\mathbf{P}(Q_1 \| Q_2) & = & \mathbf{P}(Q_1) \cup \mathbf{P}(Q_2) \\
\mathbf{P}(Axis::Test/Q) & = & Axis::Test/\mathbf{P}(Q) \\
\mathbf{P}(Axis::Test[C]/Q) & = & Axis::Test[\mathbf{C}(C)]/\mathbf{P}(Q) \cup Axis::Test/\mathbf{S}(C)
\end{array}
$$

*where:*

$$
\begin{array}{rcll}
\mathbf{C}(P) & = & P & \textit{if } \mathbf{P}(P) = \{P\} \\
\mathbf{C}(P) & = & \texttt{self::node} & \textit{if } \mathbf{P}(P) \neq \{P\} \\
\mathbf{C}(C_1 \; or \, C_2) & = & \mathbf{C}(C_1) \; or \, \mathbf{C}(C_2) & \\
\mathbf{C}(C_1 \; and \, C_2) & = & \mathbf{C}(C_1) \; and \, \mathbf{C}(C_2) & \\
\mathbf{C}(C) & = & \texttt{self::node} & \textit{otherwise}
\end{array}
$$

*and:*

$$
\begin{array}{rcll}
\mathbf{S}(P) & = & \varnothing & \textit{if } \mathbf{P}(P) = \{P\} \\
\mathbf{S}(P) & = & \mathbf{P}(P) & \textit{if } \mathbf{P}(P) \neq \{P\} \\
\mathbf{S}(C_1 \; or \, C_2) & = & \mathbf{S}(C_1) \cup \mathbf{S}(C_2) & \\
\mathbf{S}(C_1 \; and \, C_2) & = & \mathbf{S}(C_1) \cup \mathbf{S}(C_2) & \\
\mathbf{S}(C_1 \; op \; C_2) & = & \mathbf{S}(C_1) \cup \mathbf{S}(C_2) & \\
\mathbf{S}(C_1 \; cmp \; C_2) & = & \mathbf{S}(C_1) \cup \mathbf{S}(C_2) & \\
\mathbf{S}(f(C_1, \ldots, C_n)) & = & \mathbf{F}(f(C_1, \ldots, C_n)) &
\end{array}
$$

*and* $\mathbf{F}(_)$ *is the approximation of built-in functions (see Figure 3 for an excerpt).* □

The most technical point in the definition above is, as expected, the approximation of conditions, implemented by the auxiliary functions $\mathbf{C}()$ and $\mathbf{S}()$. To be precise, we differentiate between purely structural paths and non structural paths. For a structural path, $\mathbf{P}()$ does not introduce any approximation and returns the singleton containing the path itself. Otherwise, a non-structural path is approximated by a set of paths. The translation is non trivial when the path contains non structural conditions. Let us illustrate the rationale of the definition first by an example. The path

$$\texttt{descendant::}a\texttt{[(count(child::}b\texttt{)>3 and child::}c\texttt{) or descendant::}b\texttt{]/child::}d$$

is approximated by the following set of two paths

$$
\begin{array}{l}
\{\, \texttt{descendant::}a\texttt{[self::node and child::}c\texttt{ or descendant::}b\texttt{]/child::}d, \\
\qquad\qquad\qquad\qquad\qquad\qquad \texttt{descendant::}a\texttt{/child::}b \,\}
\end{array}
$$

The first is generated by an application of the function $\mathbf{C}()$, while the second derives from the application of $\mathbf{S}()$. As we see, the arithmetic expression $\texttt{count(child::}b\texttt{)>3}$ is approximated by the function $\mathbf{C}()$ into the `self::node` path occurring in the first path of the set. This condition is always true and therefore it is a sound approximation of the Boolean value of the expression (since the result is always true the type-inference algorithm will never be able to deduce an empty output type for this sub-path and therefore the type-projector inference algorithm will keep the rules associated with this node). However this is not sufficient to ensure the safety of type projection. Indeed for this test to be possible at run-time, the projected document must have the "*b*" nodes that were below *a* nodes in the original document. This approximation is made via the second path by the $\mathbf{S}()$ function and, in particular, by the $\mathbf{F}(\texttt{count(child::}b\texttt{)})$. Of course, what the function actually does depends on the semantics of the built-in function. For instance, `count(`$P$`)` returns the number of nodes selected by $P$, thus a projector keeping the type of the nodes selected by $P$ is sound. On the contrary, the function `string(`$P$`)` when applied to a node set returns the concatenation of the string-value of all the nodes in the set. The string-value of a node is the concatenation of all the **PCDATA** elements occurring below it. Therefore a suitable approximation for `string(`$P$`)` is not $P$ but rather $P$`/descendant::text`. Giving an approximation for all the functions of the XPath Core Library is a tedious task. Although our prototype implements approximation for all functions, in Figure 3 we just give an excerpt that completely covers all the different techniques we used in our prototype to approximate built-in functions.

## 6.2 Other XPath features

We purposely left out from our definitions some features of XPath that would have led to a much more intricate formalization process, in particular for what concerns definitions of the algorithms and the proofs of the theorems. Here we illustrate how these features can be either encoded or approximated within our framework.

#### 6.2.1 `descendant-or-self` and `ancestor-or-self` axes

These axes—that we used in Figure 3—can be encoded exactly by using the "$\mid$" operator. Precisely

$$P/\texttt{descendant-or-self}::\mathit{Test}\,[\,\mathit{Cond}\,]\,/P'$$

can be equivalently written as

$$P/(\texttt{descendant}::\mathit{Test}\,[\,\mathit{Cond}\,] \mid \texttt{self}::\mathit{Test}\,[\,\mathit{Cond}\,])/P'$$

#### 6.2.2 Sibling axes

We could have defined a sibling relation over node identifiers in the same way as we defined the edge relation ***Edg*** in Section 4, and used it to deal with the `following-sibling` and `preceding-sibling` axes natively. However we can also approximate these axes using only "vertical" moves. So for instance

$$P/\texttt{following-sibling}::\mathit{Test}\,[\,\mathit{Cond}\,]\,/P'$$

becomes:

$$P/\texttt{parent}::\texttt{node}/\texttt{child}::\mathit{Test}\,[\,\mathit{Cond}\,]\,/P'$$

The transformation above approximates the following siblings of a node by all its siblings, including itself. Our experiments showed that, as far as type-projection is concerned, this kind of approximation does not yield any noticeable loss of precision in practice.

#### 6.2.3 `preceding` and `following` axes

For these axes, we can directly use the W3C recommendation [35] and encode `following` accordingly. That is,

$$P/\texttt{following}::\mathit{Test}\,[\,\mathit{Cond}\,]\,/P'$$

becomes

$$P/\texttt{ancestor}::\texttt{node}/\texttt{following-sibling}::\texttt{node}/\texttt{descendant-or-self}::\mathit{Test}\,[\,\mathit{Cond}\,]\,/P'$$

#### 6.2.4 Document node

The XPath data model enforces the presence of a *document node*, the real root of the document which has no label and is selected by the initial "/" of an XPath expression. It is of course possible to represent such documents in our framework but we preferred to omit it here since it would cause many presentation issues with little theoretical interest. In particular, the document node is never referenced by the schema of the document.

#### 6.2.5 Absolute paths

Absolute paths are paths with a leading $/$. They do not start their evaluation from the current context node but from the root of the document. Our formalism easily allows us to encode absolute paths. First, if an absolute path occurs outside of a predicate, as in:

$$P/(P_1 \mid /P_2)/P'$$

then we can simply rewrite it as:

$$(P/P_1/P') \,\|\, (P_2/P')$$

Second, if the path $/P$ occurs in a predicate, then we can replace it with `self::node` (as if it was a non structural condition) and add $\mathbf{P}(/P)$ to the global approximation. Direct treatment of absolute paths would have further complicated Definition 6.2, where we would have had to maintain a set of absolute approximations, modified only by absolute paths and propagated at each function call. We chose not to clutter this definition (but absolute paths are handled by our implementation).

#### 6.2.6 `attribute` axis and attributes in the data-model and schema

Conceptually, the `attribute` axis is not very different from the `child` axis, and could be encoded as such. For instance a possible solution would be to encode an element

```
<e att="value" id="34" ><a/><b/></e>
```

as the tree:

$$e[\ @[att[\text{"}value\text{"}]\ id[\text{"}34\text{"}]\ ]\ a[\ ]\ b[\ ]\ ]$$

by introducing a phony node with label @. If such a solution were retained then we would also need to update the definitions of `child` and `descendant` to ignore @ nodes, and add an `attribute` axis selecting only the content of such nodes.

As far as schemas are concerned, they need to reflect the *uniqueness* and *unorderedness* of a sequence of attributes within an element node. This can be done with a union type. For instance, the document above could have type:

$$\begin{array}{rcl} E & \to & e[\ ATTS\ A\ B\ ] \\ ATTS & \to & @[(ATT\ ID) \mid (ID\ ATT)] \\ ATT & \to & att[String] \\ ID & \to & id[String] \\ A & \to & a[\ ] \\ B & \to & b[\ ] \end{array}$$

this encoding however incurs an exponential blow-up in the size of the sequence of attributes. Our implementation follows a much more pragmatic approach. Precisely, even though attributes could be encoded in our approach we preferred to add an unordered attribute construct directly at the grammar level and specialize type-inference and type-projector inference rules for attributes.

#### 6.2.7 *id()* function

The `id()` function of XPath is peculiar in the sense that unlike other functions, it does not take the context node as implicit argument (e.g. the `position()` function returns the position of the context node within the current result set). Rather, the expression “`id("foo")`” returns the node whose *id* is `"foo"` if it exists (a node has *id* `"foo"` if it has an attribute named `id` whose value is `"foo"` *and* if this attribute has been declared with type `ID` in the Schema, [35]). We choose to approximate this function in two steps. First, we rewrite it as an absolute path. Then we can let our approximation algorithm handle the absolute path (with the technique described in Section 6.2.5). For instance an expression such as

`id("item34")/child::`*name*

can be rewritten has

`/descendant::` $*$ `[@id="item34"]/child::`*name*

This rewrite technique was used in particular to handle queries C5-C7 of the XPathMark benchmark (see Section 9).

# 7 Extension to XQuery

In this section we extend our technique to XQuery.

**Definition 7.1 (XQuery)**

$$
\begin{array}{lcl}
\mathit{FLOWR} & ::= & \mathit{FORLET}\ \texttt{return}\ \mathit{ExprS} \mid \mathit{ExprS} \\
\mathit{FORLET} & ::= & \mathit{FOR} \mid \mathit{LET} \\
\mathit{FOR} & ::= & \texttt{for}\ \$x\ \texttt{in}\ \ \mathit{ExprS} \\
\mathit{LET} & ::= & \texttt{let}\ \$x\ \texttt{:=}\ \ \mathit{ExprS} \\
\mathit{ExprS} & ::= & \texttt{if}\ \mathit{ExprS}\ \texttt{then}\ \mathit{Cond}\ \texttt{else}\ \mathit{Cond} \mid \mathit{Cond} \\
\mathit{Cond} & ::= & \mathit{Cond}\ \texttt{or}\ \mathit{Cond} \mid \mathit{Cond}\ \texttt{and}\ \mathit{Cond} \mid \mathit{Expr} \\
\mathit{Expr} & ::= & \mathit{Expr}\ \mathit{cmp}\ \mathit{Expr} \mid \mathit{Arith} \\
\mathit{Arith} & ::= & \mathit{Arith}\ \mathit{op}\ \mathit{Arith} \mid \mathit{Atom} \\
\mathit{Atom} & ::= & f(\mathit{Expr},\ldots,\mathit{Expr}) \mid \mathit{FLOWR}/P \mid x \mid v \\
 & \mid & \mathit{FLOWR}, \mathit{FLOWR} \mid \texttt{<tag>}\mathit{FLOWR}\texttt{</tag>} \mid \texttt{()} \\
P & ::= & \mathit{Step}[\mathit{Cond}]/P \mid \mathit{Step}/P \mid \mathit{Path}
\end{array}
$$

where $x$ ranges over identifier names, $v$ ranges over values (such as integer and strings), *cmp* ranges over { `=`, `!=` ,<, >, >=, <=}, *op* ranges over { `+`,`-`,`*`, `div`, `mod`} and *Path* and *Step* are the same as in Definition 6.1, that is they denote step and path expressions free of any XQuery construct.

□

For the sake of clarity and concision we only considered formally a subset of the XQuery grammar ([36]). In a nutshell, the definition of *Atom* (given in Section 6) is extended with two new constructs: *variables* (ranged over by $x$, $y$, $z$ in what follows) and path applications $\mathit{FLOWR}/P$.

Note that XQuery constructs may occur inside a path expression (production $P$) or not (production *Path*). Also, we consider neither queries that first construct new elements and then navigate on them (these are rarely used in practice) nor queries containing "`order by`", "`switch case`", etc. constructs. XQuery queries are ranged over by $q$. In order to apply the previous analysis to infer a projector for an XQuery query $q$, we first extract a set of full XPath expressions from $q$, denoting the data needs for $q$. Then, we apply to each of these extracted paths the approximation function **P**(_) given in Definition 6.2 to obtain an XPath$^\ell$ expression. We can finally use the projector inference algorithm of Section 5.2 on the set of approximated paths, which is a sound type projector for the original XQuery query $q$.

Path extraction is performed by the extraction function **E**(_,_,_), whose definition is given in Figure 4. The extraction function has the form **E**($q$,$\Gamma$,$m$) and performs a straightforward recursive descent over its first parameter $q$ which is the query at issue. The second parameter $\Gamma$ is an environment, that keeps track of bindings of the form $(x;P)$ in whose scope $q$ occurs. Finally, $m$ is a flag indicating whether $q$ is a query that serves to materialise the full content of the queried elements ($m = 1$) or if the query just selects a set of nodes whose descendants are not needed ($m = 0$). Before explaining in details the rules in Figure 4, we introduce two auxiliary functions. The first one is $\mathbf{M}(_,_)$ (Figure 5) which given a built-in XPath function and the position of one of its arguments, returns a suitable value for the parameter $m$ (intuitively, $\mathbf{M}(f,i)$ returns 1 if $f$ needs the full content of its $i^{\text{th}}$arguments and 0 otherwise). This function is similar to the function $\mathbf{F}(_)$ introduced in Section 6, Figure 3.

The second one, **E'**(_,_,_) is defined mutually, together with **E**(_,_,_) and allows to recursively traverse XQuery expressions and resolves the variable names they contain. It works similarly to **E**(_,_,_) but do not returns sets of XPath paths, but sets of particular XQuery expressions which do not contain any variables.

Now that we have introduced environments and the auxiliary functions, we can easily describe the rules in Figure 4 First, rules 1 and 2 form the basic case of the recursive descent and return the empty set if the whole query consists of a constant. Rules 3 and 4 straightforwardly apply the extraction recursively for the content of sequence (Rule 3) and element (Rule 4) constructors. Rules 5 and 6 handle the case of variable bound in the environment Rules 7 and 8 add a constant path to the set of extracted path, according to the value of the parameter $m$. Note that in those rules, *Path* refers to the corresponding entry in the grammar of Definition 6.1, that is it does not contain any XQuery construct and only pure XPath ones. Path containing XQuery expressions are handled in the subsequent rules. Rule 9 handles the application of a path *FLOWR* expression with a path $P$. Note that as previously the notations $S_1/S_2$ where $S_1$ and $S_2$ are sets of paths stands for:

$$\bigcup_{P \in S_1} \bigcup_{P' \in S_2} \{P/P'\}$$

The case of a simple step composed with a path expression is handled similarly by Rule 10 and we recall that the notation $Step/S$ where $S$ is a set of path is syntactic sugar for the set:

$$\bigcup_{P \in S} \{Step/P\}$$

Rule 11 is more intricate, but its complexity is only bureaucratic. This rule, which did not exist in the previous version of our work ([5]), allows the extraction process to extract Full XPath expressions. In the present work, path extraction and path approximation are two separate processes. Path extraction only occurs at the level of XQuery terms and returns sets of full XPath expressions (which reflects the exact path that may be evaluated during query execution). Approximation from XPath to XPath$^{\ell}$ is handled at the XPath level. The issues solved by Rule 11 is to recursively traverse an XQuery expression, using a recursive call to the auxiliary function **E'**(_,_,_) which builds a set of XPath conditions into which all variable bindings have been resolved. Therefore what we obtain after **E'**(_,_,_) is a set of XPath conditions free of any XQuery construct (especially variables). We can now explain how **E'**(_,_,_) works. In Rules $1'$ to $3'$ use a recursive descent into the production of the XQuery grammars, starting at the condition levels and reconstruct Boolean XPath condition ($1'$), relational XPath expressions ($2'$) or arithmetic XPath expressions ($3'$). More interesting is Rule $4'$ which traverses the arguments of a function call and uses the auxiliary function **M**(_) to determine a suitable value for $m$. Lastly, if the input matches any other constructs Rule $5'$ applies and recursively applies **E**(_,_,_) to construct a set of XPath paths.

We can resume our description of **E**(_,_,_) for the remaining cases, the high level constructs "`if then else`", "`let return`" and "`for return`" handled by Rules 12, 13 and 14 respectively. Rule 12 recursively extracts paths on the Boolean test $q$, the "`then`" case $q_1$ and the "`else`" case $q_2$. The only point of interests is that the Boolean test cannot generate a result and therefore can be called with parameter $m = 0$. The `let` binding handled by Rule 13 augments the environment $\Gamma$ with the path extracted from $q_1$ and extracts the paths of query $q_2$ in this augmented environment. Note that the path bound to $x$ are added to the final results by Rule 5 or 6 only if the variable is used. On the contrary in Rule 14, `for` loops will perform their iterations even if the bound variable is never used, as long as the paths extracted from $q_1$ yield a non-empty result. It is therefore mandatory to add the paths extracted from $q_1$ to the final result.

These rules subsume and enhance the technique of Marian and Siméon [27]. In particular, ($i$) the technique we use to exclude useless intermediate paths is simpler and more compact, ($ii$) we do not need to distinguish between two kinds of extracted paths but, more simply, we always manage a unique set of path expressions, and last but not least, ($iii$) our path extractor can be used even if the user cannot access an XQuery to XQuery-Core compiler, which is necessary for [27].

Before applying the extraction function **E**(_,_,_) to some query $q$ we apply some heuristics that rewrite $q$ so as to improve the pruning capability of the inferred paths. Among these heuristics the most important is the one that rewrites

```
for y in Q/descendant-or-self::node
  return if C(y) then q else ()
```

into

```
for y in
  Q/descendant-or-self::node[C(self::node)]
return q
```

whenever $C(y)$ is a condition referring only to $y$ and does not use external functions ($C($`self ::` `node`$)$ is obtained by replacing `self :: node` for all occurrences of $y$ free in $C$). If we apply $\mathbf{E}(_,_,_)$ to the first query, then a path ending by the step `descendant-or-self::node` is extracted thus annulling further pruning: the entire forest selected by $Q$ is loaded in main memory. This also happens with the approaches of Bressan *et al.* [13] and of Marian and Siméon [27]. In ours and Marian and Siméon's approach the query can be rewritten as above, while this is not possible with Bressan *et al.* formalisms since their subset of XQuery does not include predicates. However, Marian and Siméon's path based pruning degenerates (no further pruning is performed) also for the second query, since the step

`descendant-or-self::node`

ends up in the set of pruner paths, thus selecting all nodes. This is because their approach cannot manage predicates. In our approach instead predicates are taken into account and therefore only nodes satisfying $C(y)$ are kept by the projector, thus yielding a very precise pruning.

It is important to stress that despite their specific form the first kind of queries is very common in practice since they are generated from XQuery→XQuery-Core compilation of a non negligible class of queries or when rewriting upward axes into downward ones. This latter observation shows that the application of rewriting rules of [31] to extend Marian and Siméon's approach to upward axes is not feasible since the rewriting may completely compromise pruning.

# 8 Extension to other typing policies

## 8.1 Handling un-typed documents

Although the usage of schema is being more and more wide-spread, it still is interesting to see how to perform type-based projection in an untyped world. A first, rather blunt, approach is to consider a fixed corpus of un-typed documents. For such sets of documents it is possible to *infer* a DTD. For instance, Bex *et. al.* propose several automata-based methods to infer DTDs [10] and even XMLSchemas ([11, 9]). Once a schema is inferred, our technique can be applied as-is.

More interestingly, this untyped problem can be reduced to a precise typing problem. Indeed, an un-typed document is nothing but a document of type $(\{X\},\{X \to Any\})$. If we apply the type inference-algorithm of Section 5.1 to such an input type, then the result would be $(\{X\},\{X \to Any\})$ itself (meaning that the nodes selected by the query have type *Any*). Therefore in this case, since none of the intermediary steps of the query results in an empty-type, the type-projector inference algorithm of Section 5.2 cannot remove any rule from the input type which remains $(\{X\},\{X \to Any\})$: the input document cannot be pruned. However, even though the input type does not contain any meaningful information, the query itself might. Imagine a query "$//a/b$". It is easy to deduce, by a simple examination of the query a projector which keeps only "b" nodes occurring below "a"-nodes. While the solution in this case is straightforward, solving this problem in general is a tricky issue. The solution for a forward fragment of XPath can be found in the last author's PhD. thesis (see Chapter 7 of [30]). Let us briefly outline it on our example. The first issue is the representation of types. For such precise algorithms, regular tree grammars are not well suited. Indeed instead of the type $(\{X\},\{X \to Any\})$, it is more desirable to have a type $(\{X\},E)$ where $E$ is the set of rules:

$$\{X \to String,\ \ X \to _[\ X*\ ]\}$$

where _ denotes the set of all possible tree labels. Then the result of a query $//a/b$ applied to a a tree of the above type (i.e., any XML tree) would be the type projector:

$$\{X \to _[\ X+\ ],\ X \to a[\ B+\ ],\ B \to b[\ Y*\ ],\ Y \to String,\ Y \to _[\ Y*\ ]\}$$

Note that this type-projector is non-deterministic top-down. It matches (and therefore keeps) any subtree $t$ if and only if there is a subtree $t'$ of $t$ with tag "a" which itself has a non-empty sequence of children tagged "b". Nodes that are children of an "a" node but whose tag is not "b" do not have any interpretation and therefore are discarded.

In [30], in order to achieve such a precise typing, the inference algorithm makes a heavy use of ℂDuce's type algebra (see [22]) in particular of intersection and negation types. Also note that the projector above is not obtained by erasing some of the rules of the original type but by the mean of set theoretic operations . In fact, three new rules were created and intuitively they were obtained by intersecting the initial "$X \to _[\ X*\ ]$" rules with a type "$a[\ B+]$" which is the constraint represented by the XPath query. Note also that contrary to our approach where new rules are only erasures of existing ones (of which there only exists a finite number), special care must be taken to not introduce infinitely many refined rules or, said differently, in this context even guarantying the termination of the algorithm is a very delicate issue.

### 8.2 Using regular tree languages as schemas

While our formal development remains in the very general case of regular tree grammars, our implementation only focuses on DTDs. The main reason is that for DTDs pruning is efficient memory-wise. For regular tree languages instead, validation (and pruning) may need to visit the whole tree before deciding which node to prune. At first, it seems that this completely defeats the purpose of pruning, but we argue that pruning can still be of practical use in these cases.

Indeed, a way of addressing this problem is to temporarily store the document in memory in the form of a succinct tree data-structure (based for instance on balanced parenthesis: a survey of the most popular succinct tree representations can be found in [2]). The final data-structure (e.g., a DOM representation) of the document can then be built from the temporary one, by replaying a sequence of SAX events while traversing the temporary data-structure and by not synthesizing events for pruned sub-trees. An alternative solution is to store on disk the sequence of SAX events and process it backward, thus simulating a bottom-up evaluation (validation of regular tree grammars and therefore projection can be done in a deterministic bottom-up fashion). Such a technique was used in [26] to efficiently evaluate node selecting queries bottom-up on documents up to 1 GB of size.

## 9 Experiments

### 9.1 Prototype

To gauge the benefits of type-based projection, we have implemented our pruning algorithm into a prototype. Our prototype takes as input an XQuery query, a DTD, and a document. It then performs the path extraction described in Section 7 and computes for each extracted path its XPath$^\ell$ approximation, applying the rewriting rules given in Section 6. Based on this set of paths, our program performs the static analysis described in Section 5.2 and computes a type projector. Once this is done, the prototype parses the input document, prunes it according to the inferred type-projector and serializes the result in a new document.

Besides what is included in the formal description of our algorithm, our prototype is extended also to support the full set of XPath axes as well as attributes. If we call $D$ the input document and $S$ the input DTD, then, assuming that $D$ is well-formed, the pruning process is performed in $O(|D|)$ time and $O(|S|)$ memory, where $|D|$ and $|S|$ denote the size of $D$ and $S$ respectively. Indeed, the type projector associated with a DTD is at most as big as the original schema (when no pruning is performed) and $O(|S|)$ space is required to store it in memory. Our prototype can also perform well-formedness check and validation while pruning, in which case time complexity remains $O(|D|)$ and memory complexity becomes $O(|S| + \log(|D|))$ (it is well-known that checking well-formedness during validation requires to keep a stack whose is size at most the height of the document, see e.g. [33]).

Our prototype is implemented in OCaml, using the PXP library for XML and DTD parsing.

## 9.2 Benchmark suite

We used the XPathMark ([21, 20]) and XMark ([32]) benchmark suites. The former consists of a large set of XPath queries while the latter provides XQuery queries to test against.

### 9.2.1 Data-set

Both XPathMark and XMark use the XMark document generators. These documents comply with the "auction" DTD representing an auction web-site. It defines 77 element types and 15 attributes. This size and complexity is comparable to "real-life" type definitions (for instance the XHTML transitional DTD also features 77 element definitions). Because the "auction" DTD falls outside the conditions of our completeness theorem (it features recursion and unguarded union), it is a very good test-case to illustrate the precision we achieve in practice even when completeness does not hold. The scalability of our approach was tested by using documents of varying size, ranging from 12MB to 3GB. An important aspect of the XMark generator is that the proportion of textual data versus tree structure stays the same, for all size of documents. We report here some statistics of interest which we use later-on to gauge the precision of our pruning algorithm. An XMark generated file consists of:

- 74% of text content (as PCDATA element or attribute value)
- 65% of all the text content (that is, 49% of the total file size) resides in a `description` element or one of its descendant.

### 9.2.2 XPathMark queries

Since its original publication ([21]), the XPathMark benchmark suite has evolved to provide a very complete set of XPath queries. It is composed of *functional test* queries, aiming at ensuring the correctness of an XPath implementation and *performance test* queries which provide computationally difficult queries. We highlight some of the main design goals of this test suite (the complete rationale can be found in [20]):

1. queries simulate *realistic* query needs of a potential user of the the auction site;
2. queries are divided into groups according to the intrinsic computational complexity of the corresponding evaluation problem. XPath language can be stratified in a number of fragments for which different complexity bounds are known [3]. Comparing the theoretical computational complexity of the query evaluation problem with the actual amount of resources consumed during query evaluation might be, at least, a stimulating and instructive exercise;

3. queries are defined to challenge data scalability of the XML processing system, that is the performance of the system as the data complexity (document size) grows. In particular, the queries talk about document sections (like open and closed auctions, items, people, descriptions) that become bigger when the XMark document scaling factor increases. Moreover, the results of the queries are small compared to the size of the target document. This avoids that the time taken to serialize the results (that may be relevant) obfuscates the pure query processing time.

These three points are exactly those we aim to address with the present work. Indeed our approach drastically increases the data scalability (*3.*) of XML processing systems for realistic queries (*1.*) and potentially complex ones (*2.*). XPathMark queries are divided in 5 groups, labeled from *A* to *E* that we briefly describe now.

**Group** *A* contains unary tree pattern queries. These queries use only child and descendant axes, node tests equal to `*` or to a tag name, and filters (predicates). Conjunctive and disjunctive Boolean operators are allowed, but negation is not. Relational and arithmetic operators and functions are disallowed. These queries fall therefore in the category of queries that we handle without any approximation.

**Group** *B* contains the so-called core or navigational XPath queries. This fragment extends XPath-A by admitting all XPath axes and negation. It mostly corresponds to queries for which our algorithm introduces a very lightweight approximation (we only need to approximate negations and those axes we did not treat formally such as `preceding`).

**Group** *C* extends Group *B* with relational operators (`=`, `!=`, `<`, `>`, `>=`, `<=`) and the `id()` function.

**Group** *D* extends Group *C* by allowing all arithmetic operators (`+`, `-`, `*`, `div`, `mod`) and functions `sum()` and `count()`.

**Group** *E* contains all XPath 1.0 queries. In particular, it extends Group *D* by allowing all functions (like `position()` and `contains()`).

XPathMark also provides a sixth group, which uses non-standard features of XPath, such that the transitive closure of a path expression. We excluded this group from our test, since neither our implementation nor the query engines that we used supported these extensions.

### 9.2.3 XMark test suite

To validate the extension of our approach to XQuery and in particular the path extraction algorithm, we use queries from the XMark benchmark suite ([32]). These queries feature "`for`" expressions guarded by "`where`" conditions and make use of element constructor to format their results. The corresponding code for the queries under consideration is given in Appendix B.

## 9.3 Protocol

We have designed two experiments, based on two different XQuery engine to validate our approach. For each engine and each query we described in the previous section we applied the following protocol. First, we tested the engine against original documents of increasing size and stopped when the query engine could not handle the input document anymore. Then we repeated the experiment a second time but used document pruned by our prototype as input for the query engine. We detail now our experimental settings for the two engine we considered: Saxon-b/XQuery and MonetDB/XQuery.

### 9.3.1 Test machine

The experiments where performed on an desktop PC, with an Intel Core 2 Xeon 3Ghz CPU, 3.5 GB of RAM and a S-ATA hard-drive. We used Ubuntu Linux 9.10 64bits (featuring a 2.6.31 kernel) as operating system. The file-system used was ext3, with default settings. The OS was allocated 6 GB of swap space and tests where done in a reduced environment, where only essential services were running concurrently with our experiments. In what follows, when timings are reported they are obtained by removing the best and worst timing of 5 runs and averaging the remaining three. Also, all the parameters we measured (running time, memory consumption, I/O operations, ...) were measured in independent runs in order to have as little impact as possible on the experiments. The memory consumption of a running program was measured by monitoring the so-called "resident set size" field of the `/proc/pid/statm` pseudo file (this fields summarizes the amount of private data mapped in physical memory by the process, excluding shared segments such as shared libraries or shared mmaped files). I/O operations where monitored using the `iotop` utility.

### 9.3.2 Saxon-b/XQuery

Saxon [25] is a popular XML library which implements various W3C standards (XPath, XQuery and XSLT) and has full schema support. We used version 9.0 of the Saxon-b XQuery engine (which is the Open Source one). Saxon being a main-memory query engine we focused on the following measurement both for pruned and unpruned documents: query answering time (excluding the parsing time of the document and serialization of the results, as reported by Saxon's debugging flags) and memory consumption. Saxon being written in Java, we used the latest version of the Sun's JVM available (1.6.0, 64 bit version) and set the amount of memory available to the JVM to the total physical memory of the machine.

### 9.3.3 MonetDB/XQuery

MonetDB/XQuery [12] is a well established native XML database with full XQuery support. Contrary to Saxon, MonetDB stores *on disk* an index allowing fast navigation and query answering. In particular since it uses the disc as secondary storage, MonetDB is not limited by the amount of physical memory (it uses as much memory as possible to answer a query efficiently and performs its own page management by mapping memory pages to the disk and reading them back when needed). Therefore for such query engine, speed is directly proportional to memory: the more memory is available, the less swapping occurs between pages on disk and pages in main memory. The three parameters we measured for MonetDB were the query answering time (again we did not consider document parsing or serialization time), the size of the generated index on disk for a given document and the amount of I/O performed to answer the query. Indeed since MonetDB tries to max out its memory use to favour query answering time, measuring memory consumption does not reflect the actual scalability improvement one could expect when pruning documents. Disk access on the contrary are the bottleneck for such an engine and their frequency directly impacts query answering time.

## 9.4 Experimental results

### 9.4.1 Pruning precision

We gauged the precision of our pruning algorithm for the full set of XPathMark and XMark queries by comparing the size of the pruned document (serialized on disk) with the size of the corresponding original document. We report in Figure 6 the pruning ratio in percent of the original file size for XPathMark (labelled A1 to E8) and XMark (labelled M1 to M20) queries.

#### 9.4.2 Saxon-b/XQuery

In our testing environment the biggest un-pruned document that the Saxon engine could handle was 671 MB large. We report in Figure 7 the original size of the largest *pruned* document Saxon could handle and the size of its projection (both in MB). Also, for a document of size 671 MB, we report the running time and memory consumption for the original and pruned version (as well as the size of the pruning). Lastly, we report the speed-up factor obtained thanks to pruning and the memory improvement we achieved (in percent of the original memory consumption) for the projected document. Due to the lack of space, we do not detail all of the XPathMark and XMark queries but rather for each category we give the "best performing query" –that is, the one for which we could achieve biggest speed-up– and the "worst performing query", the one for which the speed-up was the smallest.

#### 9.4.3 MonetDB/XQuery

Since MonetDB makes use of the secondary storage (disk) to query arbitrarily large documents, we chose a different approach to validate our pruning algorithm. We fixed the size of the input document to 3363 MB and then indexed it into the MonetDB document repository, yielding an index (on disk) of 4644 MB (as reported by the MonetDB administrative interface). Then for each query, we pruned the 3363 MB document with respect to the input query and indexed it. We summerize the results in Figure 8. The first line in the table reports the size in MB of the index corresponding to the pruned document. The second line reflects the ratio between the amount of I/O operation performed by the MonetDB server for the pruned file and the amount of I/O performed on the original file. We only take into account of amount of data *read* from disk which helps us gauge the amount of data fetched from the index on disk into main-memory. In this same figure, the graphics represent the absolute query answering time in seconds for the original and pruned document.

Finally, the third line gives the speed-up in query answering achieved through pruning. We were not able to run the query E5 on our version of MonetDB (the server segfaulted at some point during the query computation).

### 9.5 Interpretation

#### 9.5.1 Pruning precision

The results from Figure 6 shows that, for the vast majority of the queries we considered, the document can be pruned to less than 10% of its original size. More precisely on the 58 queries we considered (20 XMark queries and 38 XPathMark queries):

- 47 queries yielded a projected document whose size was less than 5% of the original
- 5 queries (M10, B3, B4, D2, E4) had a pruning ratio between 5% and 10% of the original file size
- 2 queries (B2, E3) had a pruning ratio of 17.035%
- 4 queries (M14, E6, E7, E8) had a pruning ratio of 27.35%

It should be noted that queries such as M14% return the content of a `description` element, consisting of almost all the textual data contained in the original XMark document. Since in these queries the value of the whole element is needed at runtime to perform string searching operations, there is little that can be done from the point of view of static pruning.

### 9.5.2 Saxon-b/XQuery

As we can see from the results in Figure 7 pruning the document before querying it always yields a speed-up and a reduction in memory use for a main memory engine such as Saxon.

On the one hand, for queries whose main bottleneck are string operations (such as calls to the `contains` function in M14 and E7), document projection gives very little speed-up. On the other hand, query A1 or C3 see a dramatic speed-up (20 and 17 times faster than the original respectively). This shows that despite the various optimizations built into the engine, a significant amount of time is often spent by iterating over "non-relevant" nodes which are discarded by the pruning process.

On the side of memory consumption, pruning the document unsurprisingly reduces memory usage drastically. Indeed, document projection reduces the number of elements and therefore simplifies the tree-structure of the XML document. This aspect is critical for main-memory engines which often (as in the case of Saxon) represent the document as a pointer-based data-structure (e.g., following the DOM model [17] where each element is represented as a node which contains a pointer to its first child to the next sibling, and to its parent). Indeed, we experienced for Saxon (but we observed similar behaviour in other main-memory query engines) a 112MB XMark document would occupy 430MB of RAM while the same document stripped of its data—amounting to only 36MB on disk—would occupy 340MB of memory. As Figure 7 illustrates, our pruning technique precisely addresses this issue, reducing in most cases the memory consumption to a few percents of what is needed to handle an un-pruned file.

### 9.5.3 MonetDB/XQuery

MonetDB is known to be one of the fastest XML database available. The efficiency of the MonetDB/XQuery engine is essentially due to the stair-case join operation ([24]) which minimizes the amount of intermediate sets constructed to answer an XPath query. Even so, the use of type-based document projection often improves query answering time. In particular, as shown in Figure 8, a smaller index often yields less I/O operations which in turn increases the speed of the query engine. On the contrary for queries such as D2, M14 and M15, the document is already optimally indexed and reducing the size of the index does not reduces the amount of I/O which explains why for these queries the gain in speed is null. Yet for some queries the speed-up can be up to twenty-folds (D1).

### 9.5.4 Comparison with related work

These results are a clear-cut improvement over current technology. While we cannot directly compare processing performances since no implementation of the other pruning approaches is publicly available, we want to stress two points. First, for XMark queries the pruning precision we achieve is equal or better than what is obtained with other approaches (with the exception of query M10 for which [27] achieves a pruning ratio of 4.5% where we could only prune the document down to 9.2% of its original size). Second, performing pruning never is a bottleneck in our case thanks to the fact that our solution consists of a single buffer-less traversal of the input document (on our test machine we were able to efficiently prune arbitrary large documents, while in case of [27] pruning can end up using as much memory as the execution of the query).

The experiments also illustrate that our approach retains a very high precision even in the presence of complex XPath features (like backward axes and external functions). While it is true that the technique of [31] could be used to allow Marian and Simeon's work to handle backward axes, it would still not be, to our sense, a satisfactory solution. The first reason is that the rewrite rules given in [31] do not support the use of data-value or negation in the filters of the original query (see [3]). For instance the query

$$\texttt{descendant}::\mathit{keyword}\,\texttt{[not(ancestor}::\mathit{item}\texttt{)]}$$

cannot be written without backward axis. Second the query generated may be exponentially bigger than the original one (and its computation takes exponential time in the size of the original) and may introduce several predicates as well as `descendant-or-self` axes. Both features degrades the pruning precision of [27].

# 10 Conclusion and future work

Our experiments show the clear advantages of applying our optimisation technique to query XML documents, and the characteristics of our solution make it profitable in all application scenarios. We discussed several aspects for which our approach improves the state of the art: for performances (better pruning, more speedup, less memory consumption), for the analysis techniques (linear pruning time, negligible memory and time consumption), for its generality (handling of all axes and of predicates), and, last but not least, for the formal foundation it provides (correctness formally proved, limits of the approach formally stated).

The present work extends and improves the preliminary version presented at VLDB 2006 [5] in several aspects. From a formal point of view, the use of regular tree grammars as schema model makes the technique applicable to the various kind of schemas currently in use. Furthermore, the closure properties that we proved ensure that type-based projection is at most as expensive as validation for a given class of schema language. We also handle a richer set of queries formally (in particular we handle nested predicates in XPath$^{\ell}$) and took special care to document how to encode or approximate several important XPath idioms that were lacking from the formal presentation. On a practical level, we have validated our approach against state of the art query engines, using realistic queries and data sets. In particular, not only did we test against an efficient main memory query engine (Saxon) but also demonstrated that our approach can be used to improve, sometimes by a double digit factor, the performances of an already very optimized disk-based XML database such as MonetDB/XQuery.

Future work will be pursued both at a formal and practical level. At a formal level, one of the main shortcoming of our approach is its reliance on XPath syntax. Indeed, even though we managed to isolate a fragment of XPath that we could formally reason with, it still leaves us with a syntax-directed approach. The problem with this is twofold. First, it makes the proofs and the specification of algorithms quite tedious and unnecessarily intricate. Second and more importantly, our pruning inference algorithm might yield different type projectors depending on the syntax of the original query. For future work, we would like to tackle a semantic based approach. In particular it seems worthwhile to consider more theoretically sound formalisms for tree queries such as, for instance, MSO formula or tree automata. The latter in particular would allow us to reuse our pruning algorithm for pattern-matching based languages (such as the ℂDuce language [1] and its pattern-based query language ℂQL [28, 6, 14]). It is also known that tree-automata (as well as MSO formula) have better closure properties than XPath expressions and support fine-grained set-theoretic operations (intersection, union, complement) that have been used with success to devise very precise type-systems for XML [22].

At a practical level we would like to see a tighter integration between document-projection and query engines. Firstly, although quite crude, our experiments show that even a carefully designed indexed system such as MonetDB can benefit from document pruning. It seems interesting to develop further such preliminary results and design a projection aware XML index. In other words we would like to be able to equip any native XML query engine optimizer with a type-projector component. In particular, one could think of an index consisting of the original document together with its projected versions. Textual data could be shared between the main document and the projected ones which would merely become a projected view of the tree structure of the document. We make the hypothesis that the overhead of such pruned tree structures would be quite small compared to the size of an XML index while providing significant speed-up in query answering time.

Secondly, coupling our projection algorithm with early query answering techniques would allow us to achieve further pruning, especially when runtime conditions are involved. For instance we could use our type-inference algorithm to determine on what type of elements a given built-in function is applied to; for instance in an expression such as

`contains(.//*, "foo").`

This information could then be used at loading time to discard elements that do not match the predicate.

# A Detailed proofs

LEMMA ((3.3)) *Let $\pi$ be a type projector for $(\mathscr{S},E)$. Then for every tree $t \in_{\mathfrak{I}} (\mathscr{S},E)$ it holds $(t \backslash_{\mathfrak{I}} \pi) \preceq t$.*

> **Proof 1** *The proof is a straightforward induction on $t$.*

LEMMA (ERASURE PRESERVES LOCALITY (3.4)) *Let $(\mathscr{S},E)$ be a local tree grammar and $(\mathscr{S}',E')$ a regular tree grammar. If $(\mathscr{S}',E') <: (\mathscr{S},E)$ then $(\mathscr{S}',E')$ is a local tree grammar.*

> **Proof 2** *By contradiction, suppose that $(\mathscr{S}',E')$ is not a local tree grammar. By Definition 2.12, $\mathscr{S}' \subseteq \mathscr{S}$ therefore, $|\mathscr{S}'| \leq |\mathscr{S}| \leq 1$.*
>
> - *Then, either there exist two competing rules $A \to l[r'_a]$ and $B \to l[r'_b]$ in $E'$. Then by definition of erasure, there exist two rules $A \to l[r_a]$ and $B \to l[r_b]$ in $E$ such that $r'_a = r_a|_{N_a}$ and $r'_b = r_b|_{N_b}$ for some $N_a \subseteq \textbf{Names}(r_a)$ and $N_b \subseteq \textbf{Names}(r_b)$. But then these two rules share the same label $l$, and therefore are competing one with the other, which contradicts the fact that $(\mathscr{S},E)$ is a local tree grammar.*
> - *Or there exists two rules $C \to l[r_1]$ and $C \to l'[r_2]$ in $E'$ with the same left hand-side (and distinct labels). But then, by definition of erasure, there exists two corresponding rules in $E$, $C \to l[r'_1]$ and $C \to l'[r'_2]$ such that $r_i = r'_i|_{N_i}, i \in \{1,2\}$ for some names $N_i$. Therefore there are two rules in $E$ with the same left-hand side, which contradicts the fact that $(\mathscr{S},E)$ is a local-tree grammar.*
>
> □

LEMMA (ERASURE PRESERVES SINGLE-TYPEDNESS (3.5)) *Let $(\mathscr{S},E)$ be a single-type tree grammar and $(\mathscr{S}',E')$ a regular tree grammar. If $(\mathscr{S}',E') <: (\mathscr{S},E)$ then $(\mathscr{S}',E')$ is a single-type tree grammar.*

> **Proof 3** *By contradiction suppose that $(\mathscr{S}',E')$ is not a single-type tree grammar and proceed by case analysis:*
>
> - *either there exist two competing non terminals $A$ and $B$ in $\mathscr{S}'$. But by definition of erasure, $\mathscr{S}' \subseteq \mathscr{S}$ and $(\mathscr{S},E)$ has two competing start symbols, which contradicts the hypothesis that $(\mathscr{S},E)$ enjoys the single-type property.*
> - *or there exists a rule $X \to l[\ r'\ ]$ and there exist two competing non terminals $A$ and $B$ in $\textbf{Names}(r')$. Since $(\mathscr{S}',E') <: (\mathscr{S},E)$, then there exists $X \to l[\ r\ ]$ such that $r' = r|_N$ for some $N \subseteq \textbf{Names}(r)$. But that means that $A$ and $B$ are in $\textbf{Names}(r)$, which implies that $(\mathscr{S},E)$ is not a single-type tree grammar, thus contradicting our hypothesis.* □

LEMMA (UNION CLOSURE OF LOCAL TYPE PROJECTORS (3.6)) *Let $(\mathscr{S},E)$ be a local tree grammar. Let $(\mathscr{S}_1,E_1)$ and $(\mathscr{S}_2,E_2)$ be two tree grammars such that $(\mathscr{S}_1,E_1) <: (\mathscr{S},E)$ and $(\mathscr{S}_2,E_2) <: (\mathscr{S},E)$. Then $(\mathscr{S}_1 \cup \mathscr{S}_2, E_1 \cup E_2)$ is a local tree grammar.*

**Proof 4** *Consider $(\mathscr{S}_1 \cup \mathscr{S}_2, E_1 \cup E_2)$ and suppose, by contradiction, that it is not local. First, remark that by definition of erasure, $\mathscr{S}_1 \subseteq \mathscr{S}$ and $\mathscr{S}_2 \subseteq \mathscr{S}$, therefore $\mathscr{S}_1 \cup \mathscr{S}_2 \subseteq \mathscr{S}$ and consequently $|\mathscr{S}_1 \cup \mathscr{S}_2| \leq |\mathscr{S}| \leq 1$. Second:*

- *either we have two rules $A \to l[r_a]$ and $B \to l[r_b]$ (with $A$ and $B$ distinct). By Lemma 3.4 we know that $(\mathscr{S}_1,E_1)$ and $(\mathscr{S}_2,E_2)$ are local tree grammars. Then it must be that one of two rules at issue is in $(\mathscr{S}_1,E_1)$ and the other in $(\mathscr{S}_2,E_2)$, otherwise one of the two grammars would not be local (it would have the competing pair in its rules). Without loss of generality we can suppose that $A \to l[r_a] \in E_1$ and $B \to l[r_b] \in E_2$. Since $(\mathscr{S}_1,E_1) <: (\mathscr{S},E)$, then by definition of erasure there exists a rule $A \to l[r'_a] \in E$ with $r_a = r'_a|_N$ for some $N \subseteq$ **Names**$(r'_a)$. Similarly, there exists $B \to l[r'_b] \in E$ with $r_b = r'_b|_{N'}$ for some $N' \subseteq$ **Names**$(r'_b)$. But then, this means that we have two competing rules in $E$, which contradicts the hypothesis that $(\mathscr{S},E)$ is a local tree grammar.*

- *or, there are two rules $C \to l[r_1]$ and $C \to l'[r_2]$ with the same left-hand side and distinct labels in $E_1 \cup E_2$. Similarly to the previous case, we must have $C \to l[r_1] \in E_1$ and $C \to l'[r_2] \in E_2$, otherwise $(\mathscr{S}_1,E_1)$ or $(\mathscr{S}_2,E_2)$ would not be local. But then by definition of erasure, it means that there exists two rules, $C \to l[r'_1] \in E$ and $C \to l'[r'_2] \in E$ such that $r_i = r'_i|_{N_i}, i \in \{1,2\}$ for some names $N_i$. This means that there are two rules in $E$ with distinct labels and the same left-hand side, which contradicts the assumption that $(\mathscr{S},E)$ is a local tree grammar.*

□

LEMMA (UNION CLOSURE OF SINGLE-TYPE TYPE PROJECTORS (3.7))

*Let $(\mathscr{S},E)$ be a single-type tree grammar. Let $(\mathscr{S}_1,E_1)$ and $(\mathscr{S}_2,E_2)$ be two tree grammars such that $(\mathscr{S}_1,E_1) <: (\mathscr{S},E)$ and $(\mathscr{S}_2,E_2) <: (\mathscr{S},E)$. Then $(\mathscr{S}_1 \cup \mathscr{S}_2, E_1 \cup E_2)$ is a single-type tree grammar.*

**Proof 5** *Consider $(\mathscr{S}_1 \cup \mathscr{S}_2, E_1 \cup E_2)$ and suppose by contradiction that it does not enjoy the single-type property. This implies that there exists a rule $X \to l[r]$, such that* **Names**$(r)$ *contains two competing non-terminals $A$ and $B$. Moreover, by Lemma 3.5 we know that $(\mathscr{S}_1,E_1)$ and $(\mathscr{S}_2,E_2)$ are single-type tree grammars. Therefore $X \to l[r]$ cannot be in $E_1$ nor in $E_2$ because they have the single-type property. The only solution is that there exists a rule $X \to l[r_1]$ in $E_1$ with $A \in$* **Names**$(r_1)$ *and $X \to l[r_2]$ in $E_2$ with $B \in$* **Names**$(r_2)$*, and that $r = r_1|r_2$ (remember that we identify two rules with the same left-hand side and the same label by merging them into a single rule). Therefore, by the definition of erasure, there exists a rule $X \to l[r'_1]$ in $E$ such that $A \in Names(r'_1)$. Similarly, there exists a rule $X \to l[r'_2]$ in $E$ such that $B \in Names(r'_2)$. Since we identify such rules, there is a rule $X \to l[r'_1|r'_2]$ in $E$. But then, this rule contains both $A$ and $B$ which are competing. This contradict the hypothesis that $(\mathscr{S},E)$ is a single-type tree grammar.* □

LEMMA (ERASURE PRESERVES LOCALITY (3.4) *Let $(\mathscr{S},E)$ be a local tree grammar*

*and $(\mathscr{S}',E')$ a regular tree grammar. If $(\mathscr{S}',E') <: (\mathscr{S},E)$ then $(\mathscr{S}',E')$ is a local tree grammar.*

**Proof 6** *By contradiction, suppose that $(\mathscr{S}',E')$ is not a local tree grammar. By Definition 2.12, $\mathscr{S}' \subseteq \mathscr{S}$ therefore, $|\mathscr{S}'| \leq |\mathscr{S}| \leq 1$.*

- *Then, either there exist two competing rules $A \to l[r'_a]$ and $B \to l[r'_b]$ in $E'$. Then by definition of erasure, there exist two rules $A \to l[r_a]$ and $B \to l[r_b]$ in $E$ such that $r'_a = r_a|_{N_a}$ and $r'_b = r_b|_{N_b}$ for some $N_a \subseteq$ **Names**$(r_a)$ and $N_b \subseteq$ **Names**$(r_b)$. But then these two rules share the same label $l$, and therefore are competing one with the other, which contradicts the fact that $(\mathscr{S},E)$ is a local tree grammar.*
- *Or there exists two rules $C \to l[r_1]$ and $C \to l'[r_2]$ in $E'$ with the same left hand-side (and distinct labels). But then, by definition of erasure, there exists two corresponding rules in $E$, $C \to l[r'_1]$ and $C \to l'[r'_2]$ such that $r_i = r'_i|_{N_i}, i \in \{1,2\}$ for some names $N_i$. Therefore there are two rules in $E$ with the same left-hand side, which contradicts the fact that $(\mathscr{S},E)$ is a local-tree grammar.*

□

LEMMA (ERASURE PRESERVES SINGLE-TYPEDNESS (3.5))
*Let $(\mathscr{S},E)$ be a single-type tree grammar and $(\mathscr{S}',E')$ a regular tree grammar. If $(\mathscr{S}',E') <: (\mathscr{S},E)$ then $(\mathscr{S}',E')$ is a single-type tree grammar.*

**Proof 7** *By contradiction suppose that $(\mathscr{S}',E')$ is not a single-type tree grammar and proceed by case analysis:*

- *either there exist two competing non terminals $A$ and $B$ in $\mathscr{S}'$. But by definition of erasure, $\mathscr{S}' \subseteq \mathscr{S}$ and $(\mathscr{S},E)$ has two competing start symbols, which contradicts the hypothesis that $(\mathscr{S},E)$ enjoys the single-type property.*
- *or there exists a rule $X \to l[\,r'\,]$ and there exist two competing non terminals $A$ and $B$ in **Names**$(r')$. Since $(\mathscr{S}',E') <: (\mathscr{S},E)$, then there exists $X \to l[\,r\,]$ such that $r' = r|_N$ for some $N \subseteq$ **Names**$(r)$. But that means that $A$ and $B$ are in **Names**$(r)$, which implies that $(\mathscr{S},E)$ is not a single-type tree grammar, thus contradicting our hypothesis.* □

LEMMA (UNION CLOSURE OF LOCAL TYPE PROJECTORS (3.6)) *Let $(\mathscr{S},E)$ be a local tree grammar. Let $(\mathscr{S}_1,E_1)$ and $(\mathscr{S}_2,E_2)$ be two tree grammars such that $(\mathscr{S}_1,E_1) <: (\mathscr{S},E)$ and $(\mathscr{S}_2,E_2) <: (\mathscr{S},E)$. Then $(\mathscr{S}_1 \cup \mathscr{S}_2, E_1 \cup E_2)$ is a local tree grammar.*

**Proof 8** *Consider $(\mathscr{S}_1 \cup \mathscr{S}_2, E_1 \cup E_2)$ and suppose, by contradiction, that it is not local. First, remark that by definition of erasure, $\mathscr{S}_1 \subseteq \mathscr{S}$ and $\mathscr{S}_2 \subseteq \mathscr{S}$, therefore $\mathscr{S}_1 \cup \mathscr{S}_2 \subseteq \mathscr{S}$ and consequently $|\mathscr{S}_1 \cup \mathscr{S}_2| \leq |\mathscr{S}| \leq 1$. Second:*

- *either we have two rules $A \to l[r_a]$ and $B \to l[r_b]$ (with $A$ and $B$ distinct). By Lemma 3.4 we know that $(\mathscr{S}_1,E_1)$ and $(\mathscr{S}_2,E_2)$ are local tree grammars. Then it* *must be that one of two rules at issue is in $(\mathscr{S}_1, E_1)$ and the other in $(\mathscr{S}_2, E_2)$, otherwise one of the two grammars would not be local (it would have the competing pair in its rules). Without loss of generality we can suppose that $A \to l[r_a] \in E_1$ and $B \to l[r_b] \in E_2$. Since $(\mathscr{S}_1, E_1) <: (\mathscr{S}, E)$, then by definition of erasure there exists a rule $A \to l[r'_a] \in E$ with $r_a = r'_a|_N$ for some $N \subseteq$ **Names**$(r'_a)$. Similarly, there exists $B \to l[r'_b] \in E$ with $r_b = r'_b|_{N'}$ for some $N' \subseteq$ **Names**$(r'_b)$. But then, this means that we have two competing rules in $E$, which contradicts the hypothesis that $(\mathscr{S}, E)$ is a local tree grammar.*

- *or, there are two rules $C \to l[r_1]$ and $C \to l'[r_2]$ with the same left-hand side and distinct labels in $E_1 \cup E_2$. Similarly to the previous case, we must have $C \to l[r_1] \in E_1$ and $C \to l'[r_2] \in E_2$, otherwise $(\mathscr{S}_1, E_1)$ or $(\mathscr{S}_2, E_2)$ would not be local. But then by definition of erasure, it means that there exists two rules, $C \to l[r'_1] \in E$ and $C \to l'[r'_2] \in E$ such that $r_i = r'_i|_{N_i}, i \in \{1,2\}$ for some names $N_i$. This means that there are two rules in $E$ with distinct labels and the same left-hand side, which contradicts the assumption that $(\mathscr{S}, E)$ is a local tree grammar.*

□

LEMMA (UNION CLOSURE OF SINGLE-TYPE TYPE PROJECTORS (3.7)) *Let $(\mathscr{S}, E)$ be a single-type tree grammar. Let $(\mathscr{S}_1, E_1)$ and $(\mathscr{S}_2, E_2)$ be two tree grammars such that $(\mathscr{S}_1, E_1) <: (\mathscr{S}, E)$ and $(\mathscr{S}_2, E_2) <: (\mathscr{S}, E)$. Then $(\mathscr{S}_1 \cup \mathscr{S}_2, E_1 \cup E_2)$ is a single-type tree grammar.*

**Proof 9** *Consider $(\mathscr{S}_1 \cup \mathscr{S}_2, E_1 \cup E_2)$ and suppose by contradiction that it does not enjoy the single-type property. This implies that there exists a rule $X \to l[r]$, such that* **Names**$(r)$ *contains two competing non-terminals $A$ and $B$. Moreover, by Lemma 3.5 we know that $(\mathscr{S}_1, E_1)$ and $(\mathscr{S}_2, E_2)$ are single-type tree grammars. Therefore $X \to l[r]$ cannot be in $E_1$ nor in $E_2$ because they have the single-type property. The only solution is that there exists a rule $X \to l[r_1]$ in $E_1$ with $A \in$* **Names**$(r_1)$ *and $X \to l[r_2]$ in $E_2$ with $B \in$* **Names**$(r_2)$*, and that $r = r_1|r_2$ (remember that we identify two rules with the same left-hand side and the same label by merging them into a single rule). Therefore, by the definition of erasure, there exists a rule $X \to l[r'_1]$ in $E$ such that $A \in Names(r'_1)$. Similarly, there exists a rule $X \to l[r'_2]$ in $E$ such that $B \in Names(r'_2)$. Since we identify such rules, there is a rule $X \to l[r'_1|r'_2]$ in $E$. But then, this rule contains both $A$ and $B$ which are competing. This contradict the hypothesis that $(\mathscr{S}, E)$ is a single-type tree grammar.* □

LEMMA (5.2) *Let $t$ be a tree $\mathfrak{I}$-valid with respect to the schema $(\mathscr{S}, E)$. For every $S \subseteq$* **Ids**$(t)$ *and type $\tau$, if $\mathfrak{I}(S) \subseteq$* **Dn**$(\tau)$*, then*

1. $\mathfrak{I}([\![Axis]\!]_t(S)) \subseteq \textbf{Dn}(\mathbf{A}_E(\tau, Axis))$
2. $\mathfrak{I}(S::_t Test) \subseteq \textbf{Dn}(\mathbf{T}_E(\tau, Test))$

**Proof 10** *The proof is done by case analysis on the possible axes (for (1.)) and tests (for (2.)).*

1. *We only need to consider the* `self` *and* `descendant` *axes. Indeed* `child` *is an instance of* `descendant`*, and* `ancestor` *and* `parent` *are the dual of* `descendant` *and* `child` *respectively.*

   `self`**:** *By Definition 4.3, we have that* $[\![\texttt{self}]\!]_t(S) = S$*, therefore* $\mathfrak{I}([\![\texttt{self}]\!]_t(S)) = \mathfrak{I}(S)$*. By Definition 5.1,* $\mathbf{A}_E(\tau, \texttt{self}) = \tau$*. Since* $\mathfrak{I}(S) \subseteq \boldsymbol{Dn}(\tau)$ *by hypothesis, we can conclude that*

   $$\mathfrak{I}([\![\texttt{self}]\!]_t(S)) \subseteq \boldsymbol{Dn}(\mathbf{A}_E(\tau, \texttt{self}))$$

   `descendant`**:** *we suppose* $\mathbf{i}' \in [\![\texttt{descendant}]\!]_t(S)$*. Let us show that* $\mathfrak{I}(\mathbf{i}') \in \boldsymbol{Dn}(\mathbf{A}_E(\tau, \texttt{descendant}))$*. If* $\mathbf{i}' \in [\![\texttt{descendant}]\!]_t(S)$ *then by Definition 4.3:*

   $$\exists \mathbf{i} \in \boldsymbol{Ids}(t) \text{ such that } (\mathbf{i}, \mathbf{i}') \in \boldsymbol{Edg}(t)^+$$

   *By definition of* ***Edg*** *this means that there exists a sequence* $\mathbf{i}_0, \mathbf{i}_1, \ldots, \mathbf{i}_n$ *such that* $t@\mathbf{i}_k = l[\ldots t' \ldots]$ *and* $\boldsymbol{RootId}(t') = \mathbf{i}_{k+1}$*, with* $\mathbf{i}_0 = \mathbf{i}$ *and* $\mathbf{i}_n = \mathbf{i}'$*. Said differently,* $\mathbf{i}_0, \mathbf{i}_1, \ldots, \mathbf{i}_n$ *is the path from* $\mathbf{i}$ *to its descendant* $\mathbf{i}'$ *in the tree t that we consider.*

   *Let us now call* $X_k = \mathfrak{I}(\mathbf{i}_k)$*. For all k, there is a rule* $X_k \to R_k \in E$ *such that* $X_{k+1} \in \boldsymbol{Names}(R_k)$ *(since t is* $\mathfrak{I}$*-valid with respect to E and by Definition 2.8). Hence, there exists a chain* $X_0 \Rightarrow_E \ldots \Rightarrow_E X_n$ *with* $X_0$ *in* $\mathfrak{I}(S)$*. By hypothesis,* $\mathfrak{I}(S) \subseteq \boldsymbol{Dn}(\tau)$*, therefore* $X_0 \in \tau$*. Since* $X_0 \Rightarrow_E \ldots \Rightarrow_E X_n$*, by Definition 4.3 we have* $X_n \to R_n \in \mathbf{A}_E(\tau, \texttt{descendant})$*. But since,* $X_n = \mathfrak{I}(\mathbf{i}')$*, we have*

   $$\mathfrak{I}(\mathbf{i}') \in \boldsymbol{Dn}(\mathbf{A}_E(\tau, \texttt{descendant}))$$

   *and therefore*

   $$\mathfrak{I}([\![\texttt{descendant}]\!]_t(S)) \subseteq \boldsymbol{Dn}(\mathbf{A}_E(\tau, \texttt{descendant}))$$

2. *By case on the test:*

   `node`**:** *By Definition 4.2 we have:*

   $$\texttt{node::}_t S = S$$

   *By Definition 5.1 we have* $\mathbf{T}_E(\tau, \texttt{node}) = \tau$*. Since* $\mathfrak{I}(S) \subseteq \boldsymbol{Dn}(\tau)$ *by hypothesis, we have that:*

   $$\mathfrak{I}(S\texttt{::}_t\texttt{node}) \subseteq \boldsymbol{Dn}(\mathbf{T}_E(\tau, \texttt{node}))$$

   $a$ **(for some element name** $a$**):** *suppose* $\mathbf{i} \in a\texttt{::}_t S$*. Let us show that* $\mathfrak{I}(\mathbf{i}) \in \boldsymbol{Dn}(\mathbf{T}_E(\tau, a))$*. By Definition 4.2, we know that* $t@\mathbf{i} = a[f]$ *for some forest* $f$*. Since t is* $\mathfrak{I}$*-valid with respect to E, then* $\mathfrak{I}(\mathbf{i}) = Y$ *and there exists a rule* $Y \to a[R]$ *in E. Since* $\mathbf{i} \in S$*, we also have that* $\mathfrak{I}(\mathbf{i}) \in \boldsymbol{Dn}(\tau)$ *(by hypothesis). By Definition 5.1, since* $Y \to a[R] \in \tau$*, then* $Y \to a[R] \in \mathbf{T}_E(\tau, a)$ *and therefore* $\mathfrak{I}(\mathbf{i}) = Y \in \boldsymbol{Dn}(\mathbf{T}_E(\tau, a)$*, hence*

   $$\mathfrak{I}(S\texttt{::}_t a \subseteq \boldsymbol{Dn}(\mathbf{T}_E(\tau, a))$$

   `text`**:** *similar to the previous case.*

□

LEMMA (TERMINATION OF TYPE INFERENCE (5.5)) *Let $(\mathscr{S}, E)$ be a type, P a path, and $\Sigma$ and $\Sigma'$ two environments. If there is a derivation for the judgment $\Sigma \vdash_E P : \Sigma'$, then this derivation is unique and finite.*

**Proof 11** *Uniqueness of the derivation is immediate, since the rules are syntax-directed: at each step, at most one of the rules applies (if no rule applies, there is no derivation and the output type is $\varnothing$). Finiteness can be shown by a simple induction on the length of the path, noted $l(P)$, that we define as follows:*

$$\begin{array}{lcl} l(Axis::Test) & = & 1 \\ l(Axis::Test\texttt{[}C\texttt{]}) & = & 1 + \sum_{P \in C} l(P) \\ l(P/P') & = & l(P) + l(P') \\ l(P|P') & = & l(P) + l(P') \end{array}$$

**Basic case:** *The query has length 1, meaning it is a single step without predicate. Then the only rules we can apply are **(down-axis)**, **(up-axis)** or **(test)**. These rules have no premise, therefore the derivation is finite and has length 1.*

**Inductive case:** *If the query has several steps, then the rule **(sequence)** applies. The lengths of the queries in the premises of the rule are strictly less than the length of the query in the goal, by definition of $l(_)$. By induction hypothesis both premises have a finite derivation, therefore the goal can be derived with a finite derivation.*

*Similarly, if the query is a top-level union, the typing rule **(union)** applies.*

*If the query is a single step with a predicate, then rule **(predicate)** applies. We should first remark that there is a finite set of rules in $\Sigma_{typ}$ and a finite number of path n in Cond. Thus, there are exactly $|\Sigma_{typ}| \times n$ premises for this rule. For each one of these premises, the path $P_{jk}$ is such that $l(P_{jk}) < l(P)$, by definition of $l(_)$. By induction hypothesis, every premise has finite derivation, therefore the judgment in the goal of the rule has a finite derivation.*

□

THEOREM (SOUNDNESS OF TYPE INFERENCE (5.6)) *Let $(\mathscr{S}, E)$ be a type and P a path. Let $E_0 = \{X \to R \mid X \to R \in E, X \in \mathscr{S}\}$. If $(E_0, E_0) \vdash_E P : (\tau, \kappa)$ then:*

$$\boldsymbol{Dn}(\tau) \supseteq \bigcup_{t \in \mathfrak{I}(\mathscr{S}, E)} \mathfrak{I}([\![P]\!]_t(\boldsymbol{RootId}(t)))$$

**Proof 12** *We consider the following, more general judgment:*

$$(\tau, \kappa) \vdash_E P : (\tau', \kappa')$$

*We show simultaneously the following properties:*

1. *Soundness : for all tree t $\mathfrak{I}$-valid with respect to $(\mathscr{S}, E)$ and all set $S \subseteq Ids(t)$, if $\mathfrak{I}(S) \subseteq \boldsymbol{Dn}(\tau)$ then:*
$$\mathfrak{I}([\![P]\!]_t(S)) \subseteq \boldsymbol{Dn}(\tau')$$
2. *Context well-formedness, if*

$$\kappa = \{Y \to R \mid \forall Z \in \boldsymbol{Dn}(\tau), X \Rightarrow_E^* Y \Rightarrow_E^* Z\}$$

*then*

$$\kappa' = \{Y \to R \mid \forall Z \in \boldsymbol{Dn}(\tau'), X \Rightarrow_E^* Y \Rightarrow_E^* Z\}$$

*Property 1 is a generalization of the soundness property we are proving. Given a set of context nodes S whose types are in $\tau$, the type of $[\![P]\!]_t(S)$ is in $\tau'$. Property 2 states that the algorithm preserves the well-formedness of contexts. We prove both properties by induction on the depth of the typing derivation, which is finite by Lemma 5.5:*

**Base case:**

**(down-axis):** *Property 1 is true by a direct application of Lemma 5.2. Property 2 holds by definition of $\mathbf{A}_E(_,_)$*

**(up-axis):** *By Lemma 5.2:*

$$\mathfrak{I}([\![Axis{::}\texttt{node}]\!]_t(S)) \subseteq \boldsymbol{Dn}(\mathbf{A}_E(\tau, Axis))$$

*We must now show that $\mathfrak{I}([\![Axis{::}\texttt{node}]\!]_t(S))$ is in $\Sigma_{ctx}$, for it to be in the intersection of both. Since $\kappa$ is a well-formed context:*

$$\kappa = \{Y \to R \mid \forall Z \in \boldsymbol{Dn}(\tau), X \Rightarrow_E^* Y \Rightarrow_E^* Z\}$$

*Let us first consider the case Axis = `ancestor`. By Definition 4.3:*

$$[\![\texttt{ancestor}{::}\texttt{node}]\!]_t(S) = \{\mathbf{i}' \mid \mathbf{i} \in S \wedge (\mathbf{i}', \mathbf{i}) \in \boldsymbol{Edg}^+(t)\}$$

*Thus*

$$\mathfrak{I}(\{\mathbf{i}' \mid \mathbf{i} \in S \wedge (\mathbf{i}', \mathbf{i}) \in \boldsymbol{Edg}^+(t)\}) = \{Y | Z \in \mathfrak{I}(S) \wedge Y \Rightarrow_E^+ Z\}$$

*Since we supposed $\mathfrak{I}(S) \subseteq \boldsymbol{Dn}(\tau)$, then clearly:*

$$\{Y \to R | Z \in \mathfrak{I}(S) \wedge Y \Rightarrow_E^+ Z\} \subseteq \kappa$$

*thus*

$$\mathfrak{I}([\![Axis :: \texttt{node}]\!]_t(S)) \subseteq \boldsymbol{Dn}(\mathbf{A}_E(\tau, Axis) \cap \kappa)$$

*which proves Property 1. The case for the "`parent`" axis is a particular instance of "`ancestor`". As for Property 2, $\kappa$ is the set of rules used to derive the context node type $\tau$. $\mathbf{A}_E(\kappa, Axis)$ is the set of all the parent rules (or ancestor rules) of the rules in $\kappa$. Consequently, the intersection is still a well formed context.*

**(test)** *: Similarly to the case of Rule **(down-axis)**, Property 1 is a direct application of Lemma 5.2. For Property 2, we can remark that*

$$\kappa' = \kappa \cap \mathbf{A}_E(T_E(\tau, Test), \texttt{ancestor})$$

*contains all the rules leading to a node in $\tau$ for which Test succeeds (including the ones of the selected node), therefore it is a well-formed context.*

**Inductive case:**

**(predicate)** *: Let us consider:*

$$[\![\texttt{self}{::}\texttt{node[}\, \texttt{Or}_j\, \texttt{And}_k\, P_{jk}\, \texttt{]}]\!]_t(S)$$

*By Definition 4.4, we have*

$$[\![\texttt{self}{::}\texttt{node[}\, \texttt{Or}_j\, \texttt{And}_k\, P_{jk}\, \texttt{]}]\!]_t(S) = \bigcup_{\mathbf{i} \in T} \mathbf{i}$$

*where $T$ is the set of ids satisfying the predicate:*

$$T = \{\mathbf{i} | \mathbf{i} \in S \wedge \bigvee_j \bigwedge_k [\![P_{jk}]\!]_t(\{\mathbf{i}\}) \neq \varnothing\}$$

*Let us consider $\mathbf{i} \in T$. We have that $\mathbf{i} \in S$, and since $\mathbf{i}$ is part of an $\mathfrak{I}$-valid tree $t$, there exists $X_i = \mathfrak{I}(\mathbf{i})$ and an associated rule $X_i \to R_i$. We apply the induction hypothesis on:*

$$(\{X_i \to R_i\}, \Sigma_{ctx}) \vdash_E P_{jk} : \Sigma^{ijk}$$

*therefore we have*

$$\mathfrak{I}([\![P_{jk}]\!]_t(\{\mathbf{i}\})) \subseteq \Sigma^{ijk}_{typ}$$

*Consequently, $\Sigma^{ijk}_{typ}$ is not empty: $X_i \to R_i \in \tau'$ (the output type). So*

$$\mathfrak{I}([\![\texttt{self}::\texttt{node}[\,\texttt{Or}_j\,\texttt{And}_k\,P_{jk}\,]]\!]_t(S)) \subseteq \textbf{\textit{Dn}}(\tau')$$

*which proves Property 1. Property 2 holds for the same argument as in rule **(test)**.*

**(sequence)** *: Property 1 is true by induction hypothesis on both premises. Property 2 is true for the first premise, by induction hypothesis. In particular, $\Sigma''_{ctx}$ is a well-formed context. We can then apply the induction hypothesis on $\Sigma''$ and we have that $\Sigma'_{ctx}$ is a well-formed context too.*

**(union)** *: is similar to the previous case.*

□

LEMMA (WITNESS OF A GRAMMAR (5.8)) *Let $(\mathscr{S}, E)$ be a non-recursive, $*$-guarded, parent-unambiguous local tree grammar. There exists a document $t$, $\mathfrak{I}$-valid with respect to $(\mathscr{S}, E)$ such that:*

$$\forall X \in \textbf{\textit{Dn}}(E), \exists \mathbf{i} \in \textbf{\textit{Ids}}(t) \text{ such that } \mathfrak{I}(\mathbf{i}) = X$$

*we call such a document a* witness *of the schema $(\mathscr{S}, E)$.*

**Proof 13** *Since the tree grammar is non recursive and parent unambiguous, we can prove the lemma by induction **(I)** on the height of the grammar, seen as a DAG.*

**Basic case:** *the grammar has height 1. It consists therefore of a single rule. The rule is either $X \to String$ and a document $s_{\mathbf{i}}$ is a suitable witness; or the rule is $X \to a[\,]$ for some label $a$ and the document $a_{\mathbf{i}}[\,]$ is a witness of the grammar.*

**Inductive case:** *Consider $(\{X\}, E)$. The rule for the start symbol $X$ is $X \to a[r_1 \cdots r_n]$ for some label $a$ (since $E$ is $*$-guarded, the rule must have this shape). We show by induction **(II)** on the structure of the regular expression $r_i$ that there is witness for this regular expression.*

**Basic case:** *Either $r_i = \varepsilon$, and therefore the empty forest $()$ is a suitable witness. Or $r_i = Z$. Then consider the grammar $(\{Z\}, E')$ where $E' = \{Y \to R \mid Y \to R \in E, Z \Rightarrow^*_E Y\}$, that is the restriction of $E$ to $Z$. Then, height$(E') <$ height$(E)$ since at least the rule associated with $X$ is not in $E'$ (and because $E$ is not recursive and parent unambiguous). Therefore, by induction hypothesis **(I)** there exists a witness $t_z$ for $Z$.*

**Inductive case:** *Either* $r_i = (r'_i | r''_i)*$ *and by induction hypothesis **(II)**, there is a witness* $t'_i$ *for* $r'_i$ *and* $t''_i$ *for* $r''_i$. *Then, the forest* $t'_i, t''_i$ *is a witness for* $r_i$ *(the first iteration of* $*$ *matches* $t_i$ *and the second one matches* $t''_i$*).*

*Or* $r_i = (r'_i)*$ *and* $r'_i$ *is not a union. Then by induction hypothesis **(II)**, there is a witness* $t'_i$ *and* $t'_i$ *is also a witness for* $r_i$. *Or* $r_i = r'_i r''_i$. *By induction hypothesis **(II)**, there is a witness* $t'_i$ *for* $r'_i$ *and* $t''_i$ *for* $r''_i$. *Then, the forest* $t'_i, t''_i$ *is a witness for* $r_i$.

*Therefore, for each* $r_i$ *there is a witness* $t_i$. *Then the tree* $a[t_1 \ \dots \ t_n]$ *is a witness of the rule* $X \to a[r_1 \dots r_n]$.

□

COROLLARY (5.9) *Let* $(\{X\}, E)$ *be a non-recursive,*$*$*-guarded, parent-unambiguous local tree grammar and t be its witness. Let* $\{Y_1 \dots, Y_n\} \subseteq \textbf{\textit{Dn}}(E)$. *If* $Y_1 \Rightarrow_E \dots \Rightarrow_E Y_n$, *then there exists* $\{\mathbf{i}_1, \dots, \mathbf{i}_n\} \subseteq \textbf{\textit{Ids}}(t)$ *such that*

$$\forall i \in \{2 \dots n\}, ((\mathbf{i}_{i-1}, \mathbf{i}_i) \in \mathscr{E}(t)) \land \Im(id_{i-1}) = Y_{i-1} \land \Im(\mathbf{i}_i) = Y_i$$

**Proof 14** *This is a direct application of Lemma 5.8. We know that for all* $Y \in \textbf{\textit{Dn}}(E), \exists \mathbf{i} \in \textbf{\textit{Ids}}(t)$ *such that* $\Im(\mathbf{i}) = Y$. *This is true in particular for* $\{Y_1, \dots, Y_n\}$. *Consider* $Y_i$ *and* $Y_{i+1}$. *We have* $Y_i \Rightarrow_E Y_{i+1}$ *which means that in E, there is a rule* $Y_i \to a[r_i]$ *for some label a and with* $Y_{i+1} \in r_i$. *Consequently,* $t@\mathbf{i}_i = a[\dots, id_{i+1}, \dots]$. *Therefore,* $(\mathbf{i}_i, \mathbf{i}_{i+1}) \in \mathscr{E}(t)$. □

THEOREM (COMPLETENESS OF TYPE INFERENCE (5.10)) *Let* $(\mathscr{S}, E)$ *be a* $*$*-guarded non-recursive and parent unambiguous local tree grammar, and P a path. Let*

$$E_0 = \{X \to R \mid X \to R \in E,\ X \in \mathscr{S}\}.$$

*If* $(E_0, E_0) \vdash_E P : (\tau, \kappa)$ *then:*

$$\textbf{\textit{Dn}}(\tau) \subseteq \bigcup_{t \in_{\Im} E} \Im([\![P]\!]_t(\textbf{\textit{RootId}}(t)))$$

**Proof 15** *Like for the proof of Theorem 5.6, we consider the following, more general judgment:*

$$(\tau, \kappa) \vdash_E P : (\tau', \kappa')$$

*let t be the witness of E. We show that if* $\textbf{\textit{Dn}}(\tau) \subseteq \Im(S)$ *then,* $\textbf{\textit{Dn}}(\tau') \subseteq \Im([\![P]\!]_t(S))$. *If this holds for the witness t then it holds for the union of all trees* $\Im$*-valid w.r.t to E (which contains t). Informally, this means that if the type* $\tau$ *"describes precisely" the nodes in S, that is, if there are no unneeded rules in* $\tau$, *then the type* $\tau'$ *describes exactly the result of the query: for each rule in* $\tau'$, *there is a node in the result of the query typed by that rule. We proceed by induction on the depth of the typing derivation:*

**Basic case:**

**(down-axis):** <u>`self` *axis:*</u> *We supposed* $\boldsymbol{Dn}(\tau) \subseteq \Im(S)$. *We have* $\tau' = \mathbf{A}_E(\tau, \texttt{self}) = \tau$. *We also have:*

$$[\![\texttt{self} :: \texttt{node}]\!]_t(S) = S$$

*by Definition 4.1. Therefore,* $\boldsymbol{Dn}(\tau') \subseteq \Im(S)$ *and so*

$$\boldsymbol{Dn}(\tau') \subseteq \Im([\![\texttt{self} :: \texttt{node}]\!]_t(S))$$

<u>`descendant` *axis:*</u> *Let us consider names,* $X \in \boldsymbol{Dn}(\tau)$ *and* $Y \in \mathbf{A}_E(\{X\}, \texttt{descendant})$*.By Definition 5.1, we have that* $X \Rightarrow_E^* Y$*. By using Corollary 5.9, we have that there exists a sequence:* $\mathbf{i}_1, \ldots, \mathbf{i}_n$ *in* $t$ *such that* $X = \Im(\mathbf{i}_1)$ *and* $Y = \Im(\mathbf{i}_n)$*. We also have that*

$$\forall i \in \{1 \ldots n-1\},\ (\mathbf{i}_i, \mathbf{i}_{i+1}) \in \mathscr{E}(t))$$

*thus* $(\mathbf{i}_1, \mathbf{i}_n) \in \mathscr{E}^+(t)$ *and therefore that*

$$\mathbf{i}_n \in [\![\texttt{descendant} :: \texttt{node}]\!]_t(\{\mathbf{i}_1\})$$

*Subsequently:*

$$\mathbf{A}_E(\{X\}, \texttt{descendant}) \subseteq \Im([\![\texttt{descendant} :: \texttt{node}]\!]_t(\{\mathbf{i}_1\}))$$

<u>`child` *axis:*</u> *is a particular instance of the previous case.*

**(up-axis)** *We only treat the case of the* `ancestor` *axis, of which the* `parent` *axis is a particular instance. This case is the symmetric of the* `descendant` *axis. Let* $X \in \boldsymbol{Dn}(\tau)$*. Let* $Y \in \mathbf{A}_E(\{X\}, \texttt{ancestor}) \cap \kappa$*. By Definition 5.1, we have that* $Y \Rightarrow_E^* X$ *(and because*$\kappa$ *is a well-formed context). By using Corollary 5.9, we have that there exists a sequence:* $\mathbf{i}_1, \ldots, \mathbf{i}_n$ *in* $t$ *such that* $Y = \Im(\mathbf{i}_1)$ *and* $X = \Im(\mathbf{i}_n)$*. We also have*

$$\forall i \in \{1 \ldots n-1\}, (\mathbf{i}_i, \mathbf{i}_{i+1}) \in \mathscr{E}(t))$$

*thus* $(\mathbf{i}_1, \mathbf{i}_n) \in \mathscr{E}^+(t)$ *and therefore*

$$\mathbf{i}_n \in [\![\texttt{ancestor} :: \texttt{node}]\!]_t(\{\mathbf{i}_1\})$$

*Thus, we have*

$$\mathbf{A}_E(\{X\}, \texttt{ancestor}) \subseteq \Im([\![\texttt{ancestor} :: \texttt{node}]\!]_t(\{\mathbf{i}_n\}))$$

*We must also show that if* $Y \in \kappa$ *then*
$Y \in \Im([\![\texttt{ancestor} :: \texttt{node}]\!]_t(\{\mathbf{i}_n\}))$
*(because the output type is intersected with the context for this rule). This is an immediate consequence of the well-formedness of contexts.* $\kappa$ *is well-formed only if* $\kappa \in \tau \cup \mathbf{A}_E(\tau, \texttt{ancestor})$*.*

**(test):** *is similar to the case* `self` *of Rule* ***(down-axis)****.*

**Inductive case:**

**(predicate)** *Suppose* $X_i \in \boldsymbol{Dn}(\tau)$ *(there is a unique rule with* $X_i$ *as left hand side, since we consider a local tree grammar). We consider the premise:*

$$(\{X_i \to R_i\}, \kappa) \vdash_E P_{jk} : \Sigma^{ijk}$$

*Let us consider a set $S_i \subseteq S$. By induction hypothesis, if $\{X_i\} \subseteq \mathfrak{I}(S_i)$ then $\Sigma_{typ}^{ijk} \subseteq \mathfrak{I}(\llbracket P_{jk} \rrbracket_t(S_i))$. There are two cases.*

*Either $\bigcup_j \bigcap_k \Sigma_{typ}^{ijk} = \varnothing$. Then, $X_i \notin \textbf{Dn}(\tau')$ (the output type). But if this is the case, because the type-system is sound (cf. Theorem 5.6), then:*
*$\forall \mathbf{i}_i$ such that $\mathfrak{I}(\mathbf{i}_i) = X_i$, $\llbracket P_{jk} \rrbracket_t(\{\mathbf{i}_i\}) = \varnothing$ and therefore*

$$\Sigma_{typ}^{ijk} = \llbracket P_{jk} \rrbracket_t(\{\mathbf{i}_i\}) = \varnothing$$

*Or $\bigcup_j \bigcap_k \Sigma_{typ}^{ijk} \neq \varnothing$ and since $\Sigma_{typ}^{ijk} \subseteq \mathfrak{I}(\llbracket P_{jk} \rrbracket_t(S_i))$, this means that $\llbracket P_{jk} \rrbracket_t(S_i) \neq \varnothing$ and therefore that*

$$S_i \subseteq \llbracket \texttt{self::node[}Cond\texttt{]} \rrbracket_t(S)$$

*Since $X_i \in \mathfrak{I}(S_i)$, $X_i \in \mathfrak{I}(\llbracket \texttt{self::node[}Cond\texttt{]} \rrbracket_t(S))$. Lastly, we remark that for each $X_i$ the set $S_i$ is not empty. This is a consequence of Lemma 5.8, for each name $X_i$, there is a node $\mathbf{i}_i$ in the witness.*

**(sequence)** *: By applying straightforwardly the induction hypothesis on the premises.*

**(union)** *: By applying straightforwardly the induction hypothesis on the premises.*

□

LEMMA (TERMINATION OF TYPE-PROJECTOR INFERENCE (5.12)) *Let $(\mathscr{S}, E)$ be a type, $P$ a path, and $\Sigma$ and $\Sigma'$ environments. The judgment $\Sigma \Vdash_E P : \Sigma'$ has a unique and finite derivation.*

**Proof 16** *The uniqueness of the derivation follows from the fact that all the rules are mutually exclusive (although not strictly syntax directed) thanks to their side conditions.*

*To prove termination, we need some more care than for the type inference algorithm. For the judgment:*

$$\Sigma \Vdash_E P : \Sigma'$$

*we give it as weight the triple $(l(P), r(P), |\Sigma_{typ}|)$ ordered lexicographically, where:*

$l(P)$ *is the length of the path $P$, as defined previously*

$r(P)$ *is the number of occurrences of a recursive step, that is the number of occurrences of* `descendant::node` *or* `ancestor::node` *in the $P$*

$|\Sigma_{\mathbf{typ}}|$ *is the number of rules in the input*

*The proof is straightforward and consists that for every rule the weight strictly decreases in the premises:*

**Basic case:** *the base of induction is an application of* ***(p-step)*** *or* ***(p-erase)*** *does not have any premises.*

**Inductive case:** *For the rules* ***(p-union)***, ***(p-test)*** *and* ***(p-predicate)***, *the weight strictly decreases in $l(P)$ in the premises. For the rule* ***(p-iterate)***, *$|\Sigma_{typ}|$ strictly decreases in the premises, since in the conclusion the weight has at least two for this component and exactly one in each of the premises. Also, $P$ is unchanged in the premises therefore $l(P)$ and $r(P)$ do not increase. For the rule* ***(p-many)***, *the $l(P)$ part is unchanged in the premises since $l(\texttt{descendant::node}/P) = l(\texttt{child::node}/P) = 1 + l(P)$ and $r(P)$ decreases strictly.*

□

LEMMA (WELL-FORMEDNESS OF TYPE-PROJECTOR INFERENCE (5.13)) *Let* $(\mathcal{S}, E)$ *be a type,* $\tau$*,* $\tau'$*, and* $\kappa$ *sets of rules, and P a path. If* $(\tau, \kappa) \Vdash_E P : \tau'$*, then* $(\tau, \kappa) \vdash_E P : (\tau'', \kappa'')$ *implies* $\kappa'' \subseteq \tau'$*.*

**Proof 17** *We use a structural induction on the derivation of* $(\tau, \kappa) \Vdash_E P : \tau'$ *which is finite by Lemma 5.12.*

**Basic case:** *The property is trivially true for the rule* ***(p-step)*** *since the result is the union of the output type and its associated context.*

*Rule* ***(p-erase)*** *can only be applied if the side conditions of the other rules fail, which means in the case where the judgment* $(\tau, \kappa) \vdash_E P : (\tau'', \kappa'')$ *does not hold. Therefore the lemma is true too in that case.*

**Inductive case:**

**(p-union)** *We suppose* $(\tau, \kappa) \vdash_E P_1 | P_2 : (\tau'', \kappa'')$*. This means that the typing rule* ***(union)*** *(*cf. *Figure 1) holds and that:*

$$(\tau, \kappa) \vdash_E P_1 : (\tau_1'', \kappa_1'')$$

*and*

$$(\tau, \kappa) \vdash_E P_2 : (\tau_2', \kappa_2'')$$

*let* $P_1$ *produce a type-projector* $\tau_1'$ *and* $P_2$ *produce* $\tau_2'$*; by induction hypothesis* $\kappa_1'' \subseteq \tau_1'$ *and* $\kappa_2'' \subseteq \tau_2'$*. But since* $\kappa'' = \kappa_1'' \cup \kappa_2''$*, we have* $\kappa'' \subseteq \tau_1' \cup \tau_2' = \tau'$

**(p-iterate)** *similar to the previous case.*

**(p-test)** *we suppose*

$$(\{Y \to R\}, \kappa) \vdash_E \texttt{self}::Test/P : (\tau'', \kappa'')$$

*According to the* ***(test)*** *typing rule, this means that:*

$$(\{Y \to R\}, \kappa) \vdash_E \texttt{self}::Test : (\tau_1, \kappa_1)$$

*and*

$$(\tau_1, \kappa_1) \vdash_E P : (\tau'', \kappa'')$$

*the induction hypothesis can be applied on the second premise of the rule* ***(p-test)*** *and we have* $(\tau_1, \kappa_1) \Vdash_E P : \tau'$ *with* $\kappa'' \subseteq \tau'$*. Since* $\tau' \subseteq \{Y \to R\} \cup \tau'$*, we have* $\kappa'' \subseteq \{Y \to R\} \cup \tau'$ *which proves this case.*

**(p-predicate)** *similar to the previous case, we can observe that the context resulting of the typing of the first step is passed as argument for the inference of the projector of the remainder of the path.*

**(p-single)** *similar to the previous case.*

**(p-many)** *we only treat the case for Axis* $=$ `descendant`*, the case for* `ancestor` *being similar. We suppose:*

$$(\{Y \to R\}, \kappa) \vdash_E \texttt{descendant::node}/P : (\tau_0, \kappa_0) \ \textbf{\textit{(*)}}$$

*and we want to show that* $\kappa_0 \subseteq \{Y \to R\} \cup \tau' \cup \tau''$*. Let us write:*

$$(\{Y \to R\}, \kappa) \vdash_E \texttt{descendant::node} : (\tau_1, \kappa_1) \ \textbf{\textit{(1)}}$$
$$(\{Y \to R\} \cup \tau_1, \kappa) \vdash_E \texttt{child::node} : (\tau_2, \kappa_2) \ \textbf{\textit{(2)}}$$

*then* $\tau_1 = \tau_2$ *an* $\kappa_1 = \kappa_2$. *Indeed:*

$$\textbf{(1)}\begin{cases} \tau_1 = \mathbf{A}_E(\{Y \to R\}, \texttt{descendant}) \\ \kappa_1 = \tau_1 \cup \kappa \end{cases}$$
$$\textbf{(2)}\begin{cases} \tau_2 = \mathbf{A}_E(\{Y \to R\} \cup \tau_1, \texttt{child}) \\ \kappa_2 = \tau_2 \cup \kappa \end{cases}$$

*by definition of* $\mathbf{A}_(_,_)$*:*

$$\textbf{(1)}\begin{cases} \tau_1 = \{Y' \to R' \mid Y \Rightarrow_E^+ Y'\} \\ \kappa_1 = \tau_1 \cup \kappa \end{cases}$$
$$\textbf{(2)}\begin{cases} \tau_2 = \bigcup_{\{Y' \to R' \mid Y \Rightarrow_E^+ Y'\} \cup \{Y \to R\}} \{Y'' \to R'' \mid Y' \Rightarrow_E Y''\} \\ \kappa_2 = \tau_2 \cup \kappa \end{cases}$$

*by transitivity of* $\Rightarrow_E$, **(2)** *becomes:*

$$\tau_2 = \bigcup_{\{Y' \to R' \mid Y \Rightarrow_E^* Y'\}} \{Y'' \to R'' \mid Y' \Rightarrow_E Y''\}$$
$$\tau_2 = \{Y' \to R' \mid Y \Rightarrow_E^+ Y'\} = \tau_1$$
$$\kappa_2 = \tau_2 \cup \kappa = \tau_1 \cup \kappa = \kappa_1$$

*therefore, for the path P*

$$(\{Y \to R\} \cup \tau_1, \kappa) \vdash_E \texttt{child::node}/P : (\tau_0, \kappa_0)$$

*by application of the* **(sequence)** *typing rule. We can finally remark that*

$$((\{Y \to R\} \cup \tau_1) \cap \tau', \kappa) \vdash_E \texttt{child::node}/P : (\tau_0, \kappa_0)$$

*all the rules ins* $\tau_1 \setminus \tau'$ *yield an empty projector. Therefore*

$$(\tau', \kappa) \vdash_E \texttt{child::node}/P : (\tau_0, \kappa_0)$$

*which allows us to apply the induction hypothesis on the third premise of* ***(p-many)****, which gives us* $\kappa_0 \subseteq \tau''$*, and therefore:* $\kappa_0 \subseteq \{Y \to R\} \cup \tau' \cup \tau''$

□

THEOREM (SOUNDNESS OF TYPE-PROJECTOR INFERENCE (5.14)) *Let* $(\mathscr{S}, E)$ *be a type and P an XPath*$^\ell$ *query. Let S be the set of rules:* $S = \{X \to R \mid X \in \mathscr{S}\}$. *If*

$$(S, S) \Vdash_E P : \tau$$

*then* $\tau$ *is a type-projector for* $(\mathscr{S}, E)$ *and for every* $t \in_{\mathfrak{I}} (\mathscr{S}, E)$ *we have:*

$$[\![P]\!]_{t \setminus_{\mathfrak{I}} \tau}(\textbf{\textit{RootId}}(t)) = [\![P]\!]_t(\textbf{\textit{RootId}}(t))$$

**Proof 18** *By simple structural induction on the path.* □

THEOREM (COMPLETENESS OF PROJECTOR INFERENCE (5.16)) *Let* $(\mathscr{S}, E)$ *be a* $*$*-guarded, non-recursive, and parent-unambiguous local tree grammar, and P a strongly-specified XPath*$^{\ell}$ *path. Let S be the set of rules:* $S = \{X \to R \mid X \in \mathscr{S}\}$. *If*

$$(S,S) \Vdash_E P : \tau$$

*then there exists* $t \in_{\mathfrak{I}} (\mathscr{S}, E)$ *such that for each* $Y \to R \in \tau$, *if* $\pi = \tau \setminus (\{Y \to R\} \cup \mathbf{A}_E(\{Y \to R\}, \mathtt{descendant}))$, *then:*

$$[\![P]\!]_{t \setminus_{\mathfrak{I}} \pi}(\textbf{\textit{RootId}}(t)) \neq [\![P]\!]_t(\textbf{\textit{RootId}}(t))$$

**Proof 19** *By induction on the length of the typing derivation which is finite. We use Theorem 5.10 to show that if we remove a name Y inferred by the type inference algorithm, then we remove nodes from the result of the query applied to the projected document. The fact that P is strongly specified is used for the treatment of predicates. Indeed, it forces any path in a predicate to be matched exactly by one node. If a path in a predicate could be matched by two (or more) nodes, then removing one of the nodes would not change the semantics of the query, since there would still be a node present to make the predicate succeed. We illustrate this in the example hereafter.* □

# B Text of the XMark and XPathMark queries

Base and induction

**(p-step)** $$\frac{\Sigma \vdash_E \mathit{Step} : (\tau, \kappa)}{\Sigma \Vdash_E \mathit{Step} : \tau \cup \kappa} \text{ if } \tau \neq \varnothing$$

**(p-union)** $$\frac{\Sigma \Vdash_E P_1 : \tau_1 \qquad \Sigma \Vdash_E P_2 : \tau_2}{\Sigma \Vdash_E P_1 | P_2 : \tau_1 \cup \tau_2}$$

**(p-erase)** $$\frac{}{\Sigma \Vdash_E P : \varnothing} \text{ if no other rule applies}$$

**(p-iterate)** $$\frac{(\{X_1 \to R_1\}, \kappa) \Vdash_E P : \tau_1 \qquad \dots \qquad (\{X_n \to R_n\}, \kappa) \Vdash_E P : \tau_n}{(\{X_1 \to R_1, \dots, X_n \to R_1\}, \kappa) \Vdash_E P : \bigcup_{i=1..n} \tau_i} \text{ if } n \geq 2$$

Path Rules

**(p-test)** $$\frac{(\{Y \to R\}, \kappa) \vdash_E \texttt{self} :: \mathit{Test} : \Sigma \qquad \Sigma \Vdash_E P : \tau}{(\{Y \to R\}, \kappa) \Vdash_E \texttt{self} :: \mathit{Test}/P : \{Y \to R\} \cup \tau} \text{ if } \Sigma_{\mathsf{typ}} \neq \varnothing$$

**(p-predicate)** $$\frac{(\{Y \to R\}, \kappa) \vdash_E \texttt{self} :: \texttt{node}[\, \mathop{\texttt{Or}}_i \mathop{\texttt{And}}_j P_{ij}\,] : \Sigma \qquad \begin{array}{l} \Sigma \Vdash_E P : \tau \\ \Sigma \Vdash_E P_{ij} : \tau_{ij} \quad \text{if } \Sigma_{\mathsf{typ}} \neq \varnothing \end{array}}{(\{Y \to R\}, \kappa) \Vdash_E \texttt{self} :: \texttt{node}[\, \mathop{\texttt{Or}}_i \mathop{\texttt{And}}_j P_{ij}\,]/P : \{Y \to R\} \cup \tau \cup \bigcup_i \bigcup_j \tau_{ij}}$$

**(p-single)** $$\frac{\begin{array}{c} (\{Y \to R\}, \kappa) \vdash_E \mathit{Axis} :: \texttt{node} : (\tau, \kappa') \\ \text{(for } i = 1..n\text{)} \quad (\{X_i \to R_i\}, \kappa') \vdash_E P : \Sigma^i \quad (\tau', \kappa') \Vdash_E P : \tau'' \end{array}}{(\{Y \to R\}, \kappa) \Vdash_E \mathit{Axis} :: \texttt{node}/P : \{Y \to R\} \cup \tau' \cup \tau''} \; (*)$$

(*) where $\mathit{Axis} \in \{\texttt{parent}, \texttt{child}\}$, $\tau = \{X_1 \to R_1, \dots, X_n \to R_n\}$, $\tau' = \{X_i \to R_i \mid i = 1..n,\ \Sigma^i_{\mathsf{typ}} \neq \varnothing\}$, $\tau \neq \varnothing$, and $\tau' \neq \varnothing$

**(p-many)** $$\frac{\begin{array}{c} (\{Y \to R\}, \kappa) \vdash_E \mathit{Axis} :: \texttt{node} : (\tau, \kappa') \\ \text{(for } i = 1..n\text{)} \quad (\{X_i \to R_i\}, \kappa') \vdash_E \mathit{Axis} :: \texttt{node}/P : \Sigma^i \quad (\tau', \kappa') \Vdash_E \mathsf{s}(\mathit{Axis}) :: \texttt{node}/P : \tau'' \end{array}}{(\{Y \to R\}, \kappa) \Vdash_E \mathit{Axis} :: \texttt{node}/P : \{Y \to R\} \cup \tau' \cup \tau''} \; (**)$$

(**) where $\mathit{Axis} \in \{\texttt{ancestor}, \texttt{descendant}\}$, $\tau = \{X_1 \to R_1, \dots, X_n \to R_n\}$, $\tau' = \{Y \to R\} \cup \{X_i \to R_i \mid i = 1..n,\ \Sigma^i_{\mathsf{typ}} \neq \varnothing\}$, $\tau \neq \varnothing$, $\tau' \neq \varnothing$, $\mathsf{s}(\texttt{descendant}) = \texttt{child}$, and $\mathsf{s}(\texttt{ancestor}) = \texttt{parent}$.

Figure 2: Inference rules for type-projectors

$$
\begin{aligned}
\mathbf{F}(\texttt{count}(Q)) &= \mathbf{P}(Q)\\
\mathbf{F}(\texttt{last}(Q)) &= \mathbf{P}(Q)\\
\mathbf{F}(\texttt{position}(Q)) &= \mathbf{P}(Q)\\
\mathbf{F}(\texttt{string}(Q)) &= \mathbf{P}(Q)/\texttt{descendant-or-self}::\texttt{node}[\texttt{text}()]\\
\mathbf{F}(\texttt{number}(Q)) &= \mathbf{P}(Q)/\texttt{descendant-or-self}::\texttt{node}[\texttt{text}()]\\
\mathbf{F}(\texttt{not}(Q)) &= \mathbf{P}(Q)\\
\mathbf{F}(\texttt{true}()) &= \{\texttt{self}::\texttt{node}\}\\
\mathbf{F}(\texttt{false}()) &= \{\texttt{self}::\texttt{node}[\texttt{self}::a \text{ and } \texttt{self}::b]\}\\
\mathbf{F}(f(v_1,\ldots,v_n)) &= \{\texttt{self}::\texttt{node}\} \quad \text{where } v_i \text{ is a value}
\end{aligned}
$$

Figure 3: Approximation of XPath functions

1. $\mathbf{E}((),\Gamma,m) = \varnothing$
2. $\mathbf{E}(v,\Gamma,m) = \varnothing$
3. $\mathbf{E}((q_1,q_2),\Gamma,m) = \mathbf{E}(q_1,\Gamma,m) \cup \mathbf{E}(q_2,\Gamma,m)$
4. $\mathbf{E}(\texttt{<tag>}q\texttt{</tag>},\Gamma,m) = \mathbf{E}(q,\Gamma,m)$
5. $\mathbf{E}(x,\Gamma,1) = \bigcup_{(x;P)\in\Gamma} \{P/\texttt{descendant-or-self}::\texttt{node}\}$
6. $\mathbf{E}(x,\Gamma,0) = \bigcup_{(x;\,P)\in\Gamma} \{P\}$
7. $\mathbf{E}(Path,\Gamma,1) = \{Path/\texttt{descendant-or-self}::\texttt{node}\}$
8. $\mathbf{E}(Path,\Gamma,0) = \{Path\}$
9. $\mathbf{E}(FLOWR/P,\Gamma,m) = \mathbf{E}(FLOWR,\Gamma,m)/\mathbf{E}(P,\Gamma,m)$
10. $\mathbf{E}(Step/P,\Gamma,m) = Step/\mathbf{E}(P,\Gamma,m)$
11. $\mathbf{E}(Step[Cond]/P,\Gamma,m) = \left(\bigcup_{q\in\mathbf{E'}(Cond,\Gamma,m)} Step[q]\right)/\mathbf{E}(P,\Gamma,m)$
12. $\mathbf{E}(\texttt{if } q \texttt{ then } q_1 \texttt{ else } q_2,\Gamma,m) = \mathbf{E}(q,\Gamma,0) \cup \mathbf{E}(q_1,\Gamma,m) \cup \mathbf{E}(q_2,\Gamma,m)$
13. $\mathbf{E}(\texttt{let \$}x \texttt{ := } q_1 \texttt{ return } q_2,\Gamma,m) = \mathbf{E}(q_2,\Gamma\cup\Gamma',m)$ where $\Gamma' = \{(x;P) \mid P \in \mathbf{E}(q_1,\Gamma,0)\}$
14. $\mathbf{E}(\texttt{for \$}x \texttt{ in } q_1 \texttt{ return } q_2,\Gamma,m) = \mathbf{E}(q_1,\Gamma,0) \cup \mathbf{E}(q_2,\Gamma\cup\Gamma',m)$ where $\Gamma' = \{(x;P) \mid P \in \mathbf{E}(q_1,\Gamma,0)\}$

1′. $\mathbf{E'}(Cond_1\ op\ Cond_2,\Gamma,m) = \bigcup_{q\in\mathbf{E'}(Cond_1,\Gamma,m)} \bigcup_{q'\in\mathbf{E'}(Cond_2,\Gamma,m)} \{q\ op\ q'\}$ where $op \in \{\texttt{and},\texttt{or}\}$

2′. $\mathbf{E'}(Expr_1\ cmp\ Expr_2,\Gamma,m) = \bigcup_{q\in\mathbf{E'}(Expr_1,\Gamma,m)} \bigcup_{q'\in\mathbf{E'}(Expr_2,\Gamma,m)} \{q\ cmp\ q'\}$ where $cmp \in \{\texttt{=},\texttt{!=},\texttt{<},\texttt{>},\texttt{>=},\texttt{<=}\}$

3′. $\mathbf{E'}(Arith_1\ op\ Arith_2,\Gamma,m) = \bigcup_{q\in\mathbf{E'}(Arith_1,\Gamma,m)} \bigcup_{q'\in\mathbf{E'}(Arith_2,\Gamma,m)} \{q\ op\ q'\}$ where $op \in \{\texttt{+},\texttt{-},\texttt{*},\texttt{div},\texttt{mod}\}$

4′. $\mathbf{E'}(f(Expr_1,\ldots,Expr_n),\Gamma,m) = \bigcup_{q_1\in\mathbf{E'}(Expr_1,\Gamma,\mathbf{M}(f,1))} \cdots \bigcup_{q_n\in\mathbf{E'}(Expr_n,\Gamma,\mathbf{M}(f,n))} \{f(q_1,\ldots,q_n)\}$

5′. $\mathbf{E'}(Atom,\Gamma,m) = \mathbf{E}(Atom,\Gamma,m) \quad Atom \neq f(q_1,\ldots,q_n)$

Figure 4: XQuery path extraction

$$
\begin{aligned}
\mathbf{M}(\texttt{count},1) &= 0 & \mathbf{M}(\texttt{string},1) &= 1\\
\mathbf{M}(\texttt{last},1) &= 0 & \mathbf{M}(\texttt{number},1) &= 1\\
\mathbf{M}(\texttt{position},1) &= 0 & \mathbf{M}(\texttt{not},1) &= 0
\end{aligned}
$$

Figure 5: Value of the parameter *m* for various built-in XPath functions

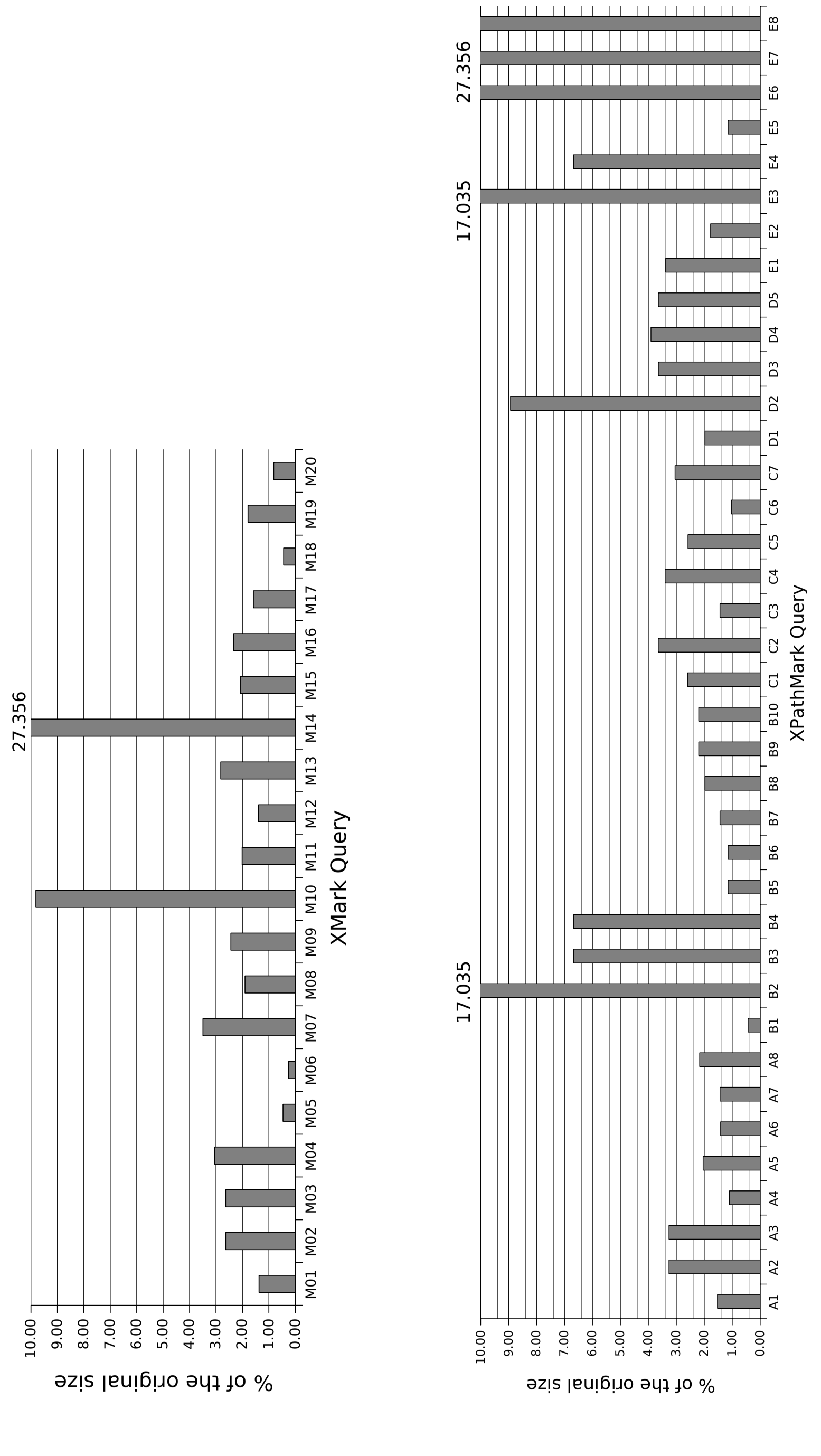

Figure 6: Pruning ratio in percent of the original file size, for XMark and XPathMark queries

| | A 1 | A 6 | B 1 | B 2 | C 3 | C 4 | D 1 | D 2 | E 5 | E 7 | M 3 | M 6 | M 7 | M 14 | M 15 |
|---|---|---|---|---|---|---|---|---|---|---|---|---|---|---|---|
| *(i)* | 3363 | 3363 | 3363 | 3363 | 3363 | 3363 | 3363 | 3363 | 3363 | 2242 | 3363 | 3363 | 3363 | 2242 | 3363 |
| *(ii)* | 447 | 67.6 | 50.5 | 571 | 68.3 | 137 | 65.5 | 297.25 | 60.8 | 605 | 92.8 | 9 | 121 | 605 | 67.6 |
| *(iii)* | 10 | 9 | 3 | 113 | 9 | 23 | 13 | 59 | 12 | 190 | 18 | 2 | 24 | 190 | 65 |
| *(iv)* | 20 | 17 | 22 | 2.9 | 17 | 5.8 | 12.1 | 5.6 | 12.3 | 2.1 | 9 | 15.8 | 14.3 | 2.19 | 7.5 |
| *(v)* | 3.7 | 5.5 | 2.2 | 22.2 | 3.7 | 8 | 5.4 | 9.7 | 7.44 | 22 | 9 | 1.8 | 4.9 | 29 | 15.2 |

*(i)*: Largest queryable document (MB). We stopped our testing at 3363 MB
*(ii)*: Pruned size (MB)
*(iii)*: Pruned size for 671 MB (MB)
*(iv)*: Speed up (× faster)
*(v)*: Memory use in % of original

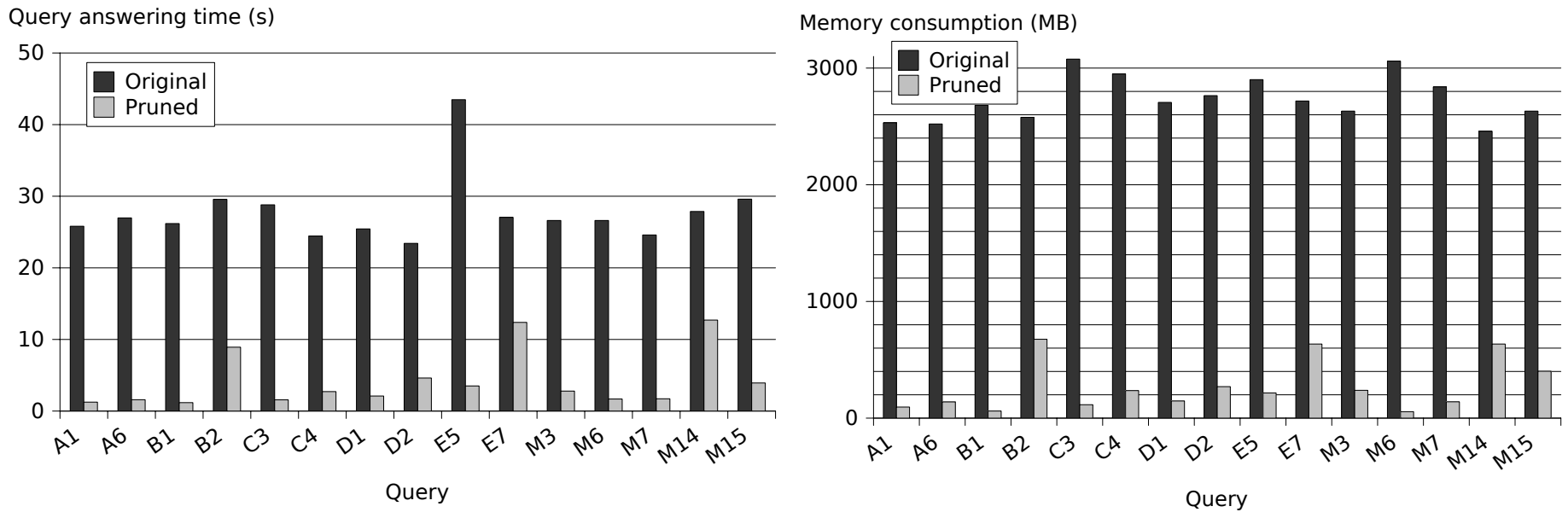

Figure 7: Experimental results for the Saxon-b/XQuery engine

| | A1 | A6 | B1 | B2 | C3 | C4 | D1 | D2 | E5 | E7 | M3 | M6 | M7 | M14 | M15 |
|---|---|---|---|---|---|---|---|---|---|---|---|---|---|---|---|
| *(i)* | 675 | 119 | 95 | 843 | 117 | 232 | 93 | 443 | 116 | 1073 | 138 | 14 | 111 | 1073 | 436 |
| *(ii)* | 0.5 | 0.5 | 0.3 | 24.8 | 0.5 | 45.4 | 10.0 | 100.0 | - | 59 | 1.1 | 84.7 | 50.8 | 54.0 | 94.0 |
| *(iii)* | 6.8 | 10.5 | 8.5 | 3.3 | 18.1 | 2.1 | 21.6 | 1.0 | - | 1.7 | 5.0 | 2.8 | 1.7 | 2.06 | 1.0 |

*(i)*: Size of the index (MB)
*(ii)*: Amount of I/O (% of the original)
*(iii)*: Speed up (× faster)

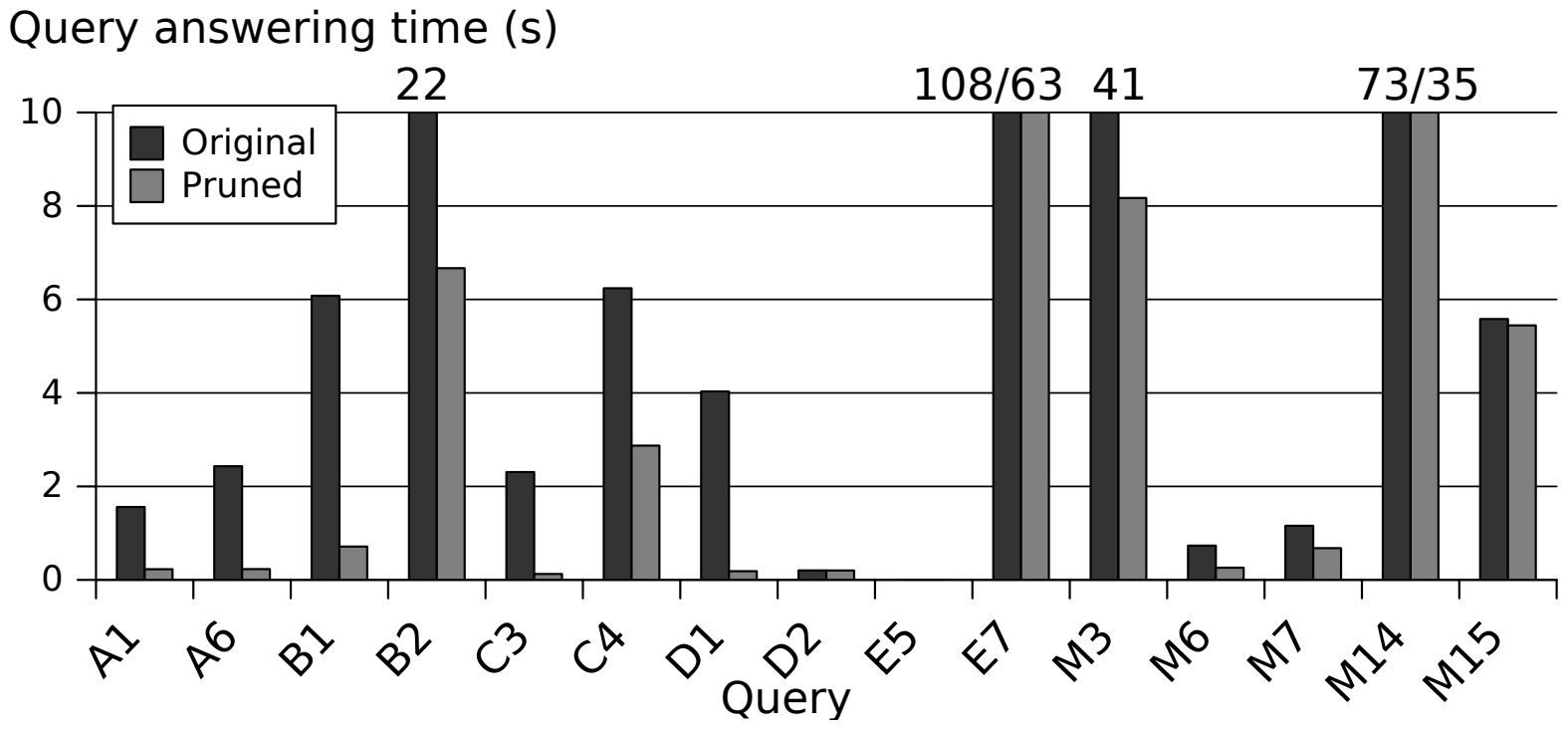

Figure 8: Experimental results for the MonetDB engine

A 1 `/site/closed_auctions/closed_auction/annotation/description/text/keyword`

A 6 `/site/people/person[profile/gender and profile/age]/name`

B 1 `/site/regions/*/item[parent::namerica or parent::samerica]/name`

B 2 `//keyword/ancestor::listitem/text/keyword`

C 3
```
/site/people/person[profile/@income =
                              /site/open_auctions/open_auction/current]/name
```

C 4 `/site/people/person[watches/watch/id(@open_auction)/seller/@person = @id]/name`

D 1 `/site/open_auctions/open_auction[(count(bidder) mod 2) = 0]/interval`

D 2 `count(//text) + count(//bold) + count(//emph) + count(//keyword)`

E 5 `/site/regions/*/item[preceding::item[100] and following::item[100]]/name`

E 7
```
/site/regions/*/item[contains(substring-before(description,'eros'),'passion')
        and contains(substring-after(description, 'eros'), 'dangerous')]/name
```

M 3
```
for $b in $doc/site/open_auctions/open_auction
 where zero-or-one($b/bidder[1]/increase/text()) * 2
                                         <= $b/bidder[last()]/increase/text()
    return
    <increase
    first="$b/bidder[1]/increase/text()"
    last="$b/bidder[last()]/increase/text()"/>
```

M 6 `for $b in $doc//site/regions return count($b//item)`

M 7
```
for $p in $doc/site
return
 count($p//description) + count($p//annotation) + count($p//emailaddress)
```

M 14
```
for $i in $doc/site//item
                 where contains(string(exactly-one($i/description)), "gold")
return $i/name/text()
```

M 15
```
for $a in
     $doc/site/closed_auctions/closed_auction/annotation/description/parlist/
                           listitem/parlist/listitem/text/emph/keyword/text()
return <text>$a</text>
```

Figure 9: XPathMark (A-E) and XMark (M) queries